\documentclass[3p,review]{elsarticle}
\usepackage{lineno}
\modulolinenumbers[1]

\usepackage{multicol}
\usepackage{color}
\usepackage{xspace}
\usepackage{pdfwidgets}
\usepackage{natbib}
\usepackage{amsmath}
\newcommand{\average}[1]{\{\!\!\{{#1}\}\!\!\}}
\newcommand{\jump}[1]{[\![{#1}]\!]}
\usepackage{graphicx}
\usepackage{array}
\usepackage{multirow}
\usepackage{color}
\usepackage[normalem]{ulem}
\usepackage{chngcntr}

\usepackage{subfigure}

\graphicspath{ {./Figures/} }

\newcommand*\patchAmsMathEnvironmentForLineno[1]{%
  \expandafter\let\csname old#1\expandafter\endcsname\csname #1\endcsname
  \expandafter\let\csname oldend#1\expandafter\endcsname\csname end#1\endcsname
  \renewenvironment{#1}%
     {\linenomath\csname old#1\endcsname}%
     {\csname oldend#1\endcsname\endlinenomath}}%
\newcommand*\patchBothAmsMathEnvironmentsForLineno[1]{%
  \patchAmsMathEnvironmentForLineno{#1}%
  \patchAmsMathEnvironmentForLineno{#1*}}%
\AtBeginDocument{%
\patchBothAmsMathEnvironmentsForLineno{equation}%
\patchBothAmsMathEnvironmentsForLineno{align}%
\patchBothAmsMathEnvironmentsForLineno{flalign}%
\patchBothAmsMathEnvironmentsForLineno{alignat}%
\patchBothAmsMathEnvironmentsForLineno{gather}%
\patchBothAmsMathEnvironmentsForLineno{multline}%
}

\newcommand{\oldrevision}[1]{\textcolor{black}{#1}}

\newcommand{\revision}[1]{\textcolor{black}{#1}}
\newcommand{\RLG}[1]{\textcolor{black}{#1}}

\begin{document}

\begin{frontmatter}
\title{Generalized Multiscale Finite-Element Method (GMsFEM) for elastic wave propagation in heterogeneous, anisotropic media}
\author[lanl]{Kai Gao\corref{corresponding}}
\ead{kaigao87@gmail.com}
\cortext[corresponding]{Corresponding author, }

\author[tamumath]{Shubin Fu}
\ead{shubinfu89@gmail.com}

\author[tamugeo]{Richard L. Gibson Jr.}
\ead{gibson@tamu.edu}

\author[cuhk]{Eric T. Chung}
\ead{tschung@math.cuhk.edu.hk}

\author[tamumath,kaust]{Yalchin Efendiev}
\ead{efendiev@math.tamu.edu}

\address[lanl]{Formerly Department of Geology and Geophysics, Texas A\&M University\\ College Station, TX 77843, U.S.A., \\ currently Geophysics Group, Los Alamos National Laboratory \\Los Alamos, NM 87545, U.S.A.}
\address[tamumath]{Department of Mathematics, Texas A\&M University\\ College Station, TX 77843, U.S.A. }
\address[kaust]{Numerical Porous Media SRI Center (NumPor) \\King Abdullah University of Science and Technology \\Thuwal, Saudi Arabia}
\address[tamugeo]{Department of Geology and Geophysics, Texas A\&M University\\ College Station, TX 77843, U.S.A. }
\address[cuhk]{Department of Mathematics, Chinese University of Hong Kong\\ Shatin, NT, Hong Kong }

\begin{abstract}
It is important to develop fast yet accurate numerical methods for seismic wave propagation to characterize complex geological structures and oil and gas reservoirs. However, the computational cost of conventional numerical modeling methods, such as finite-difference method and finite-element method, becomes prohibitively expensive when applied to very large models. We propose a Generalized Multiscale Finite-Element Method (GMsFEM) for elastic wave propagation in heterogeneous, anisotropic media, where we construct basis functions from multiple local problems for both the boundaries and interior of a coarse node support or coarse element. The application of multiscale basis functions can capture the fine scale medium property variations, and allows us to greatly reduce the degrees of freedom that are required to implement the modeling compared with conventional finite-element method for wave equation, while restricting the error to low values. We formulate the continuous Galerkin and discontinuous Galerkin formulation of the multiscale method, both of which have pros and cons. Applications of the multiscale method to three heterogeneous models show that our multiscale method can effectively model the elastic wave propagation in anisotropic media
with a significant reduction in the degrees of freedom in the modeling system.
\end{abstract}


\end{frontmatter}


\section{Introduction}
Seismic wave propagation has long been a fundamental research field in both global scale seismology and reservoir exploration scale seismics. There are two basic categories of methods to investigate the propagation of waves through the Earth media,  approximate methods and the full wavefield (exact) methods. 
Approximate methods rely on either the simplification of the Earth media, or the approximation of the wave equation, which include, for instance, the ray tracing method \cite{Cerveny-Hron_1980,Beydoun-Keho_1987,Gibson-etal_2005}, the Gaussian beam method \cite{Hill_1990,Gray-Bleistein_2009}, the one-way wave equation approach \cite{Claerbout_1985,Zhang-etal_2005}, the reflectivity method \cite{Kennett_1985}, etc.. These methods are generally fast and computationally affordable. However, they are intrinsically incomplete and therefore may fail in complex geology, where steep dips, faults, salt bodies, irregular interfaces, or fractures exist. The direct methods on the other hand, consist of many different numerical methods to solve various forms of the wave equation directly without approximations and simplifications, including, for example, the finite-difference method \cite{Dablain_1986,Virieux_1986,Saenger-etal_2000}, the finite-element method \cite{Marfurt_1984,Drake-Bolt_1989,Komatitsch-Tromp_2002,Chung-Engquist_2006}, the pseudo-spectral method \cite{Fornberg_1990}, and so on, and are essential fundamentals of full-wavefield based seismic imaging and inversion methods, such as reverse-time migration \cite{McMechan_1983,Symes_2007} and full waveform inversion \cite{Tarantola_1984,Virieux-Operto_2009,Shipp-Singh_2002}. However, the applications of full wavefield methods are also computationally expensive, where the computation costs are directly proportional to the number of discrete elements that are required to represent the geological model, and this makes the wide applications of full-wavefield based imaging and inversion methods infeasible for realistic large 2D and 3D geological models. Moreover, the Earth medium should be considered as a complex system that is heterogeneous at different spatial scales. To include the influence of heterogeneities at finer scales when simulating the wave propagation on coarser scale, researchers tend to apply various effective medium theories \cite{Backus_1962,Schoenberg-Muir_1989,Sayers_2002} to get a set of equivalent parameters that is supposed to best approximate the properties of the heterogeneous media. However, all of these effective medium theories rely on a long wavelength assumption, i.e., size of the heterogeneities is much smaller than the dominant wavelength of the wavelet, and when such assumptions fail, wave reflections and scattering become important, which cannot be correctly modeled by the effective medium approach. 

In this paper, we are interested in developing fast yet accurate full wavefield modeling method for elastic wave propagation in heterogeneous, anisotropic media. 
The most straightforward way to model various types of wave equations is the finite-difference method due to its simplicity in implementation, where we have the conventional central finite-difference method (FDM) \cite{Alterman-etal_1968,Alford-etal_1974,Kelly-etal_1976,Dablain_1986,Liu_2013}, the staggered-grid finite-difference method \cite{Virieux_1986,Levander_1988}, the rotated staggered-grid method \cite{Saenger-etal_2000,Saenger-Bohlen_2004}, etc.. However, FDM enjoys less flexibility in handling unstructured meshes, hanging nodes, and non-conforming meshes, and free surface topography problem, and only recently, the mimetic finite-difference method \cite{Lipnikov-Huang_2008,Puente-etal_2014} claims be able to achieve this goal, yet there are corresponding increases in computational costs and  decreases in required time step size due to the distortion of grids. 
The finite-element methods (FEM), on the other hand, provide an effective solution to deal with the unstructured mesh of the geological model, which can honor the curved interfaces of the geological bodies, or the complex fault systems. The FEM also introduces great benefits for dealing with free surface topography that can be naturally satisfied through the weak formulation of the FEM. Various FEM techniques have been developed. Some of the earliest efforts to solve the wave equation with the FEM are  conventional continuous Galerkin (CG) FEMs \cite{Bolt-Smith_1976,Marfurt_1984,Hughes_1987,Drake-Bolt_1989}. However, CG-FEM can be quite computationally expensive due to the requirement of inverting the global mass matrix, which is not diagonal or block diagonal without mass lumping. This problem is removed with the spectral-element method (SEM) \cite{Patera_1984,Komatitsch-Vilotte_1998,Komatitsch-etal_1999,Komatitsch-Tromp_1999,Komatitsch-Tromp_2002,Komatitsch-etal_2010,Cohen_2002,Cohen-Fauqueux_2005}, which adopts Gauss-Lobatto-Legendre (GLL) integration points to obtain a strictly diagonal global mass matrix. Nevertheless, CG-FEM requires the continuity of wavefield solutions at the edges of elements, and is therefore less accurate when describing the wave propagation across high-contrast interfaces or discontinuities in the model. Besides, CG-FEM is unable to handle mesh discretization that is composed of different types of elements, non-conforming mesh or hanging nodes. 
These problems are naturally solved with the discontinuous Galerkin (DG) FEM initially developed for the transport equation \cite{Reed-Hill_1973} and elliptic partial differential equations \cite{Wheeler_1978,Riviere-etal_1999,Arnold-etal_2002}. The DG-FEM has gradually gained broader applications in time-dependent problems such as wave equations  \cite{Grote-etal_2006,Chung-Engquist_2006,Chung-Engquist_2009,Kaser-Dumbser_2006,Dumbser-Kaser_2006,DeBasabe-etal_2008,Puente-etal_2008,Dupuy-etal_2011,Wilcox-etal_2010}. Importantly, DG-FEM has the advantage over CG-FEM that the global mass matrix is block diagonal, and the support of elements is distinct, a feature that favors straightforward parallel implementation, and this is quite important for \oldrevision{wave equation simulations in} large models. However, DG-FEM also suffers from some drawbacks, such as more complicated error and dispersion analyses, the requirement of tuning penalty parameters and \oldrevision{much} more degrees of freedom. 

Regardless of the implementation complexity, neither FDMs nor FEMs addressed the common issue of high computational cost when solving the wave equation in large models. One approach to reduce such costs is the so-called multiscale method. The multiscale method was originally designed for elliptic partial differential equations \cite{Hou-Wu_1997}. Unlike all the above mentioned FEMs, the multiscale FEM (MsFEM) seeks special basis functions, i.e., the multiscale basis functions, to include the influence of fine-scale heterogeneity when solving the PDEs on the coarse scale, and the usage of the multiscale basis functions enables the MsFEM to consider high contrasts in medium properties that may vary by several orders of magnitudes spatially. These multiscale basis functions are not predefined polynomials like those in conventional FEMs \cite[e.g.,][]{Bengzon-Larson_2013}. Instead, they are solved from appropriately defined local problems \cite{Hou-Wu_1997,Efendiev-Hou_2009,Jiang_etal_2009}. Chung~et.~al. \cite{Chung-etal_2011a,Chung-etal_2011b} and Gibson~et.~al. \cite{Gibson-etal_2014} applied the idea of the multiscale basis functions and designed a multiscale method for mixed-form (pressure-velocity form) acoustic wave equation. To improve the accuracy of the MsFEM, Efendiev~et.~al. \cite{Efendiev-etal_2011,Efendiev-etal_2013a} proposed to utilize multiple multiscale basis functions solved from local spectral problem, which is the generalized multiscale finite-element method (GMsFEM). These basis functions are constructed from the eigenfunctions that correspond to the first several smallest eigenvalues of the local spectral problem, and are therefore correspond to the local eigenmodes with lowest frequencies. Chung~et.~al. \cite{Chung-etal_2013a} proposed a discontinuous Galerkin (DG) GMsFEM for the second-order acoustic wave equation, where they constructed so-called interior basis and boundary basis functions to capture fine-scale media heterogeneity information for the wavefield simulation on coarse scale. This DG-GMsFEM was also strictly analyzed \oldrevision{by} Chung~et.~al. \cite{Chung-etal_2013b}. There are other methods titled ``multiscale'', yet they begin with different assumptions and methodologies, for instance, the operator-based upscaling for the acoustic wave equation \cite{Arbogast-etal_1998,Vdovina-etal_2005}. Korostyshevskaya and Minkoff \cite{Koro-Minkoff_2006} and Vdovina and Minkoff \cite{Vdovina-Minkoff_2008} analyzed the error and convergence characteristics of this approach. However, in their approach, local problems have to be solved at each time step, whereas in the multiscale approach by  Chung~et.~al. \cite{Chung-etal_2011b,Chung-etal_2013b}, the local problems only need to be solved once before the time stepping to get the multiscale basis functions. Vdovina~et.~al. \cite{Vdovina-etal_2009} developed a similar operator-based upscaling approach for elastic wave equation. Owhadi and Zhang \cite{Owhadi-Zhang_2005,Owhadi-Zhang_2007,Owhadi-Zhang_2008} proposed the multiscale method for the wave equation based on the global change of coordinates. E and Engquist \cite{E-Engquist_2002,E-Engquist_2005} proposed the heterogeneous multiscale method (HMM) \RLG{that was later} developed in finite-difference and finite-element formulations \cite{Engquist-etal_2007,Engquist-etal_2011,Abdulle-etal_2011}. The HMM also requires evaluations of local problem in each time step, which \RLG{is expensive}. Capdeville~et.~al. \cite{Capdeville-etal_2010} proposed a numerical homogenization method for non-periodic heterogeneous elastic media, which extracts the microscopic part of medium properties, followed by a homogenization expansion. However, this method assumes scale separation of the media, which cannot always be satisfied in practice.

Based on previous \RLG{work} for the elliptic partial differential equations, the acoustic wave equation and the isotropic linear elasticity equation \cite{Efendiev-etal_2011,Efendiev-etal_2013a,Efendiev-etal_2013b,Chung-etal_2013a,Chung-etal_2013b,Gibson-etal_2014,Chung_etal_2014}, we propose a GMsFEM to simulate the wave propagation in heterogeneous, anisotropic elastic media on the coarse mesh. The essence of our GMsFEM is to construct multiscale basis functions with appropriately defined local problems, which will be used in both CG and DG formulations of the GMsFEM.
We investigated two types of related yet different multiscale basis functions \revision{for GMsFEM}. For the first type of multiscale basis function, we solve a linear elasticity eigenvalue problem in the support of a node on the coarse mesh, or in the region of a coarse element. By selecting the eigenfunctions correspond to the first several smallest eigenvalues, we construct a finite-dimensional basis function space for GMsFEM. For the second type of multiscale basis function, we construct a basis space which is composed of two orthogonal subspaces, and these two subspaces consist of multiscale functions defined with different local spectral problems. The first subspace is spanned by the basis functions that are solved directly from the local eigenvalue problem of linear elasticity for the interior nodes of the coarse node support or coarse element, while the second subspace consists of the basis functions solved from a local spectral problem which is related to the boundaries of the coarse node support or the coarse element in GMsFEM. For both of these spaces, we select the eigenfunctions that correspond to the first several smallest eigenvalues. These basis functions correspond to the local eigenmodes with lowest frequencies. The resulting GMsFEM allows us to utilize these multiscale basis functions to capture the fine scale information of the heterogeneous media, while effectively reducing the degrees of freedom that are required to implement the modeling compared with conventional method such like CG- \oldrevision{ and DG-}FEM. 

Our paper is organized as follows. We first introduce the CG and DG formulations of GMsFEM for the elastic wave equation in heterogeneous, anisotropic media. Specifically, we define the appropriate bilinear forms for the elastic wave equation, then we introduce two approaches to construct the multiscale basis functions with appropriately defined local problems, as well as the oversampling technique to reduce the influence of prescribed boundary conditions, and an adaptive way to assign different numbers of basis functions for coarse elements in DG-GMsFEM. We then present three numerical results to verify the effectiveness of our multiscale method, including a heterogeneous model composed of \revision{\RLG{vertical transversely isotropic (VTI), tilted transversely isotropic (TTI)} and isotropic} layers, a heterogeneous model 
\revision{composed of curved layers and random heterogeneities}.
The last numerical example is devoted to verify the adaptive assignment of number of basis functions. Finally, we give a brief discussion of limitations of our current work and propose some possible improvements. 

\section{Theory}
We will develop both the CG- and DG-GMsFEM in this section. We will first give the weak forms of the elastic wave equation in CG and DG formulations, then we will show how to construct the multiscale basis functions using appropriately defined local spectral problems. Although the formulations of CG- and DG-GMsFEM are different, the multiscale basis functions for these two formulations can be constructed in the same way. In other words, the construction of the multiscale basis functions is independent of the CG or DG formulations for the elastic wave equation. In fact, we will see that the construction of the multiscale basis function is only related to the spatial part of the elastic wave equation. 
We remark that we present the definitions, equations and derivations in this part in a general style, and therefore they are valid for both 2D and 3D cases. However, we will present only 2D examples in this part, as well as the next part of numerical results.

\subsection{Weak form of the elastic wave equation}

\subsubsection{Elastic wave equation}
We begin with the elastic wave equation in the form \oldrevision{\cite[e.g.,][]{Carcione_2007}}
\begin{subequations}\label{eq:ewe}
\begin{align}
\rho \partial_{t}^{2}\mathbf{u} & = \nabla\cdot\boldsymbol{\sigma}+\mathbf{f}, \\
\boldsymbol{\sigma}& =\mathbf{c}:\boldsymbol{\varepsilon}, \\
\boldsymbol{\varepsilon} &=\frac{1}{2}[\nabla \mathbf{u}+(\nabla \mathbf{u})^{\mathrm{T}}]
\end{align}
\end{subequations}
where $\mathbf{u}=\mathbf{u}(\mathbf{x},t)$ is the displacement wavefield we aim to solve with our multiscale method in the spatial domain $\Omega$, which could be 2D or 3D in general, and \RLG{the} temporal domain $[0, T]$.
Also $\boldsymbol{\sigma}=\boldsymbol{\sigma}(\mathbf{u})$ is the stress tensor, $\boldsymbol{\varepsilon}=\boldsymbol{\varepsilon}(\mathbf{u})$ is the strain tensor, 
$\mathbf{f}$ is the external source term, \oldrevision{$\mathbf{c}=\mathbf{c}(\mathbf{x})=c_{ijkl}(\mathbf{x})$ is the fourth-order elasticity tensor  where $i,j,k,l=1,2,3$ \cite[]{Carcione_2007}} and $\rho=\rho(\mathbf{x})$ is the density of the medium. 

In our approach, the elasticity tensor $\mathbf{c}$ can be generally anisotropic, i.e., all the 21 independent elasticity parameters in $\mathbf{c}$ can be non-zero in the 3D case. However, since we will present only 2D results in this paper, \oldrevision{we only consider the elasticity tensor with $i,j,k,l=1,3$. Therefore, with Voigt notation \cite[e.g.,][]{Carcione_2007}, we can express the elasticity tensor $\mathbf{c}$ in the following matrix form}:
\begin{equation}
\mathbf{C}=\left(\begin{array}{ccc}
C_{11} & C_{13} & C_{15}\\
C_{13} & C_{33} & C_{35}\\
C_{15} & C_{35} & C_{55}
\end{array}\right), \label{eq:elasticity_voigt}
\end{equation}
which can describe the elastic wave propagation in anisotropic media with symmetry up to hexagonal anisotropy with tilted symmetry axis in the $x_1-x_3$ plane (transversely isotropy with tilted axis, TTI), and monoclinic anisotropy (assuming the symmetry plane is the $x_1-x_3$ plane), where $C_{15}$ and $C_{35}$ are possibly nonzero. 

\subsubsection{CG formulation}
\begin{figure}
\begin{center}
\includegraphics[width=0.4\textwidth]{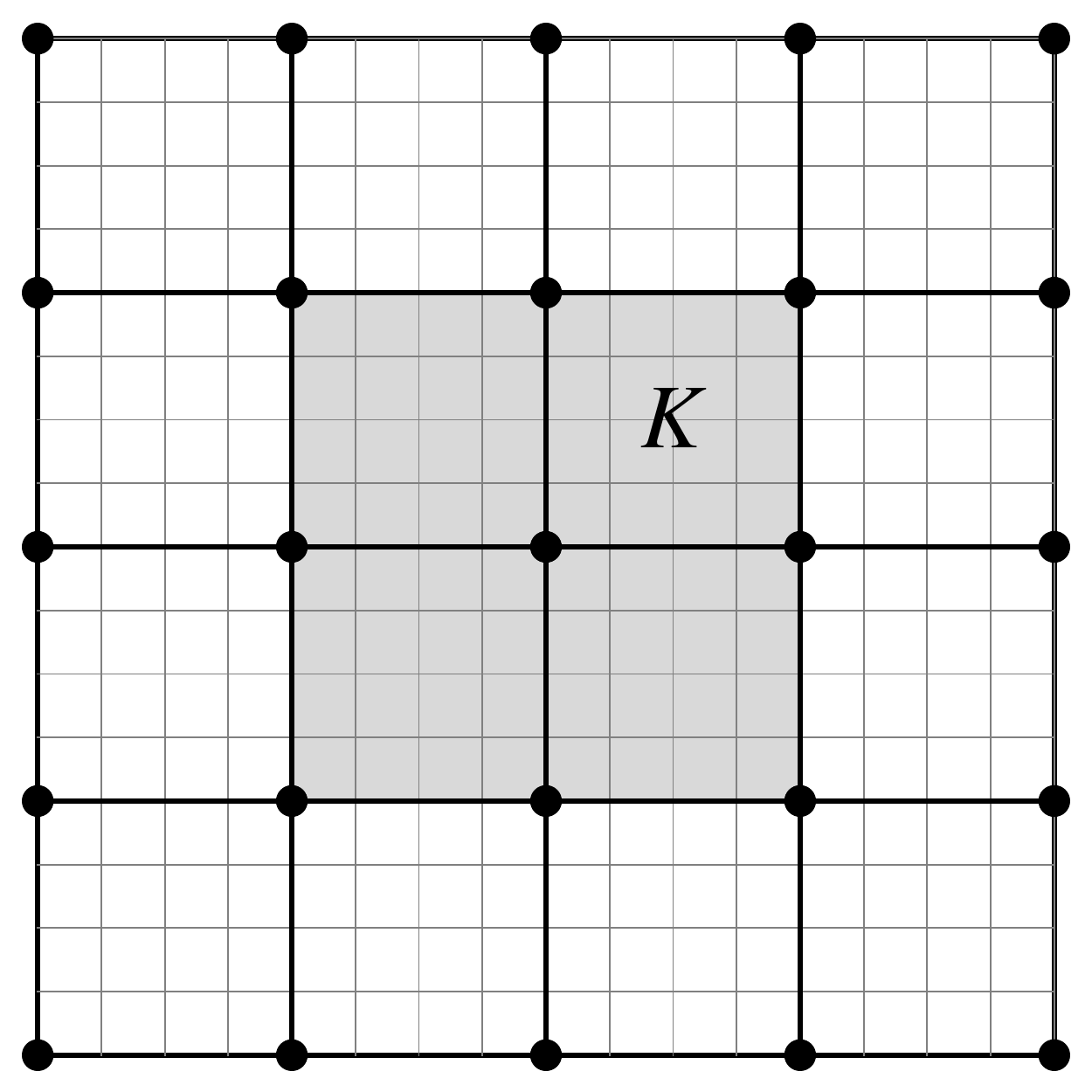}
\caption{A sketch of the fine mesh $\mathcal{T}_h$, denoted by gray mesh, and coarse mesh $\mathcal{T}_H$, denoted by black mesh, in CG formulation of GMsFEM. Gray rectangle labeled $K$ represents the support of the $i$-the coarse node. $K$ contains many finer elements which might have high contrasts in medium properties. }
\label{fig:mesh_CG}
\end{center}
\end{figure}

We first formulate the multiscale method in the CG framework 
for 2D \revision{elastic wavefield} simulations with applications to higher-order cases of anisotropy.
For the CG formulation, we first discretize the whole computational domain $\Omega$ with a coarse mesh $\mathcal{T}_H$ overlying a finer mesh $\mathcal{T}_h$. Figure \ref{fig:mesh_CG} illustrates this mesh design, where we use the black lines to represent the coarse mesh, and gray lines to represent the finer mesh. The support of a coarse node can be denoted as $K$, which contains many finer elements. The mesh can be unstructured, though we assume structured elements in the theory development to develop the current results. Nevertheless, the following derivations are equally valid for an unstructured mesh. 

We express the displacement wavefield $\mathbf{u}$ on the coarse mesh $\mathcal{T}_H$ as 
\begin{equation}
\mathbf{u}_H(\mathbf{x},t)=\sum\limits_{i=1}^{N} \mathbf{d}_i(t) \boldsymbol{\Phi}_i(\mathbf{x}), \label{eq:cg_uH}
\end{equation}
where $\boldsymbol{\Phi}_i(\mathbf{x})$ are the spatial basis functions of $\mathbf{u}_H(\mathbf{x},t)$, and $\boldsymbol{\Phi}_i$ belong to the finite-dimensional function space $V_H=\{\boldsymbol{\Phi}_i\}_{i=1}^{N}$. Note that each $\boldsymbol{\Phi}_i$ is piecewise continuous in $\Omega$. Space $V_H$ is our multiscale basis function space, which will be defined in the next section. We multiply the elastic wave equation \ref{eq:ewe} with a test function $\mathbf{v} \in V_H$, integrate over $\Omega$, apply Gauss's theorem, and get the weak form of the elastic wave equation as
\begin{equation}
\int_{\Omega} \rho \partial_{t}^{2}\mathbf{u}_H \cdot \mathbf{v} d\mathbf{x}+a_{\text{CG}}(\mathbf{u}_H,\mathbf{v})=\int_{\Omega} \mathbf{f} \cdot \mathbf{v} d\mathbf{x}, \label{eq:weak_form}
\end{equation}
where the bilinear form $a_{\text{CG}}$ is 
\begin{equation}
a_{\text{CG}}(\mathbf{u},\mathbf{v})=\int_{\Omega}\boldsymbol{\sigma}(\mathbf{u}):\boldsymbol{\varepsilon}(\mathbf{v})d\mathbf{x}+ \int_{\partial \Omega}[\boldsymbol{\sigma}(\mathbf{u})\cdot  \mathbf{n}] \cdot \boldsymbol{\mathbf{v}} ds.
\end{equation}
Also, $\mathbf{n}$ is the outward pointed normal of $\partial \Omega$. We have set homogeneous Neumann boundary condition, i.e., $\boldsymbol{\sigma}(\mathbf{u})\cdot  \mathbf{n}=\mathbf{0}$, for simplicity. 

\subsubsection{DG formulation}
\begin{figure}
\centering
\includegraphics[width=0.4\textwidth]{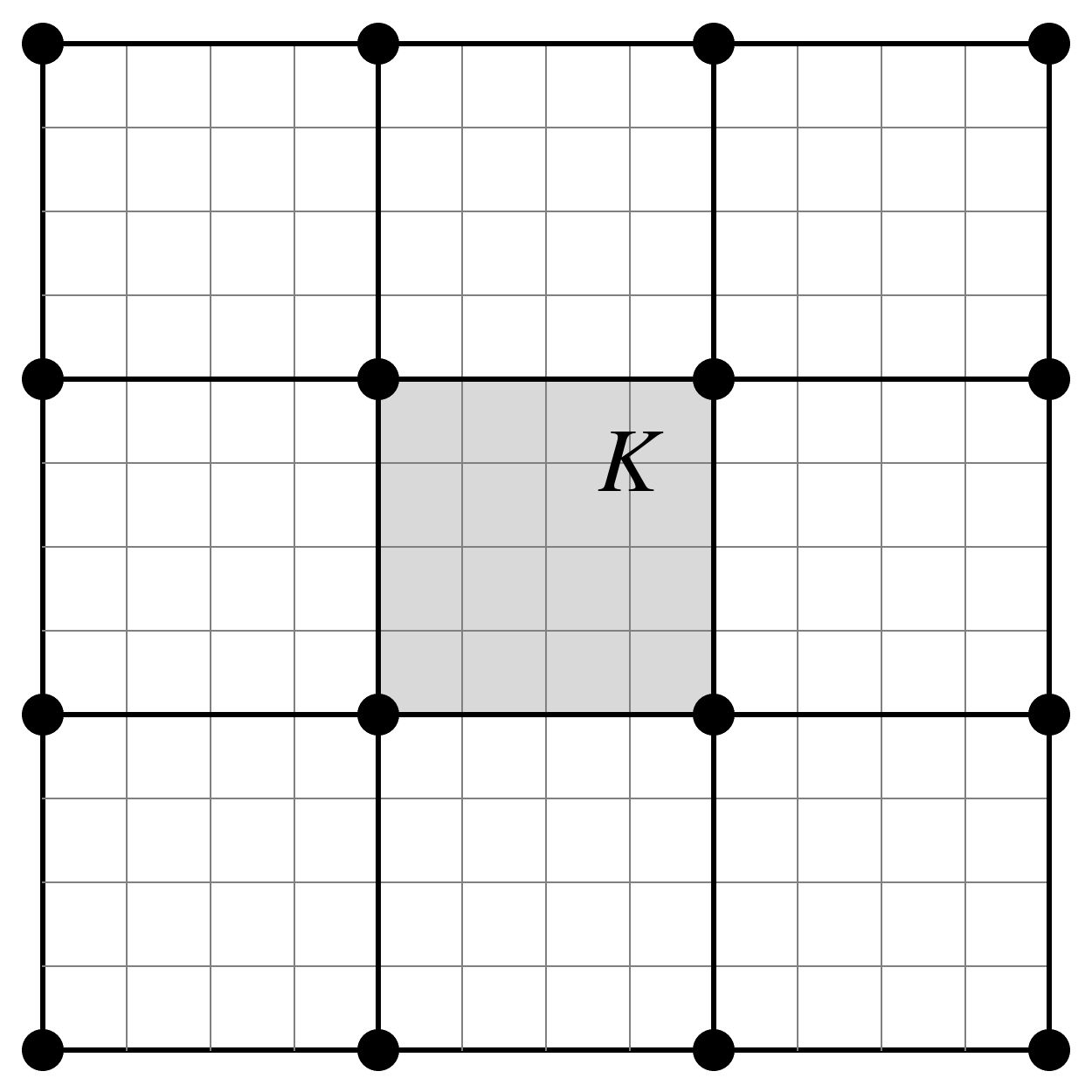}
\caption{A sketch of the fine mesh $\mathcal{P}_h$, denoted by gray mesh, and coarse mesh $\mathcal{P}_H$, denoted by black mesh in DG formulation of GMsFEM. Gray rectangle labeled $K$ represents the $i$-th coarse element. Same with that in CG-GMsFEM, coarse block $K$ contains many finer element which might have high contrasts in medium properties. }
\label{fig:mesh_DG}
\end{figure}

The discontinuous Galerkin formulation of our multiscale method is a natural choice if a non-conformal mesh is taken into consideration. For DG formulation, we discretize $\Omega$ with a set of coarse mesh cells $\mathcal{P}_H$, each coarse element containing more finely discretized elements in the  finer mesh $\mathcal{P}_h$, as is shown in 
Figure \ref{fig:mesh_DG} for a 2D meshing case. Again, the solution of the wave equation \ref{eq:ewe} can be expressed as
\begin{equation}
\mathbf{u}_H(\mathbf{x},t)=\sum\limits_{i=1}^{N} \mathbf{d}_i(t) \boldsymbol{\Psi}_i(\mathbf{x}), \label{eq:dg_uH}
\end{equation}
where the basis functions $\boldsymbol{\Psi}_i \in W_H$. The multiscale basis function space $W_H$ will be defined in the next section. We assume that the basis functions $\boldsymbol{\Psi}_i$ are continuous within each coarse element $K$, but generally discontinuous at the coarse element boundaries $\partial K$.

As is true in general for discontinuous Galerkin finite-element methods \cite[e.g.,][]{Grote-etal_2006,Arnold-etal_2002,Wihler_2006}, we define some terms related to the boundaries of the coarse element. 
Letting $\mathcal{E}_H$ be the set of all interior coarse element edges in the 2D case (the set of all interior coarse element faces in 3D), then we define the average of a tensor $\boldsymbol{\sigma}$ on $E\in \mathcal{E}_H$ as
\begin{equation} 
\average{\boldsymbol{\sigma}}=\frac{1}{2}(\boldsymbol{\sigma}^{+}+\boldsymbol{\sigma}^{-}),
\end{equation}
where $\boldsymbol{\sigma}^{\pm} = \boldsymbol{\sigma}|_{K^{\pm}}$, and $K^{\pm}$ are the two coarse elements having the common $E$. Meanwhile, the jump of a vector $\mathbf{v}$ on $E\in \mathcal{E}_H$ is given by:
\begin{equation}
\jump{\mathbf{v}}=\mathbf{v}^{+}-\mathbf{v}^{-}.
\end{equation}
We also have a matrix jump term resulting from the outer product of \RLG{a} vector with edge or face normals, which is defined as 
\begin{equation}
\underline{\jump{\mathbf{v}}}=\mathbf{v}^{+}\otimes\mathbf{n}^{+}+\mathbf{v}^{-}\otimes\mathbf{n}^{-},
\end{equation}
where $\mathbf{n}^{\pm}$ is the unit outward normal vector on the boundary of $K^{\pm}$.
In addition, for the edges on the computation domain boundary $\partial \Omega$, the above average and jump terms can be defined as
\begin{equation}
\average{\boldsymbol{\sigma}}=\boldsymbol{\sigma}, \quad \jump{\mathbf{v}}=\mathbf{v}, \quad \underline{\jump{\mathbf{v}}}=\mathbf{v}\otimes\mathbf{n},
\end{equation}
where $\mathbf{n}$ is the outward pointed normal of coarse element $K$.

We multiply the elastic wave equation \ref{eq:ewe} with some arbitrary test function $\mathbf{v}\in W_H$, and get the weak form
\begin{equation}
\int_{\Omega}\rho \partial_{t}^{2}\mathbf{u}_H \cdot \mathbf{v}d\mathbf{x} +a_{\text{DG}}(\mathbf{u}_H,\mathbf{v})=\int_{\Omega}\mathbf{f}\cdot\mathbf{v}d\mathbf{x},
\end{equation}
where the bilinear form $a_{\text{DG}}(\mathbf{u},\mathbf{v})$ is defined as
\begin{align}
a_\text{DG}(\mathbf{u},\mathbf{v}) &=\sum_{K\in\mathcal{P}_{H}}\int_{K}\boldsymbol{\sigma}(\mathbf{u}):\boldsymbol{\varepsilon}(\mathbf{v})d\mathbf{x} \nonumber \\
&  -\sum_{E\in\mathcal{E}_{H}}\int_{E}(\average{\boldsymbol{\sigma}(\mathbf{u})}:\underline{\jump{\mathbf{v}}}+\eta \underline{\jump{\mathbf{u}}}:\average{\boldsymbol{\sigma}(\mathbf{v})})ds \nonumber \\
&  +\sum_{E\in\mathcal{E}_{H}}\frac{\gamma}{|E|}\int_{E}(\underline{\jump{\mathbf{u}}}:\average{\mathbf{c}}:\underline{\jump{\mathbf{v}}}+\jump{\mathbf{u}}\cdot\average{\mathbf{D}}\cdot\jump{\mathbf{v}})ds, \label{eq:a_dg}
\end{align}
with $\mathbf{D}=\mathrm{diag}(C_{11},C_{22},C_{33})$, $C_{IJ}$ are components of the fourth-order elasticity tensor $\mathbf{c}$ in Voigt notation \oldrevision{\cite[e.g.,][]{Carcione_2007}}. $\eta$ is a parameter that takes values $-1$, $0$ or $1$, and we choose $\eta=1$, which makes our method the classical symmetric interior penalty Galerkin (SIPG) method \cite[]{Arnold-etal_2002,Chung-etal_2013b,DeBasabe-etal_2008}. $\gamma$ is the penalty parameter, and we set $\gamma > 0$. We have omitted the terms related to the boundary edges, since we assume \RLG{a} homogeneous Neumann boundary condition. This bilinear form is inspired by those defined for linear elasticity problem \cite[]{Wihler_2006} and isotropic elastic wave  equation \cite[]{DeBasabe-etal_2008}, however, we have used non-constant matrix penalty parameters and two different penalty terms, i.e., $\average{\mathbf{c}}=\average{\mathbf{c}(\mathbf{x})}$ and $\average{\mathbf{D}}=\average{\mathbf{D}(\mathbf{x})}$. 
We find that such penalty terms can better guarantee the stability of the DG scheme. Meanwhile, we use a fixed $\gamma$ for all boundaries for convenience, which can alternatively vary from edge to edge. It should be remarked that the bilinear form \ref{eq:a_dg}, which is essentially the time-independent part of the elastic wave equation \ref{eq:ewe}, is not unique, and there are some other similar  choices which may be equally good \oldrevision{\cite[e.g.,][]{Riviere_2008,Kaufmann-etal_2008,Hansbo-Larson_2011}}. 

\subsection{Multiscale basis functions}
The key task in our multiscale method, given the choice of one of the above weak forms of the elastic wave equation, is to construct appropriate multiscale basis functions $\boldsymbol{\Phi}_i$ or $\boldsymbol{\Psi}_i$ to form the function space $V_H$ or $W_H$ for CG- or DG-GMsFEM. In this section, we will introduce two methods to construct the multiscale basis functions, both are solved from appropriately defined local problems, and both can be taken to form the basis function space for the wave equation. We note that the same basis functions are used for both the CG- and the DG-GMsFEM simulations and the selection of basis functions is therefore independent of the coarse scale formulation.

\subsubsection{Type I}
The first way to define a set of multiscale basis functions for GMsFEM is solving a local linear elasticity eigenvalue problem. Specifically, suppose $K$ is the support of a coarse node in CG formulation, or the coarse element in DG formulation, 
then we solve the following eigenvalue problem in $K$: 
\begin{subequations}\label{eq:local_eig}
\begin{align}
-\nabla\cdot\boldsymbol{\sigma} & = \zeta \rho \mathbf{u}, \\
\boldsymbol{\sigma} & = \mathbf{c}:\boldsymbol{\varepsilon}, \\
\boldsymbol{\varepsilon} &=\frac{1}{2}[\nabla \mathbf{u}+(\nabla \mathbf{u})^{\mathrm{T}}], 
\end{align}
\end{subequations}
with zero Neumann boundary condition $\boldsymbol{\sigma} \cdot \mathbf{n} = \mathbf{0}$ on $\partial K$, where $\zeta$ is the eigenvalue, and $\mathbf{n}$ is the outward pointed normal of $K$.  The elasticity tensor $\mathbf{c}$ can be spatially heterogeneous. This local problem corresponds to the following discrete system:
\begin{equation}
\mathbf{A}\mathbf{U}=\zeta \mathbf{M} \mathbf{U},
\end{equation}
where the global stiffness and global mass matrices $\mathbf{A}$ and $\mathbf{M}$ are computed from 
\begin{align}
\mathbf{A}=\int_{K}\boldsymbol{\sigma}(\boldsymbol{\gamma}): \boldsymbol{\varepsilon}(\boldsymbol{\eta}) d\mathbf{x}, \label{eq:local_A}\\
\mathbf{M}=\int_{K} \rho \boldsymbol{\gamma}\cdot \boldsymbol{\eta} d\mathbf{x}, \label{eq:local_M}
\end{align}
for the coarse node support or coarse element 
$K$, with $\boldsymbol{\gamma},\boldsymbol{\eta}\in V_h$, 
and they can be  discretized and calculated with appropriate quadrature and integration rules \cite[]{Hughes_1987,Bengzon-Larson_2013} for calculation of eigenvectors. 

The above linear elasticity eigenvalue problem can be solved with a conventional solver 
without difficulties, since normally the dimension of the above system is not large due to the limited size of a coarse element. To ensure stability, we can add to $\mathbf{A}$ a value $10^{-8}$ to $10^{-9}$ times the maximum on the diagonal of $\mathbf{A}$. Solutions of the eigenvalue problem for the displacement $\mathbf{u}$ are labeled as $\boldsymbol{\psi}_k$, denoting the $k$-th eigen-displacement in the coarse block $K$. Physically, they are the standing modes in $K$ with frequencies $\omega_k=\sqrt{\zeta_k}$. 

Depending on the dimension of the coarse block $K$, there can be many eigenfunctions associated with the local problem \ref{eq:local_eig}.
The analyses for elliptic partial differential equation \cite{Efendiev-etal_2013a} and for acoustic wave equation \cite{Chung-etal_2013a,Chung-etal_2013b,Fu-etal_2013} indicate that it is adequate to select only a few of the eigenfunctions as the basis functions for $\mathbf{u}_H$. The criterion for selecting  eigenfunctions is to chose those representing most of the energy in the eigenmodes $\boldsymbol{\psi}_k$. Correspondingly,  the sum of the inverse of selected eigenvalues $\sum_{l=1}^m \zeta_l^{-1}$ should be a large portion of the sum of all the inverse of eigenvalues $\sum_{l=1}^L \zeta_l^{-1}$ ($L$ is the number of eigenfunctions). 
We can select the first $m$ eigenfunctions $\boldsymbol{\psi}_1,\boldsymbol{\psi}_2,\cdots,\boldsymbol{\psi}_m$ corresponding to the $m$ smallest eigenvalues $0\leq \zeta_1\leq \zeta_2 \leq \cdots \leq \zeta_m$ of the above local problem, and construct the multiscale basis function space \revision{for DG-GMsFEM} as
\begin{equation}
W_H(K)=\mathrm{span}\{\boldsymbol{\psi}_1,\boldsymbol{\psi}_2,\cdots,\boldsymbol{\psi}_m\}. \label{eq:type1_space_dg}
\end{equation}
\revision{For CG-GMsFEM, the basis functions space can be constructed from the first $m$ eigenfunctions as}
\begin{equation}
V_H(K)=\mathrm{span}\{\boldsymbol{\chi}_K  \boldsymbol{\psi}_1,\boldsymbol{\chi}_K \boldsymbol{\psi}_2,\cdots,\boldsymbol{\chi}_K  \boldsymbol{\psi}_m\}, \label{eq:type1_space_cg}
\end{equation}
\oldrevision{where $\boldsymbol{\chi}_i$ is the partition of unity that is} defined as a collection of smooth and nonnegative functions in the appropriate space $M$ that satisfy $\sum_K\chi_K(\mathbf{x})=1$ for any $\mathbf{x}\in M$. Thus $\chi_K$ could be understood as the standard FEM basis functions that are defined for various kinds of elements and various orders. For example, in one dimension, $\chi_K$ are the standard linear basis functions, i.e., $\chi_K=\{1-x,x\}$, in the lowest order case. \oldrevision{The above choice} applies the bases \oldrevision{ corresponding to the} most dominant wave modes, i.e., the wave modes with the lowest several frequencies. Due to the limited resolution of the coarse block $K$, higher frequencies cannot be accurately represented.

It is clear that the basis functions solved from the local eigenvalue problem \ref{eq:local_eig} are  influenced by the anisotropic and heterogeneous properties in the region $K$, and they are different for different local $\mathbf{c}(\mathbf{x})$ and $\rho(\mathbf{x})$. This is the most distinct difference between our multiscale basis functions and the high order basis functions in various finite-element methods \cite{Marfurt_1984,Hughes_1987,Komatitsch-etal_1999}, where the basis functions are predefined polynomials and are independent of the earth model. 

Three examples help to illustrate the behavior of these basis functions.

\begin{figure}
\centering
\subfigure[]{
\label{fig:typeI_isohomo_u1_1}
\includegraphics[trim=50 50 30 0,width=0.415\textwidth]{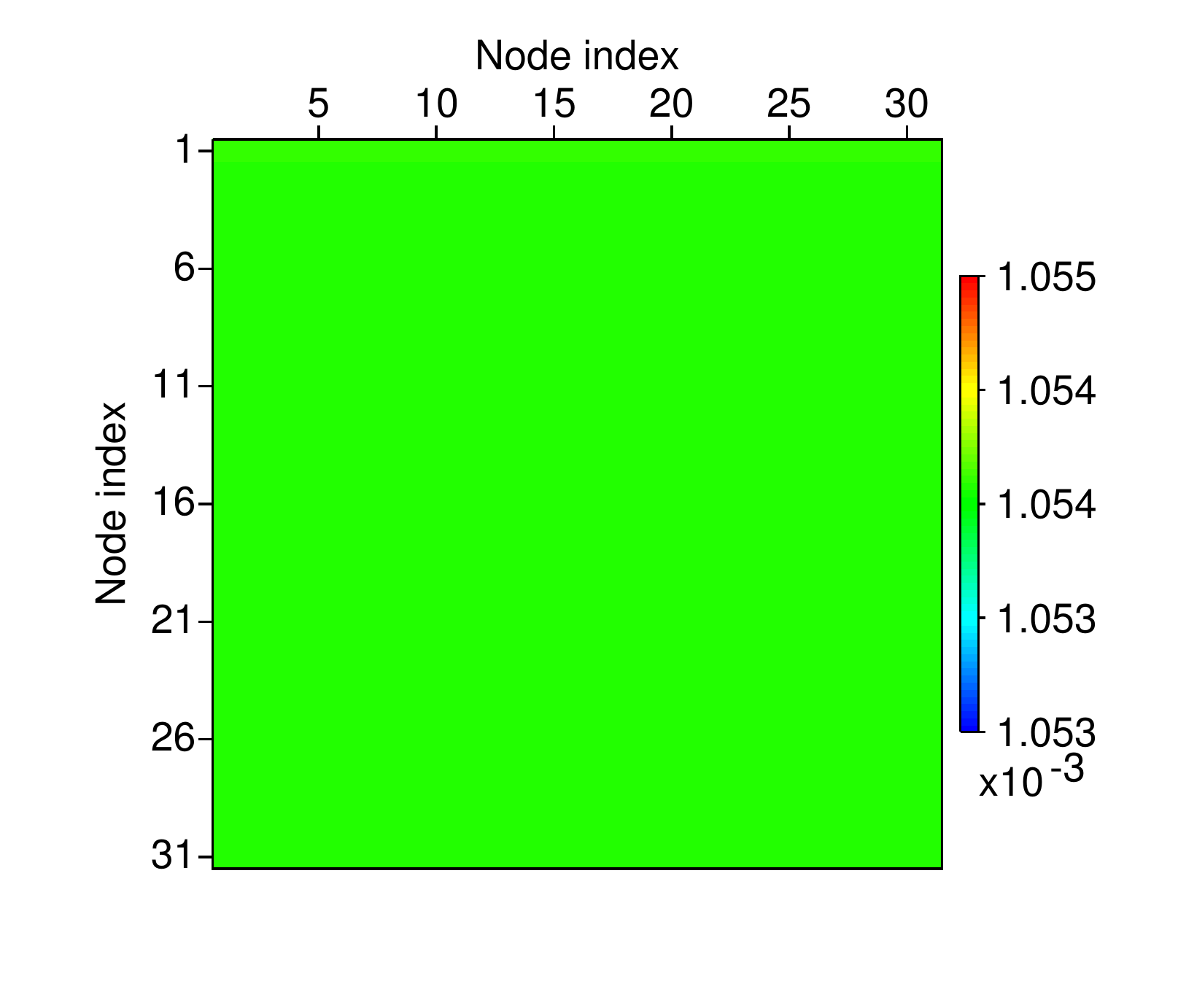}
}
\subfigure[]{
\label{fig:typeI_isohomo_u1_2}
\includegraphics[trim=50 50 30 0,width=0.415\textwidth]{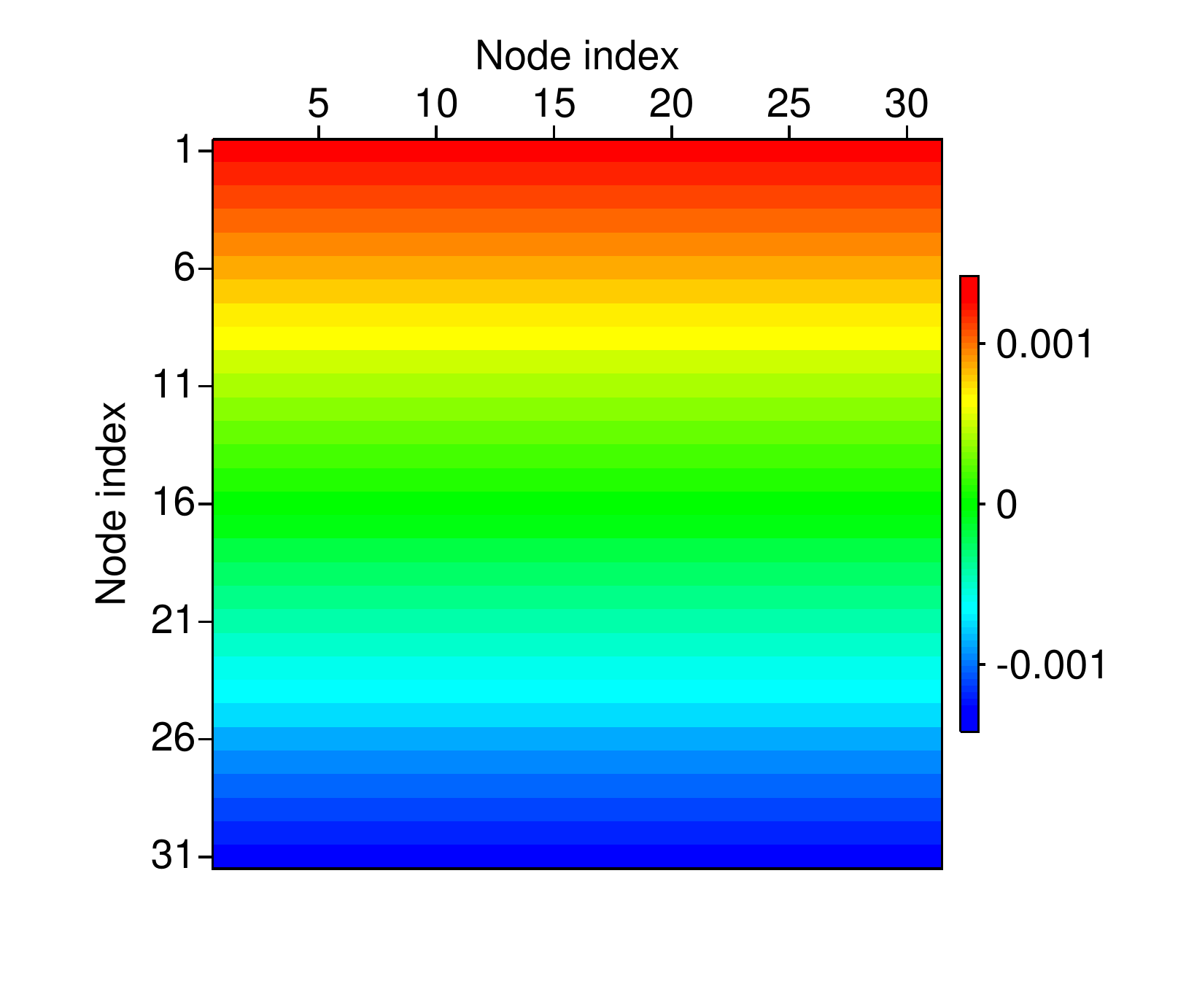}
}
\subfigure[]{
\label{fig:typeI_isohomo_u1_3}
\includegraphics[trim=50 50 30 0,width=0.415\textwidth]{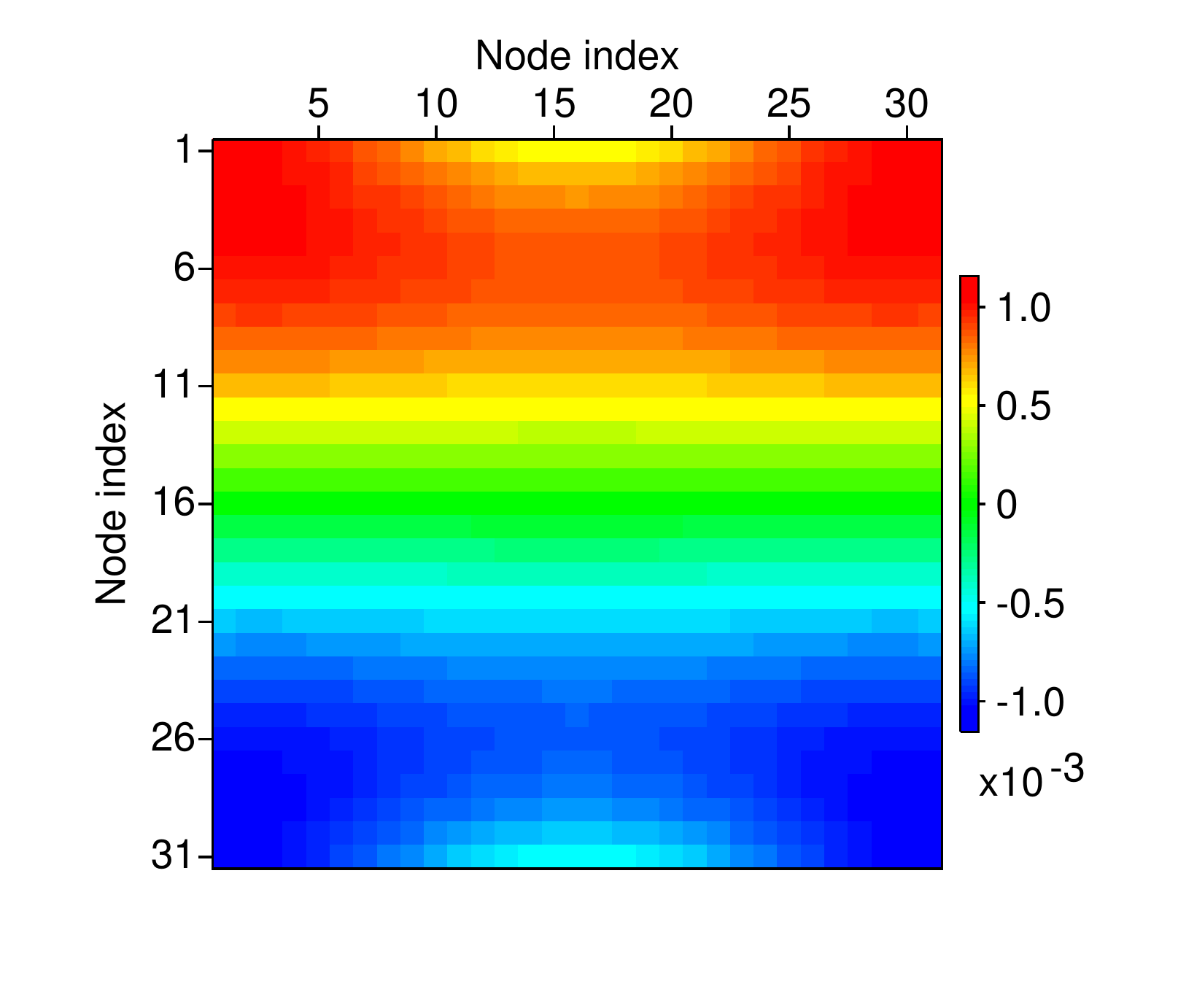}
}
\subfigure[]{
\label{fig:typeI_isohomo_u1_4}
\includegraphics[trim=50 50 30 0,width=0.415\textwidth]{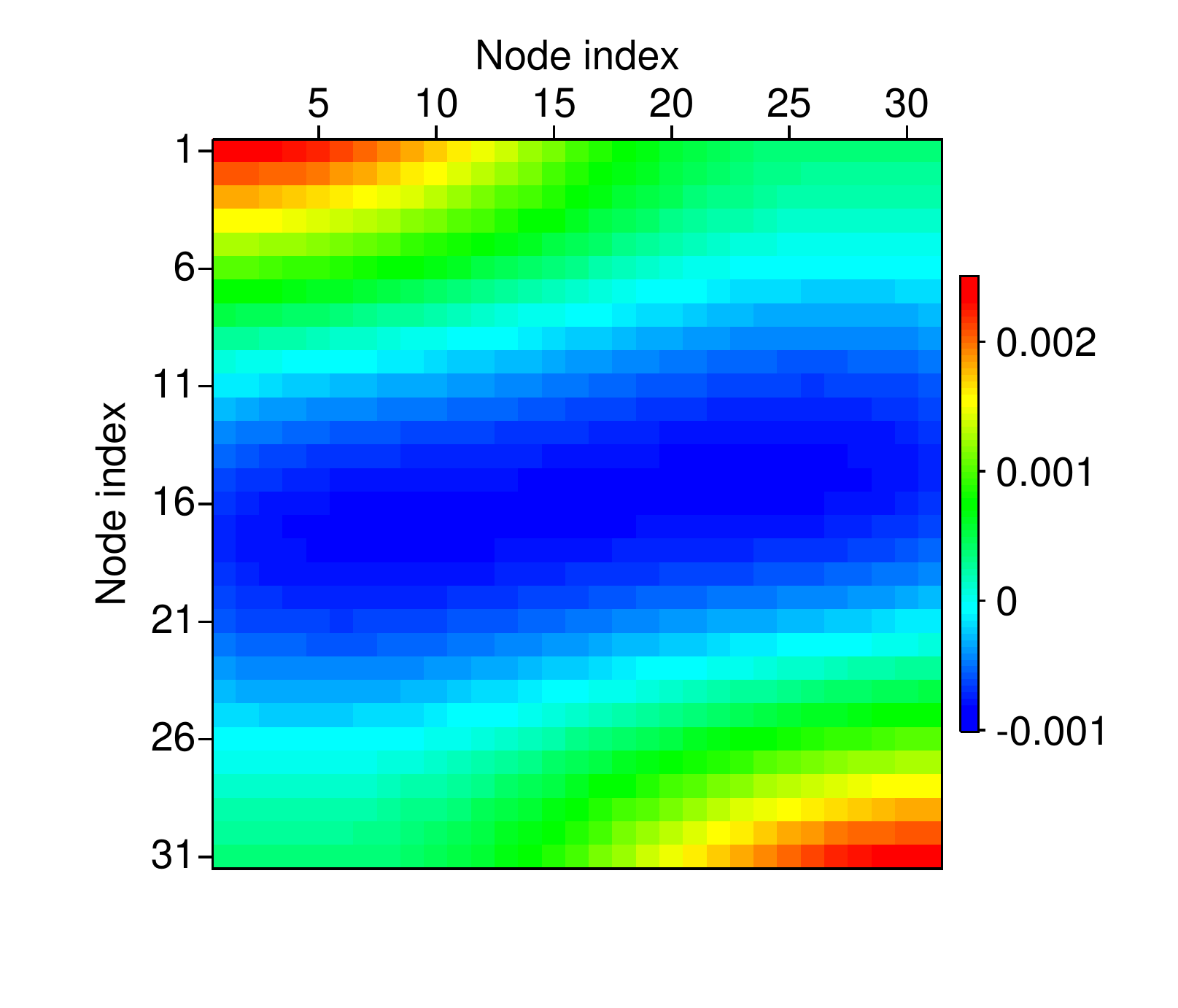}
}
\subfigure[]{
\label{fig:typeI_isohomo_u1_5}
\includegraphics[trim=50 50 30 0,width=0.415\textwidth]{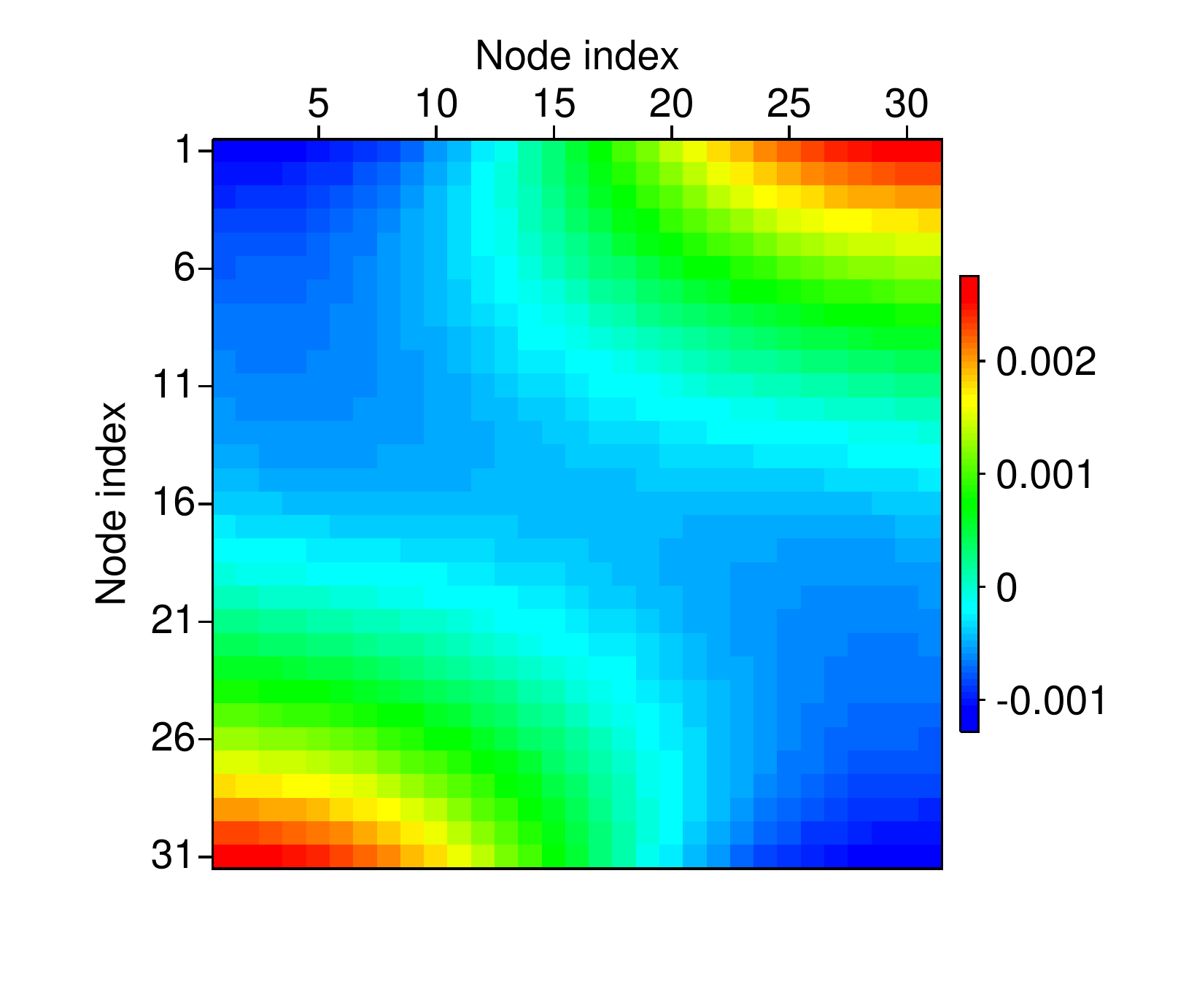}
}
\subfigure[]{
\label{fig:typeI_isohomo_u1_6}
\includegraphics[trim=50 50 30 0,width=0.415\textwidth]{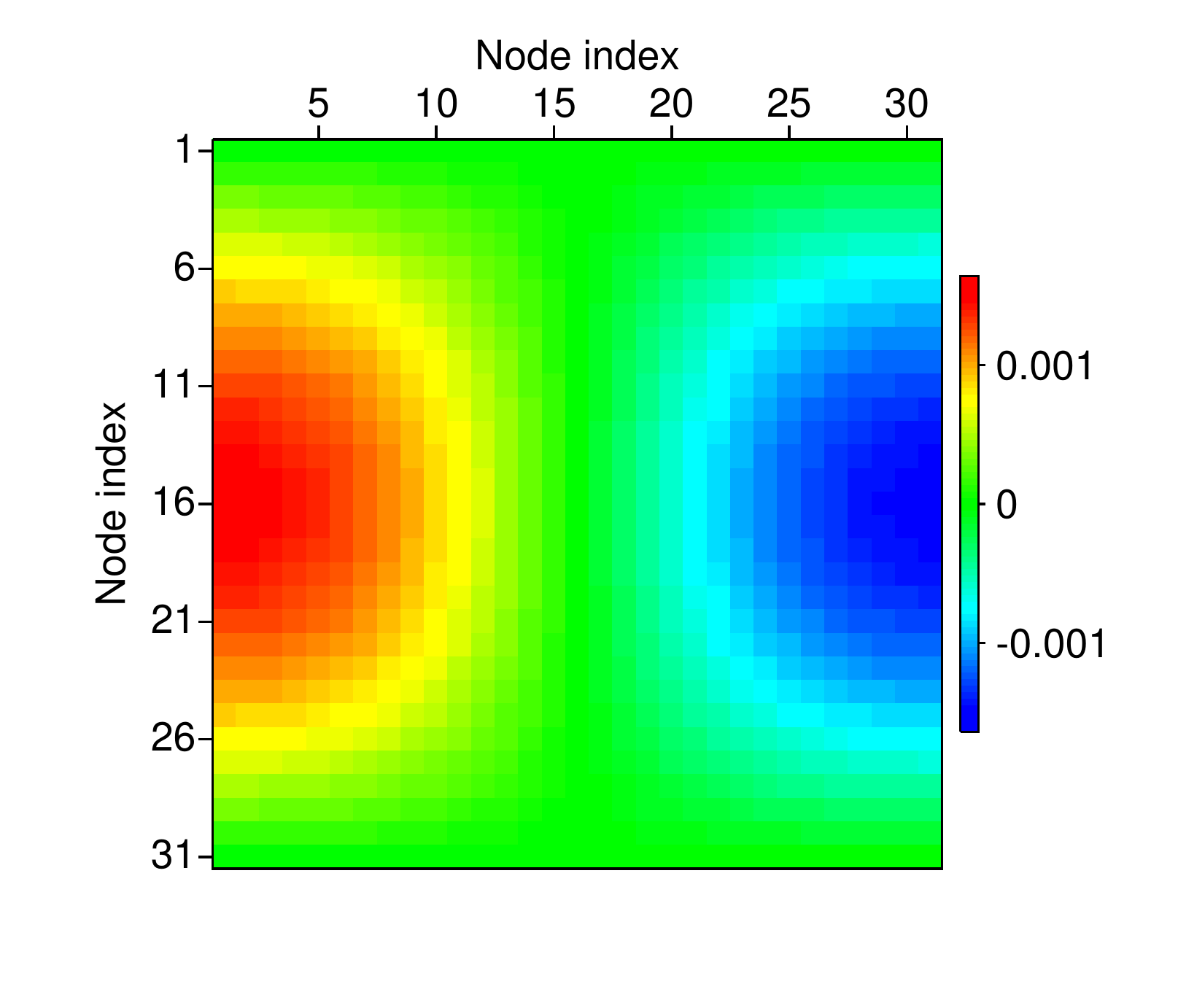}
}
\caption{(a)--(f) represent the $u_1$ component of the first 6 \revision{type I multiscale} basis functions corresponding with the first 6 smallest eigenvalues for an isotropic homogeneous subgrid model.}
\end{figure}

Figures \ref{fig:typeI_isohomo_u1_1}--\ref{fig:typeI_isohomo_u1_6} represent the $u_1$ component of the first 6 eigenfunctions corresponding to the first 6 smallest eigenvalues obtained by solving the local eigenvalue problem for an isotropic homogeneous subgrid model, with elastic parameters $C_{11}$=10.0~GPa and $C_{55}$=4.0~GPa, $C_{33}=C_{11}$, $C_{13}=C_{11}-2C_{55}$, $C_{13}=C_{15}=0$, and density $\rho$=1000~kg/m$^3$. Note that the first eigenfunction in Figures \ref{fig:typeI_isohomo_u1_1} is constant, 
corresponding to the constant solution that satisfies local problem \ref{eq:local_eig} by default. 

\begin{figure}
\centering
\subfigure[]{
\label{fig:typeI_anisohomo_u1_1}
\includegraphics[trim=50 50 30 0,width=0.415\textwidth]{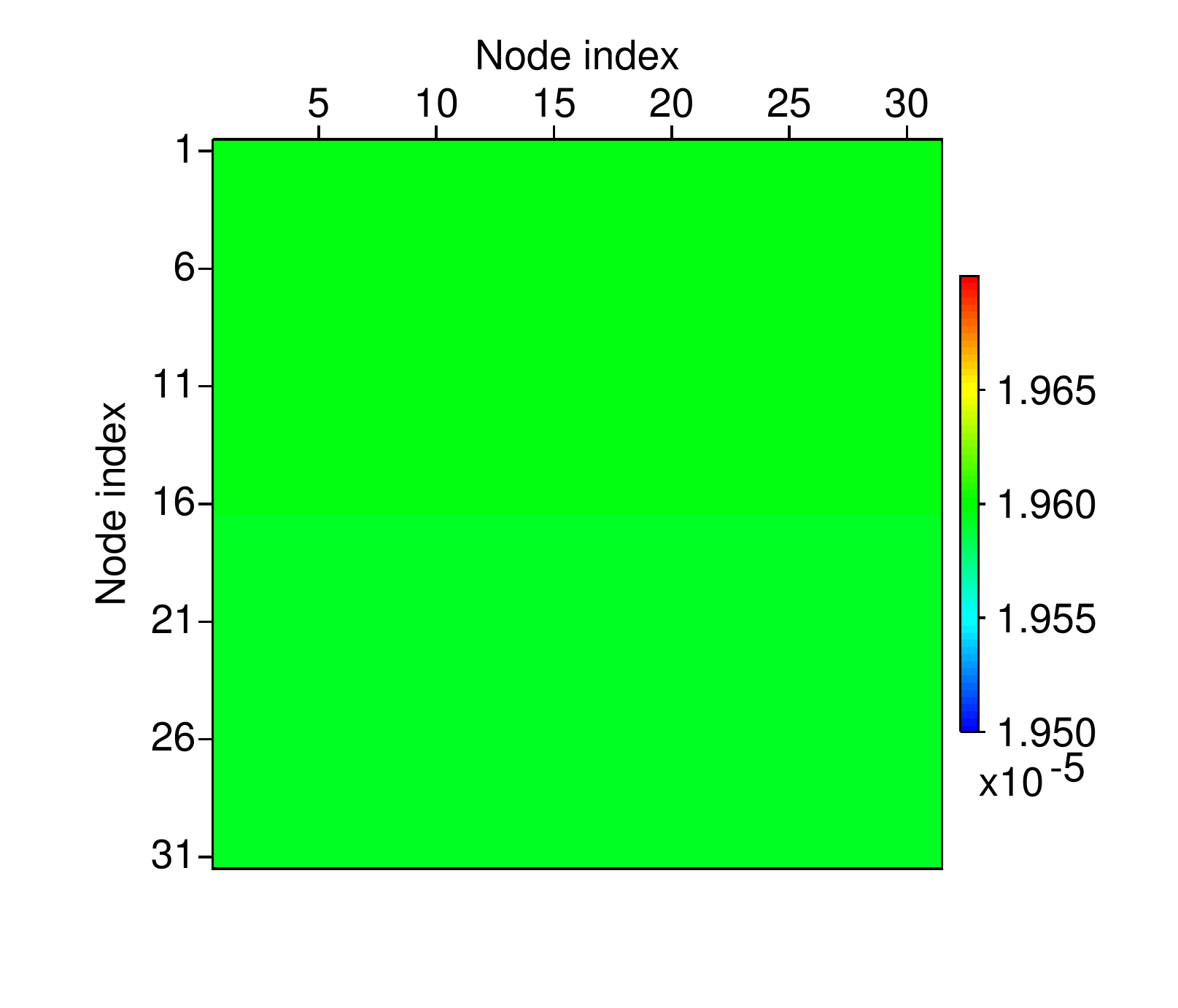}
}
\subfigure[]{
\label{fig:typeI_anisohomo_u1_2}
\includegraphics[trim=50 50 30 0,width=0.415\textwidth]{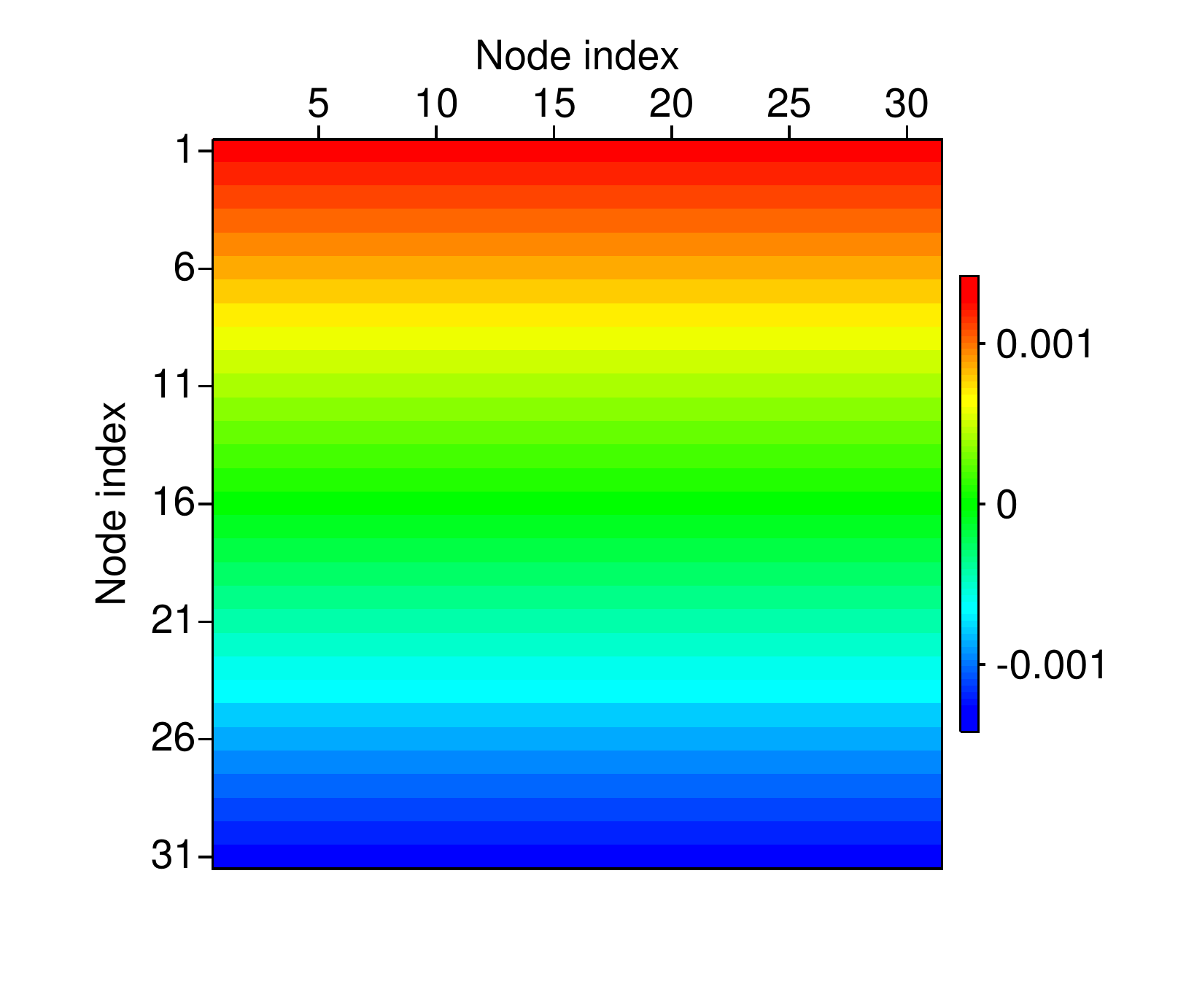}
}
\subfigure[]{
\label{fig:typeI_anisohomo_u1_3}
\includegraphics[trim=50 50 30 0,width=0.415\textwidth]{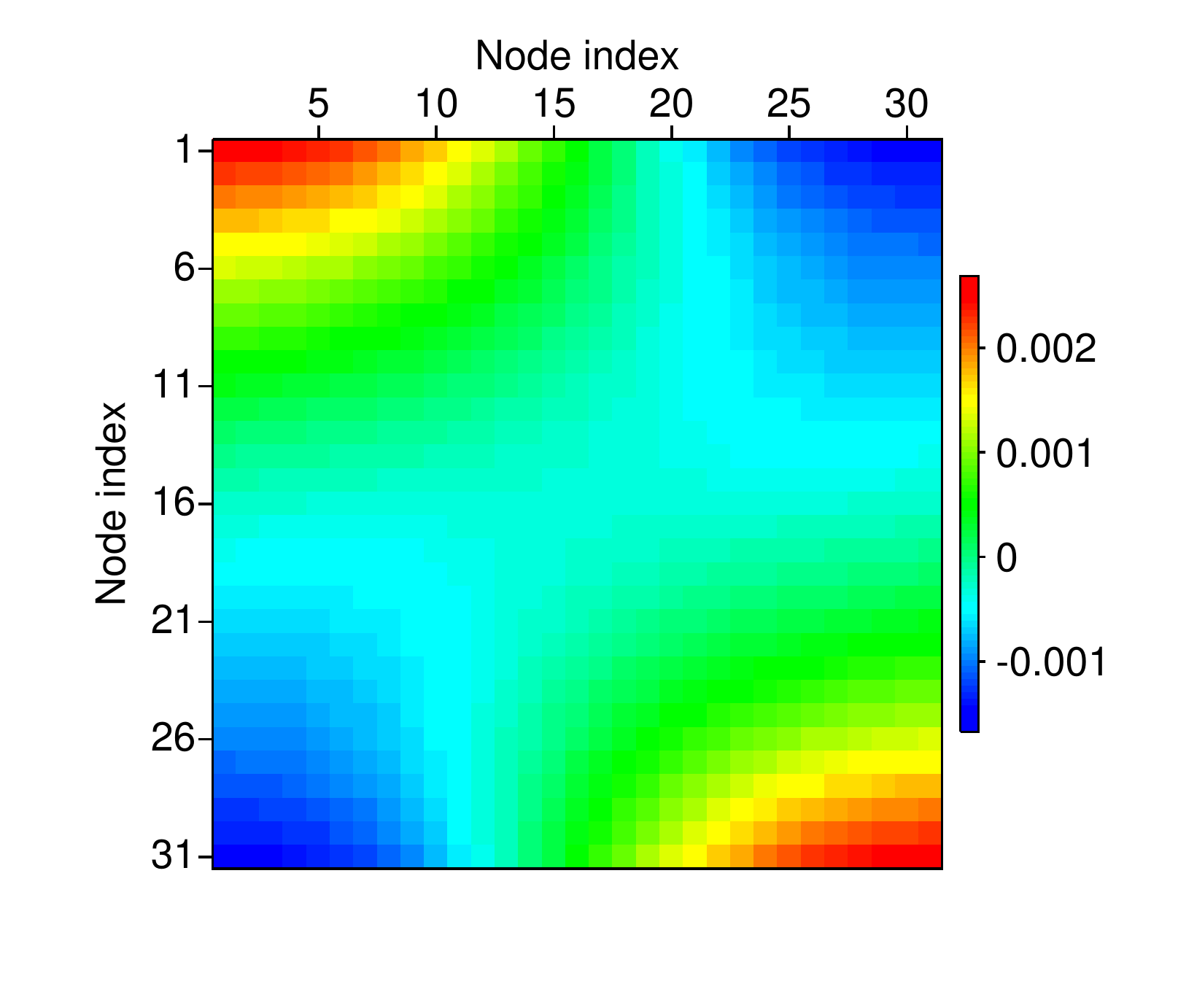}
}
\subfigure[]{
\label{fig:typeI_anisohomo_u1_4}
\includegraphics[trim=50 50 30 0,width=0.415\textwidth]{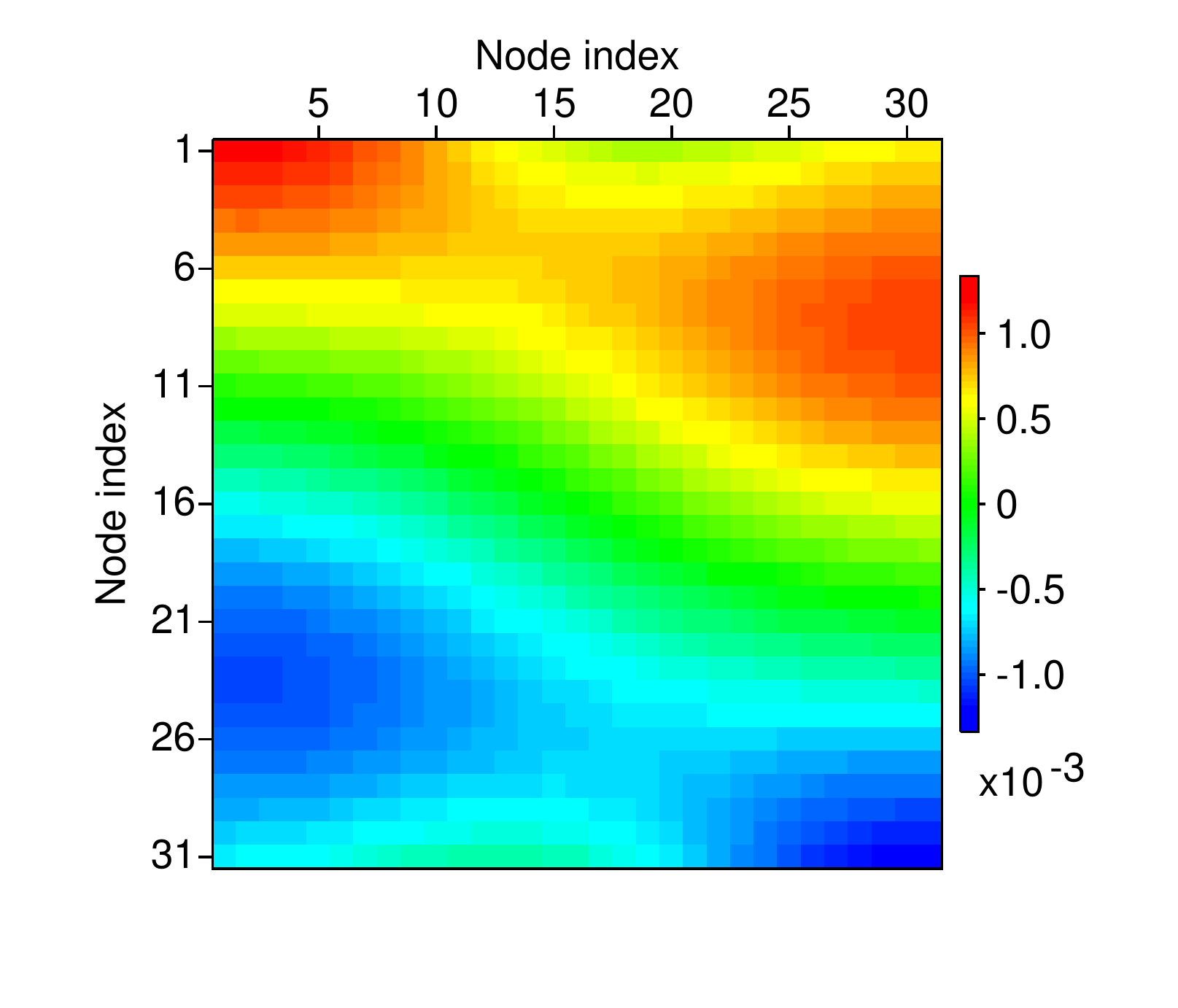}
}
\subfigure[]{
\label{fig:typeI_anisohomo_u1_5}
\includegraphics[trim=50 50 30 0,width=0.415\textwidth]{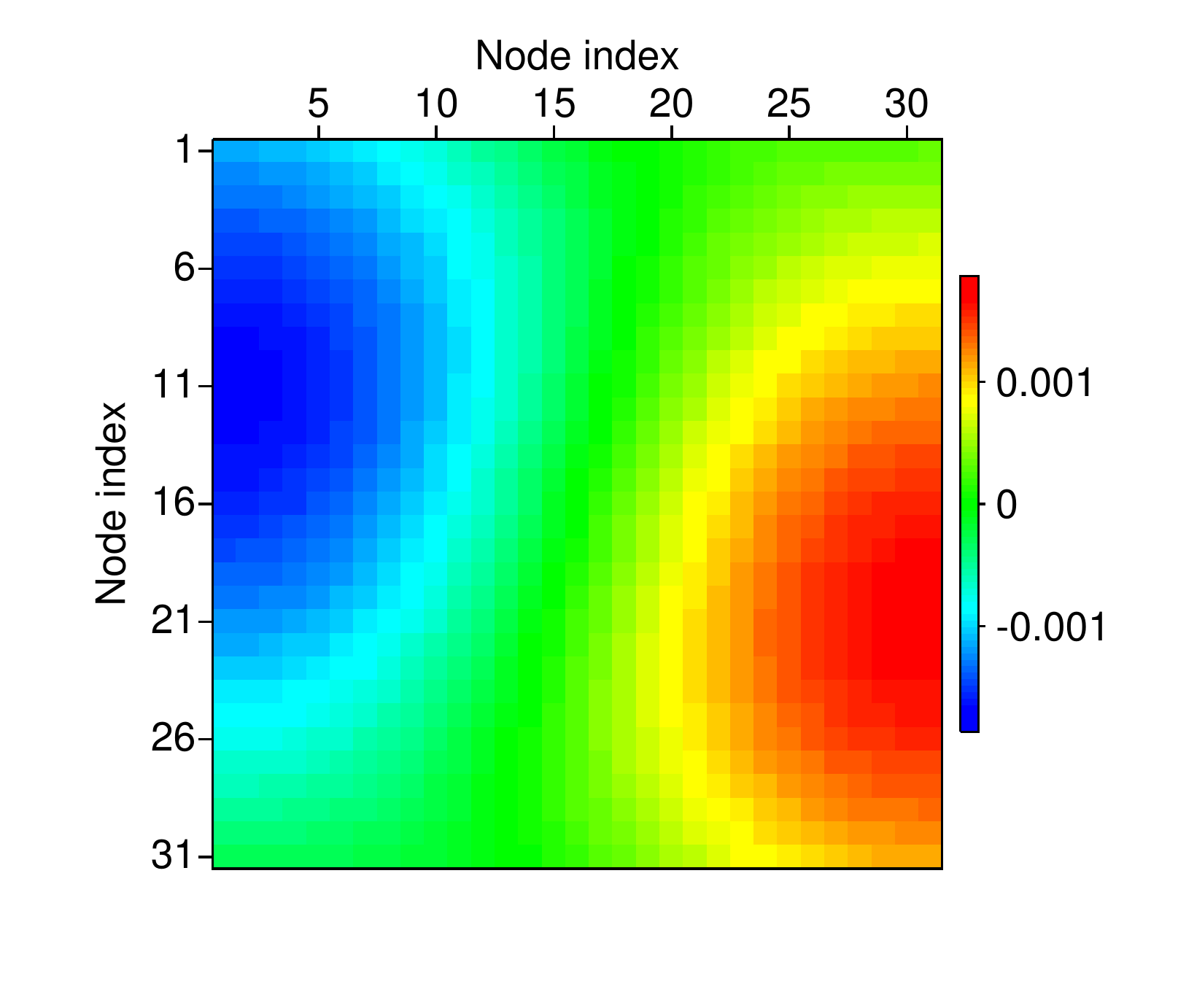}
}
\subfigure[]{
\label{fig:typeI_anisohomo_u1_6}
\includegraphics[trim=50 50 30 0,width=0.415\textwidth]{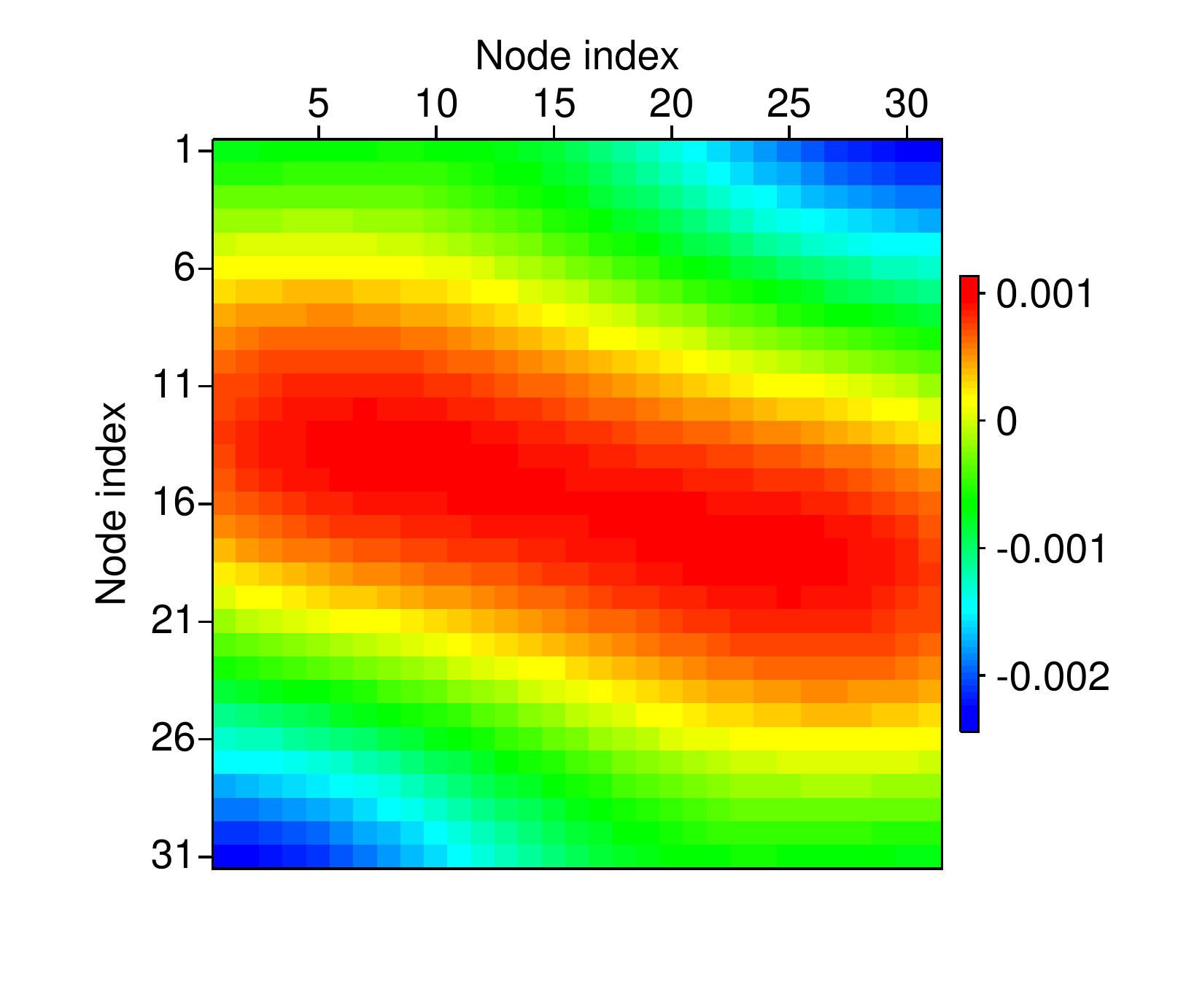}
}
\caption{(a)--(f) represent the $u_1$ component of the first 6 \revision{type I multiscale} basis functions corresponding with the first 6 smallest eigenvalues for an anisotropic homogeneous subgrid model. These basis functions are clearly different from those \revision{in Figure \ref{fig:typeI_isohomo_u1_1}--\ref{fig:typeI_isohomo_u1_6}, an important characteristic of the basis functions in our GMsFEM that they are affected by the medium properties.}}
\end{figure}

In contrast, Figures \ref{fig:typeI_anisohomo_u1_1}--\ref{fig:typeI_anisohomo_u1_6} show an example of selecting the first 6 eigenfunctions for a 2D TTI homogeneous subgrid model, with elasticity constants $C_{11}$=10.5~GPa, $C_{13}$=3.25~GPa, $C_{15}$=-0.65~GPa, $C_{33}$=13.0~GPa, $C_{35}$=-1.52~GPa and $C_{55}$=4.75~GPa, and density $\rho$=1000~kg/m$^3$. The spectral basis functions clearly have different \revision{spatial} patterns than those in isotropic homogeneous medium, and it is this difference that results in the different kinetic, dynamic and anisotropy patterns in the seismic wavefields.

\begin{figure}
\centering
\subfigure[]{
\label{fig:hetero_c11}
\includegraphics[trim=50 50 30 0,width=0.415\textwidth]{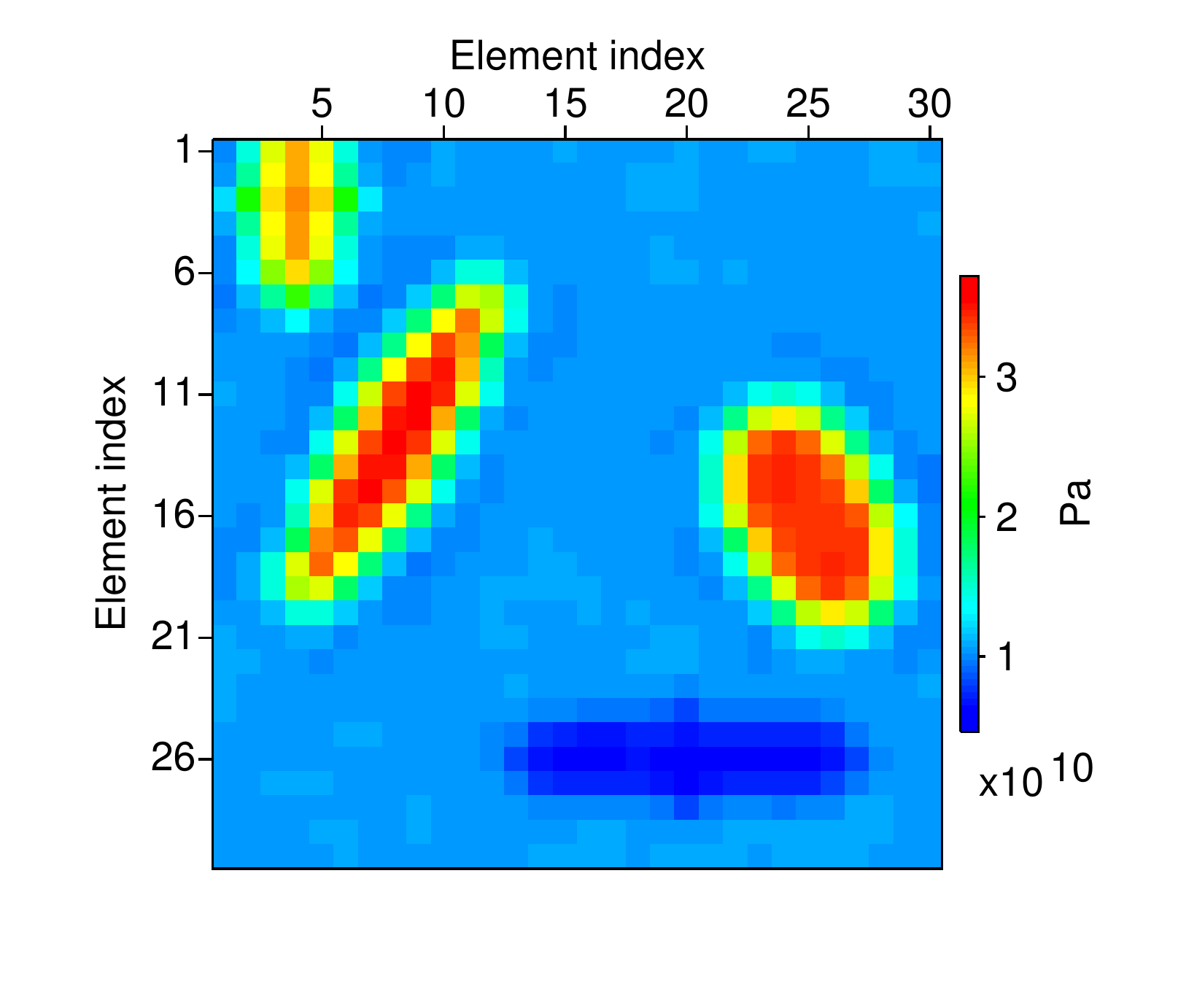}
}
\subfigure[]{
\label{fig:hetero_c55}
\includegraphics[trim=50 50 30 0,width=0.415\textwidth]{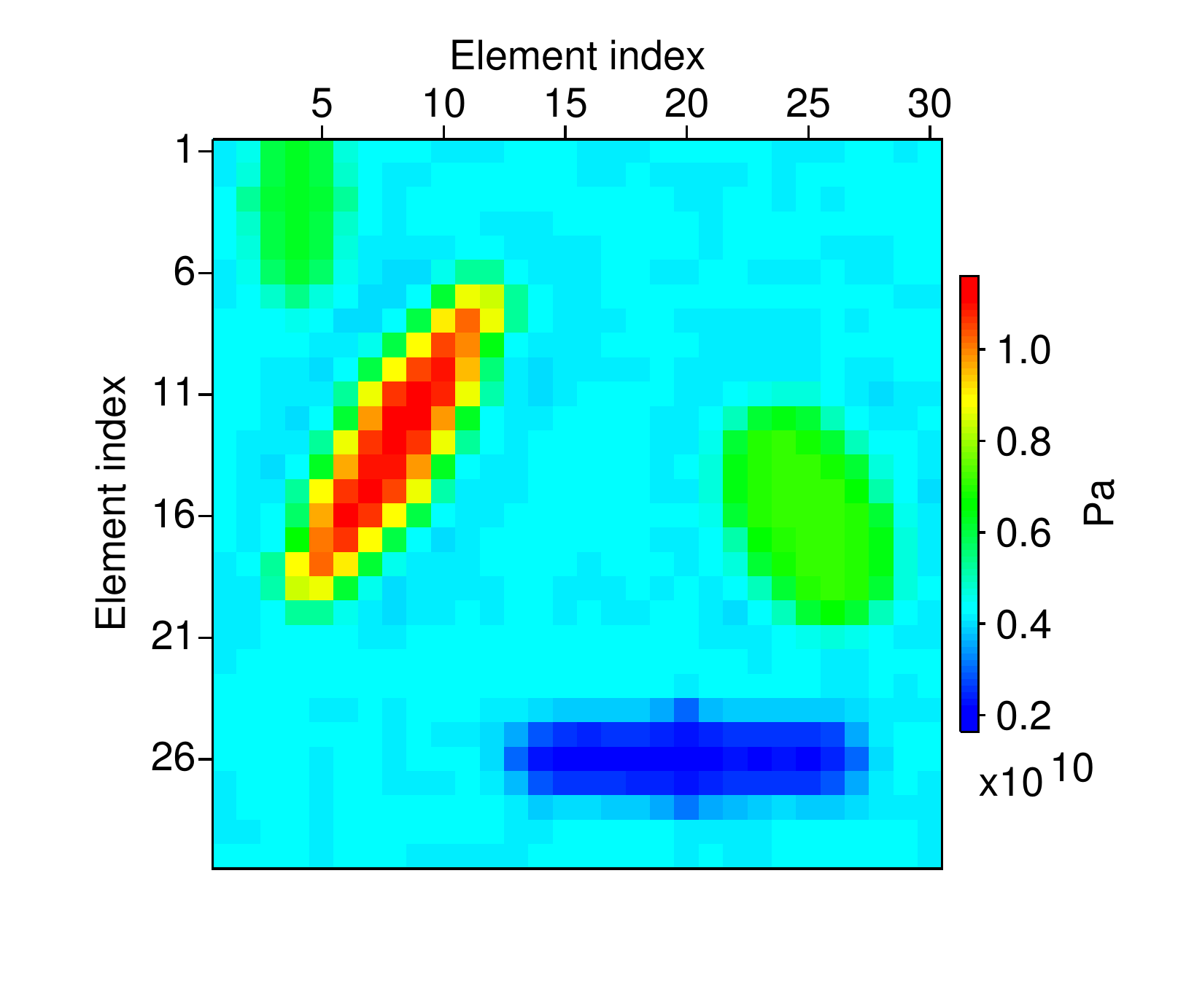}
}
\caption{Elasticity parameter variations within one coarse block. (a) and (b) represents $C_{11}$ and $C_{55}$, respectively. }
\end{figure}

\begin{figure}
\centering
\subfigure[]{
\label{fig:typeI_isohetero_u1_1}
\includegraphics[trim=50 50 30 0,width=0.415\textwidth]{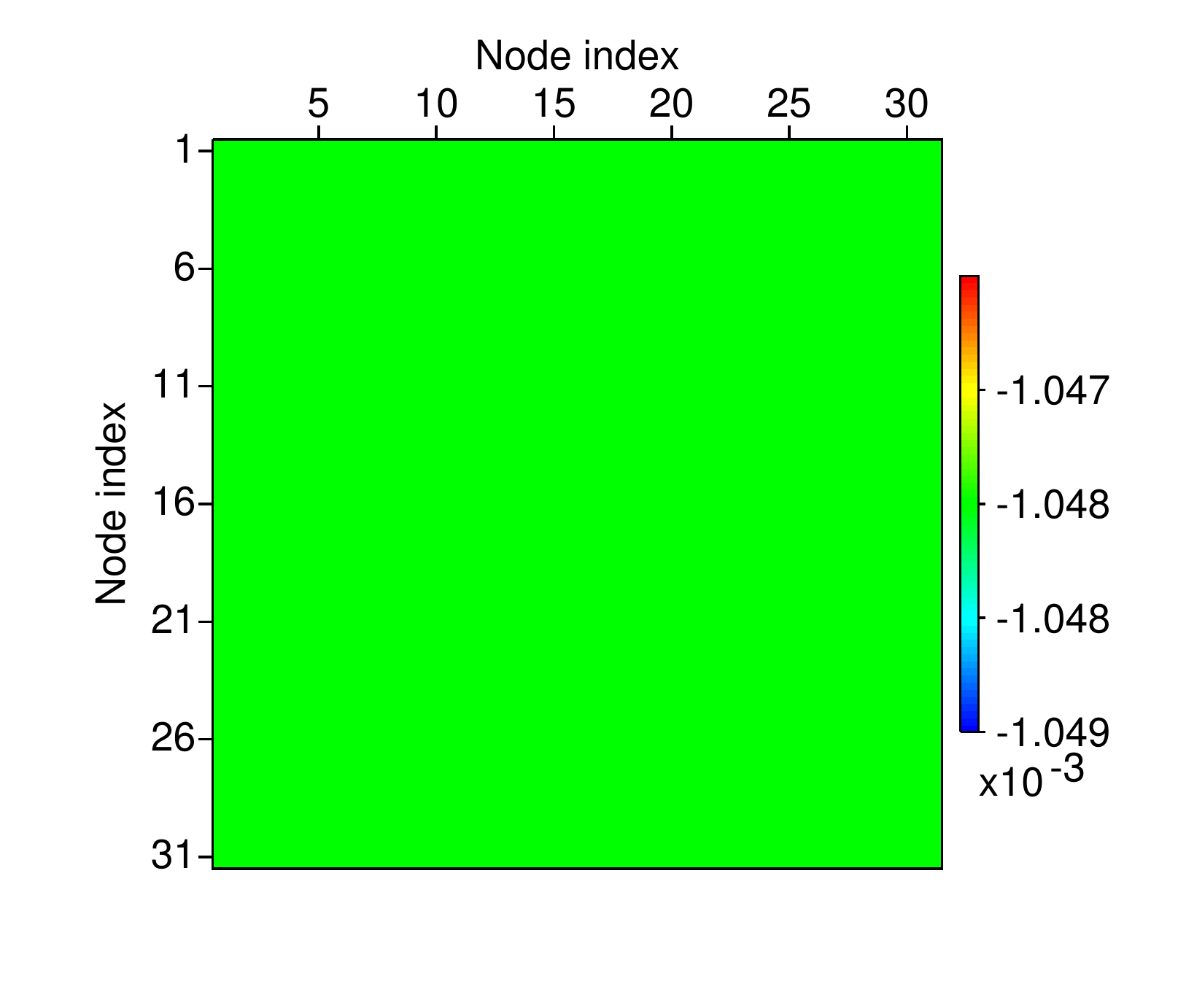}
}
\subfigure[]{
\label{fig:typeI_isohetero_u1_2}
\includegraphics[trim=50 50 30 0,width=0.415\textwidth]{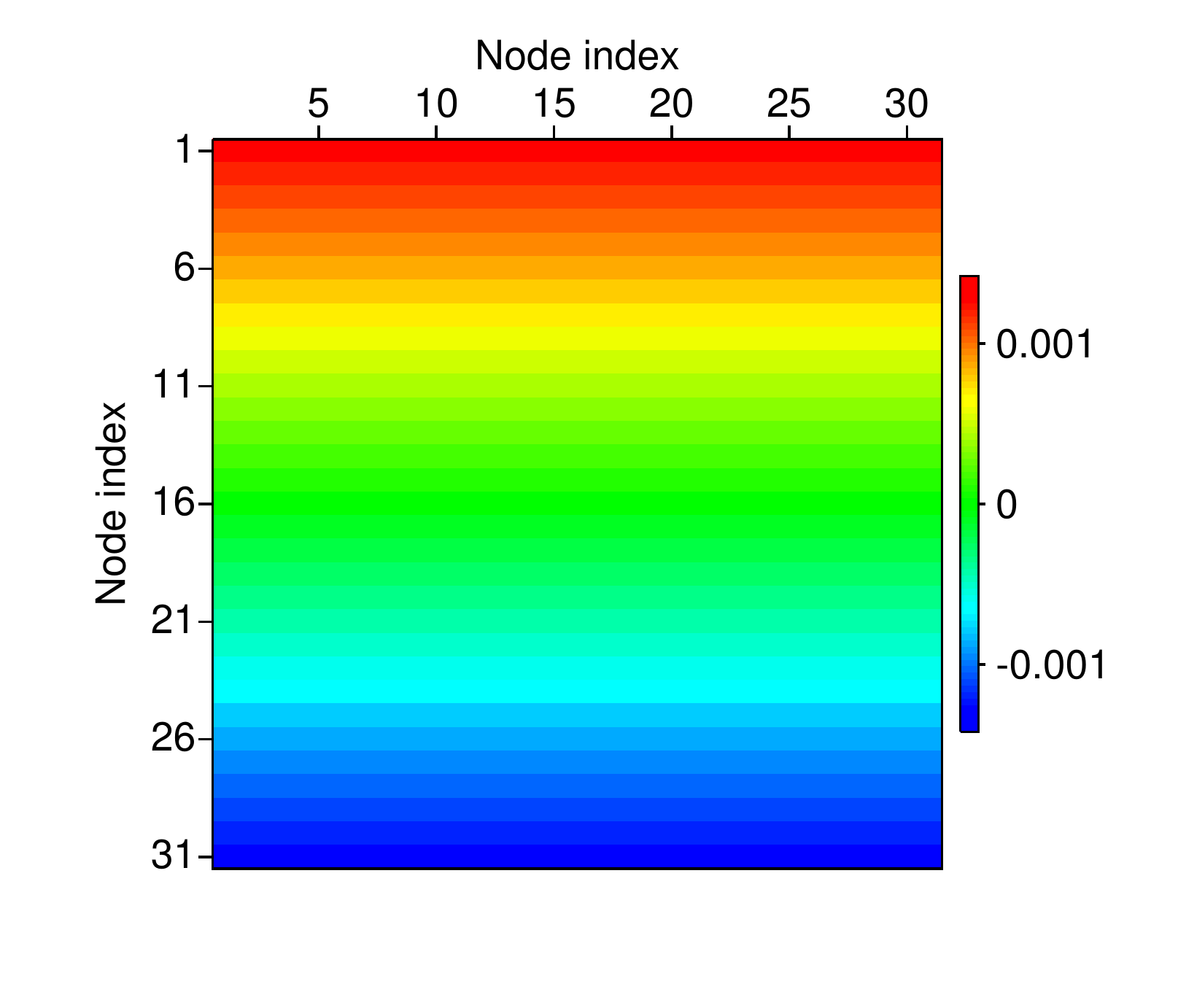}
}
\subfigure[]{
\label{fig:typeI_isohetero_u1_3}
\includegraphics[trim=50 50 30 0,width=0.415\textwidth]{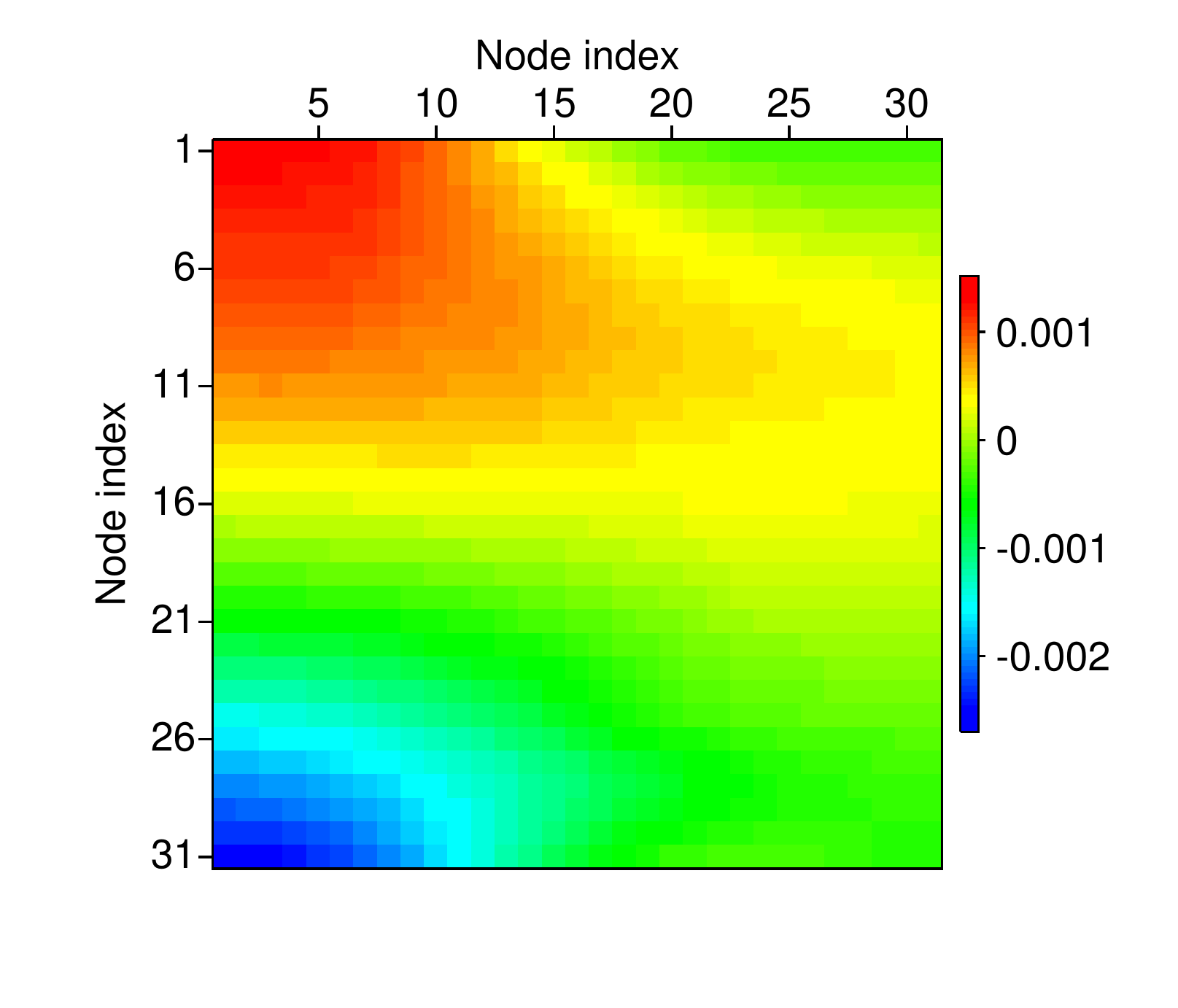}
}
\subfigure[]{
\label{fig:typeI_isohetero_u1_4}
\includegraphics[trim=50 50 30 0,width=0.415\textwidth]{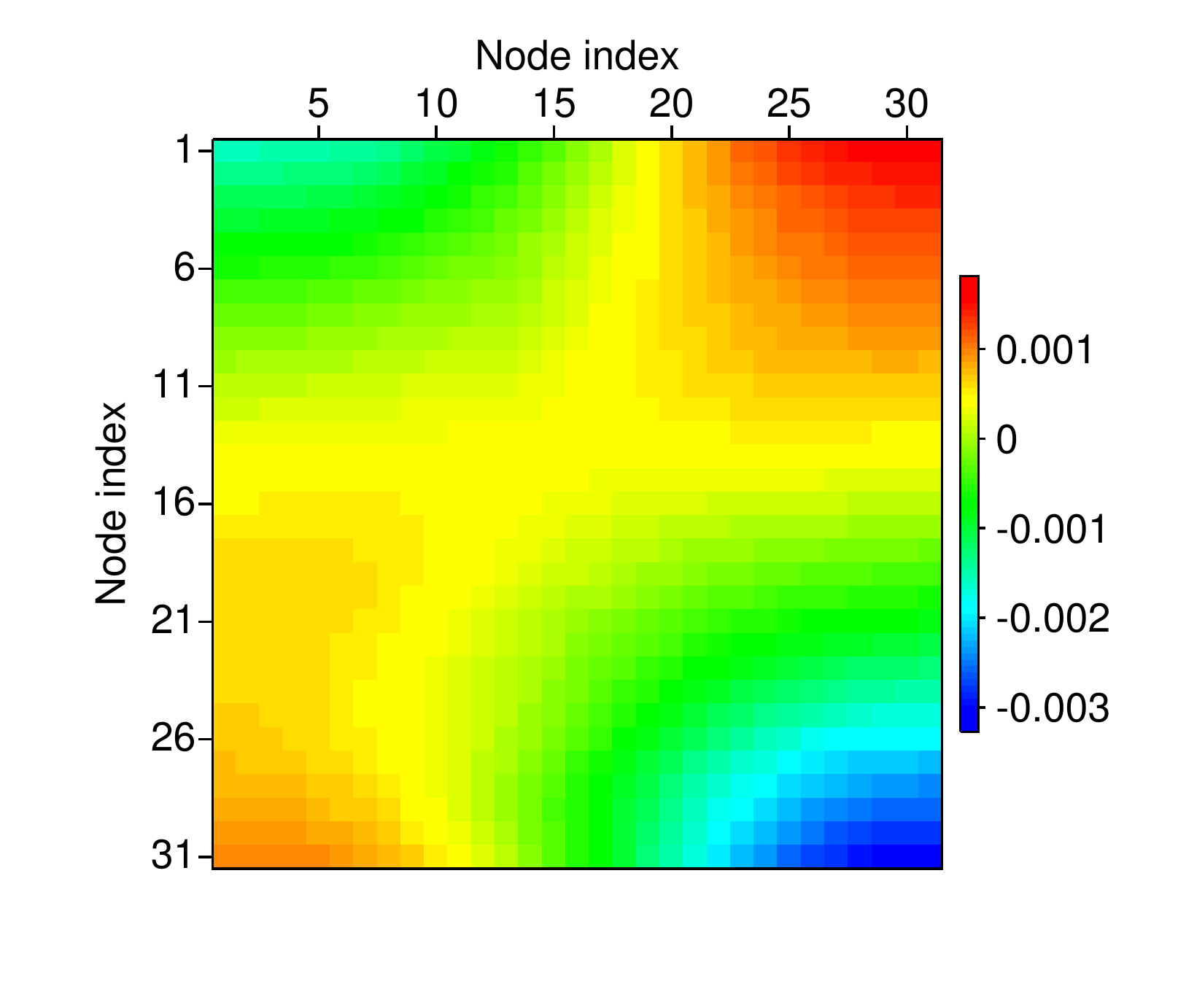}
}
\subfigure[]{
\label{fig:typeI_isohetero_u1_5}
\includegraphics[trim=50 50 30 0,width=0.415\textwidth]{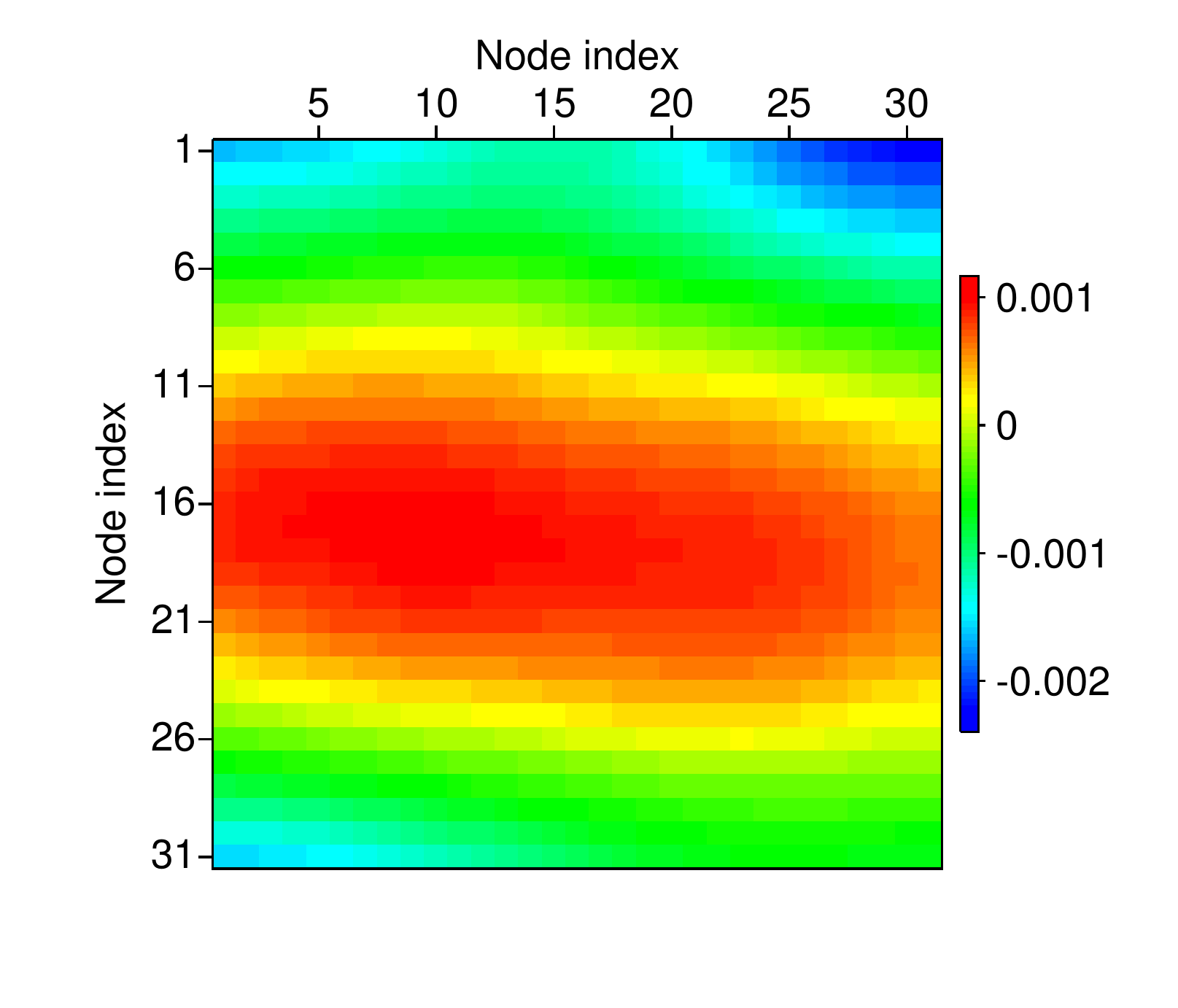}
}
\subfigure[]{
\label{fig:typeI_isohetero_u1_6}
\includegraphics[trim=50 50 30 0,width=0.415\textwidth]{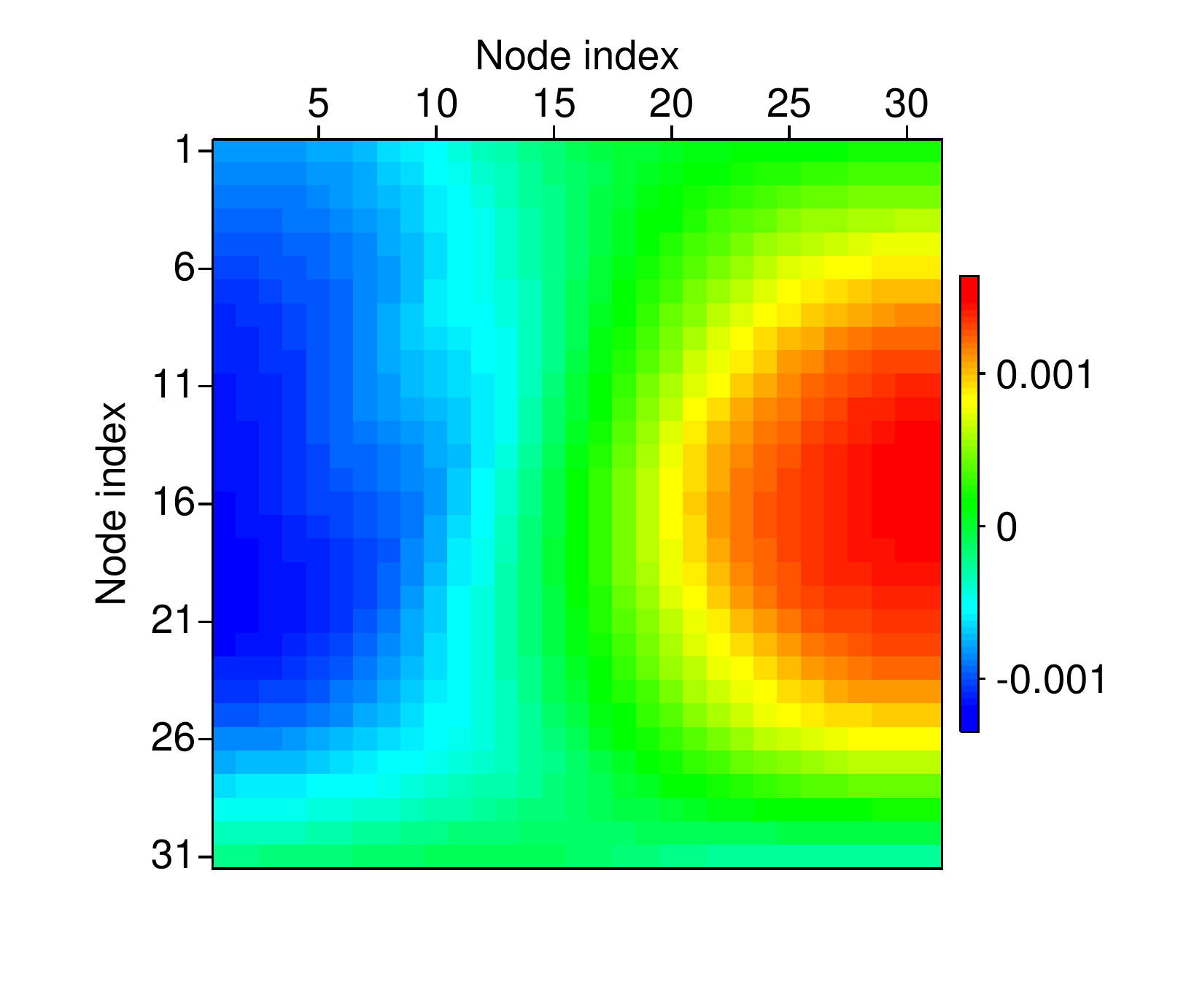}
}
\caption{(a)--(f) represent the $u_1$ component of the first 6 \revision{type I multiscale} basis functions corresponding with the first 6 smallest eigenvalues for an isotropic heterogeneous subgrid model in Figure \ref{fig:hetero_c11} and \ref{fig:hetero_c55}. \revision{Note the asymmetric spatial pattern of these basis functions (except the first basis function), which is a result of the heterogeneities in the subgrid model. }}
\end{figure}

Complex heterogeneities will also introduce variations in the local spectral basis functions. Figures \ref{fig:hetero_c11} and \ref{fig:hetero_c55} show a subgrid model that contains several elliptic inclusions and some random heterogeneities on a homogeneous isotropic elastic background. Figures \ref{fig:typeI_isohetero_u1_1}--\ref{fig:typeI_isohetero_u1_6} show the first 6 eigenfunctions for this subgrid model. Patterns of the eigenfunctions in this model are no long symmetric as in Figures \ref{fig:typeI_isohomo_u1_1}--\ref{fig:typeI_isohomo_u1_6}, but contain spatial variations that are related to the shape and elastic properties of the heterogeneous inclusions.

\subsubsection{Type II}
\revision{Another way} to construct the multiscale basis functions \revision{for GMsFEM is to} decompose the basis function space into two parts,  i.e., \oldrevision{$W_H=W_H^1 \oplus  W_H^2$} \revision{for DG formulation or $V_H=V_H^1 \oplus  V_H^2$ for CG formulation}, which is an elastic extension of the acoustic wave equation case \cite{Chung-etal_2013a,Chung-etal_2013b}.

\begin{figure}
\centering
\subfigure[]{
\label{fig:typeII_isohetero_u1_interior_1}
\includegraphics[trim=50 50 30 0,width=0.415\textwidth]{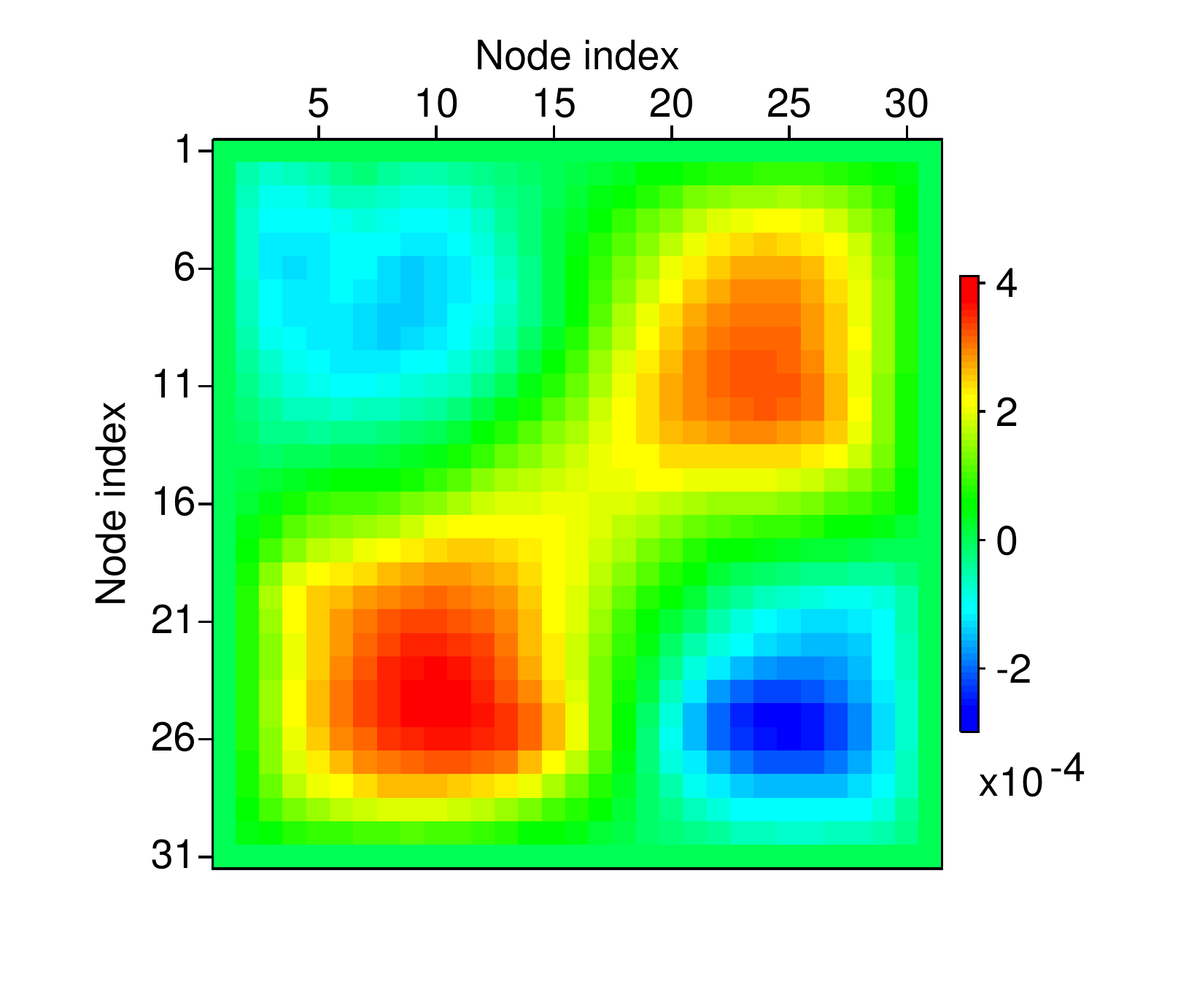}
}
\subfigure[]{
\label{fig:typeII_isohetero_u1_interior_2}
\includegraphics[trim=50 50 30 0,width=0.415\textwidth]{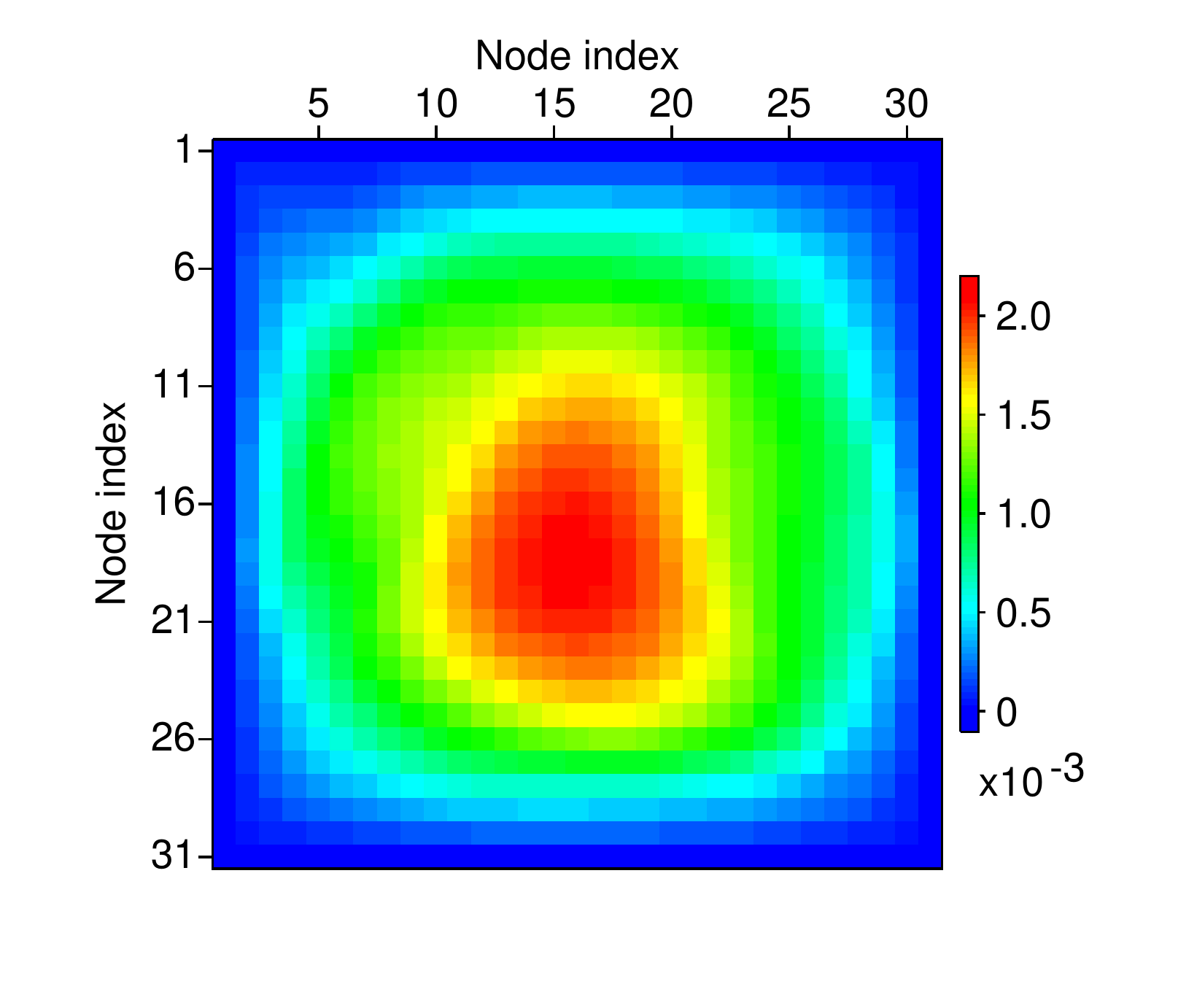}
}
\subfigure[]{
\label{fig:typeII_isohetero_u1_interior_3}
\includegraphics[trim=50 50 30 0,width=0.415\textwidth]{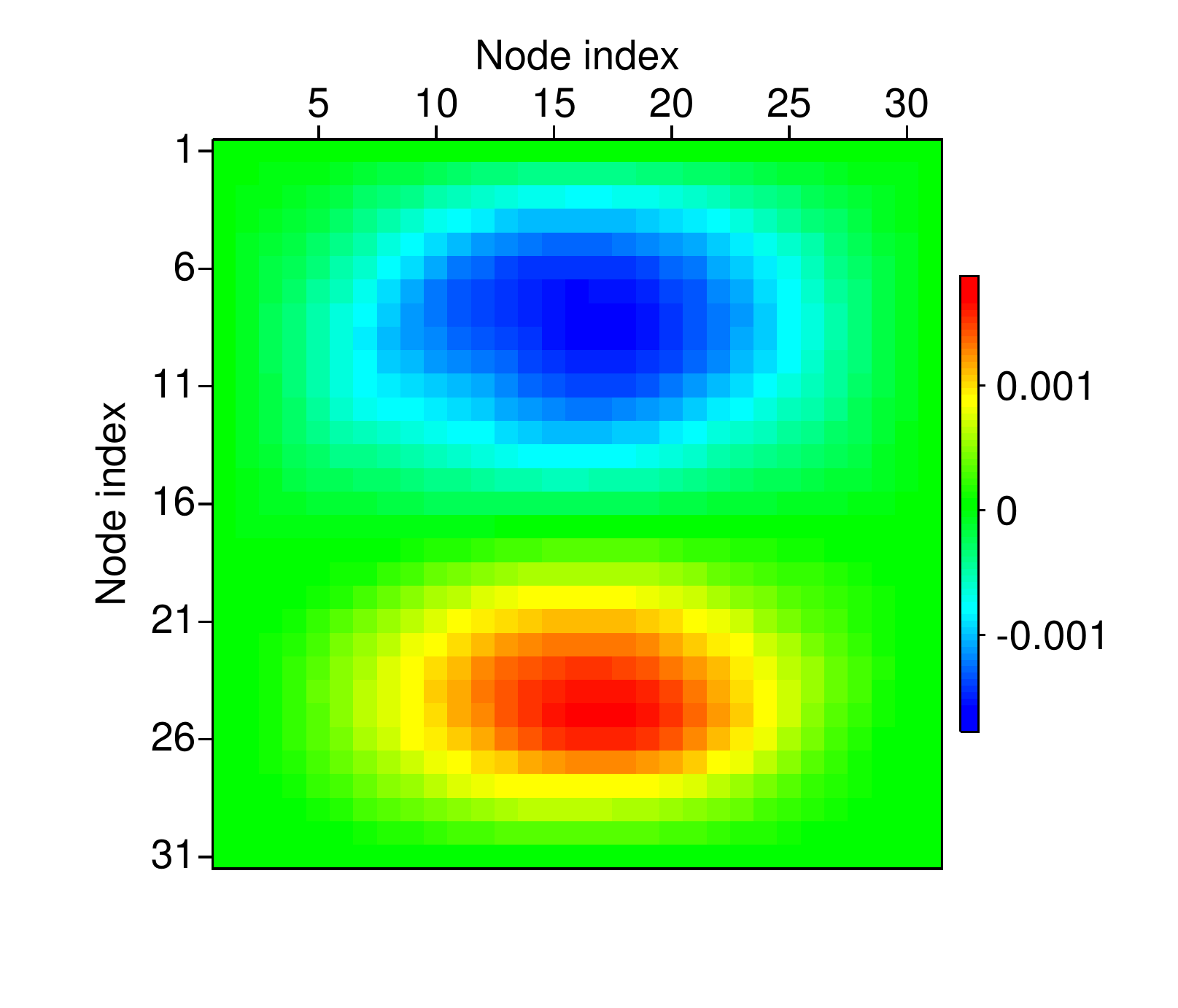}
}
\subfigure[]{
\label{fig:typeII_isohetero_u1_interior_4}
\includegraphics[trim=50 50 30 0,width=0.415\textwidth]{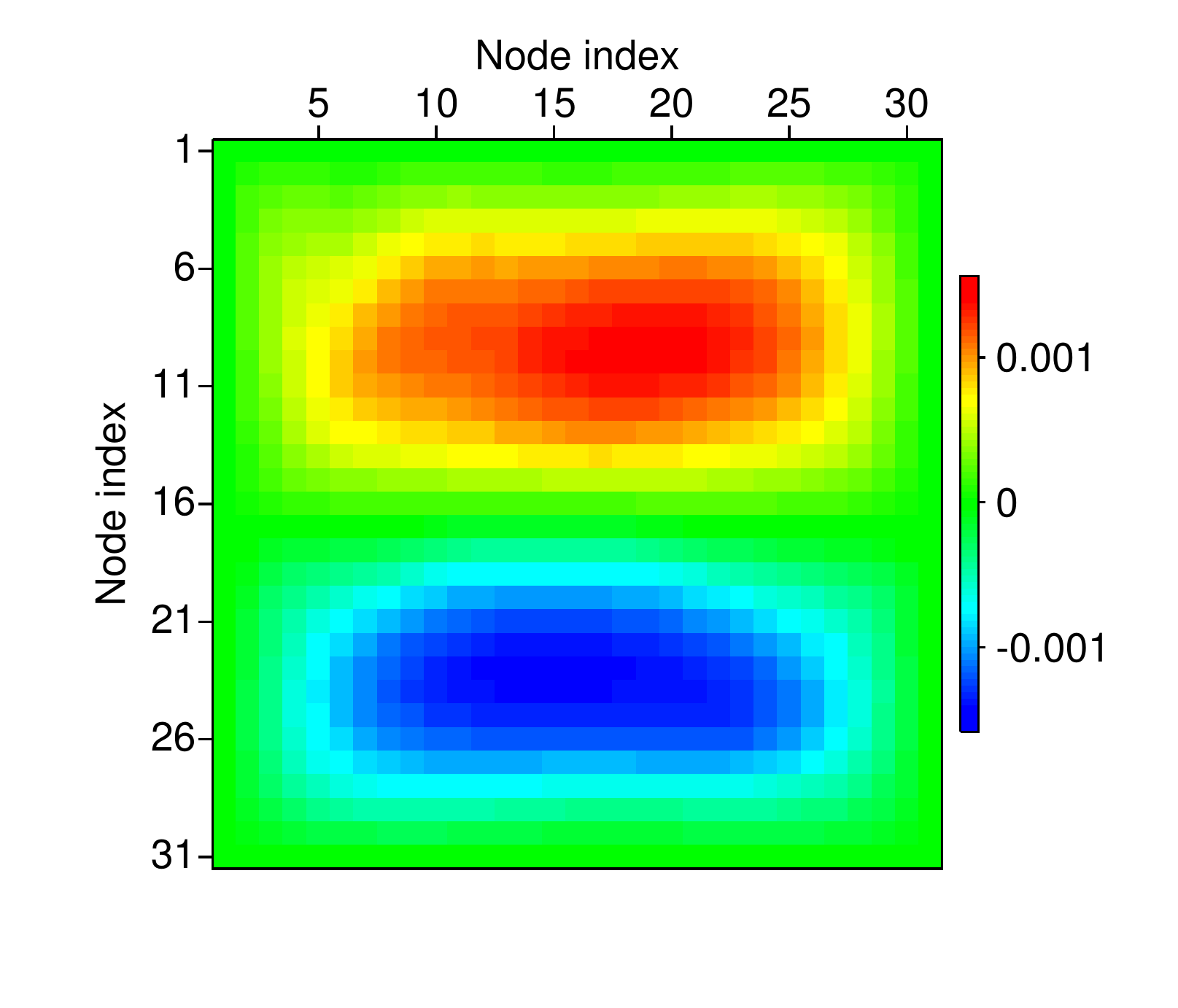}
}
\subfigure[]{
\label{fig:typeII_isohetero_u1_interior_5}
\includegraphics[trim=50 50 30 0,width=0.415\textwidth]{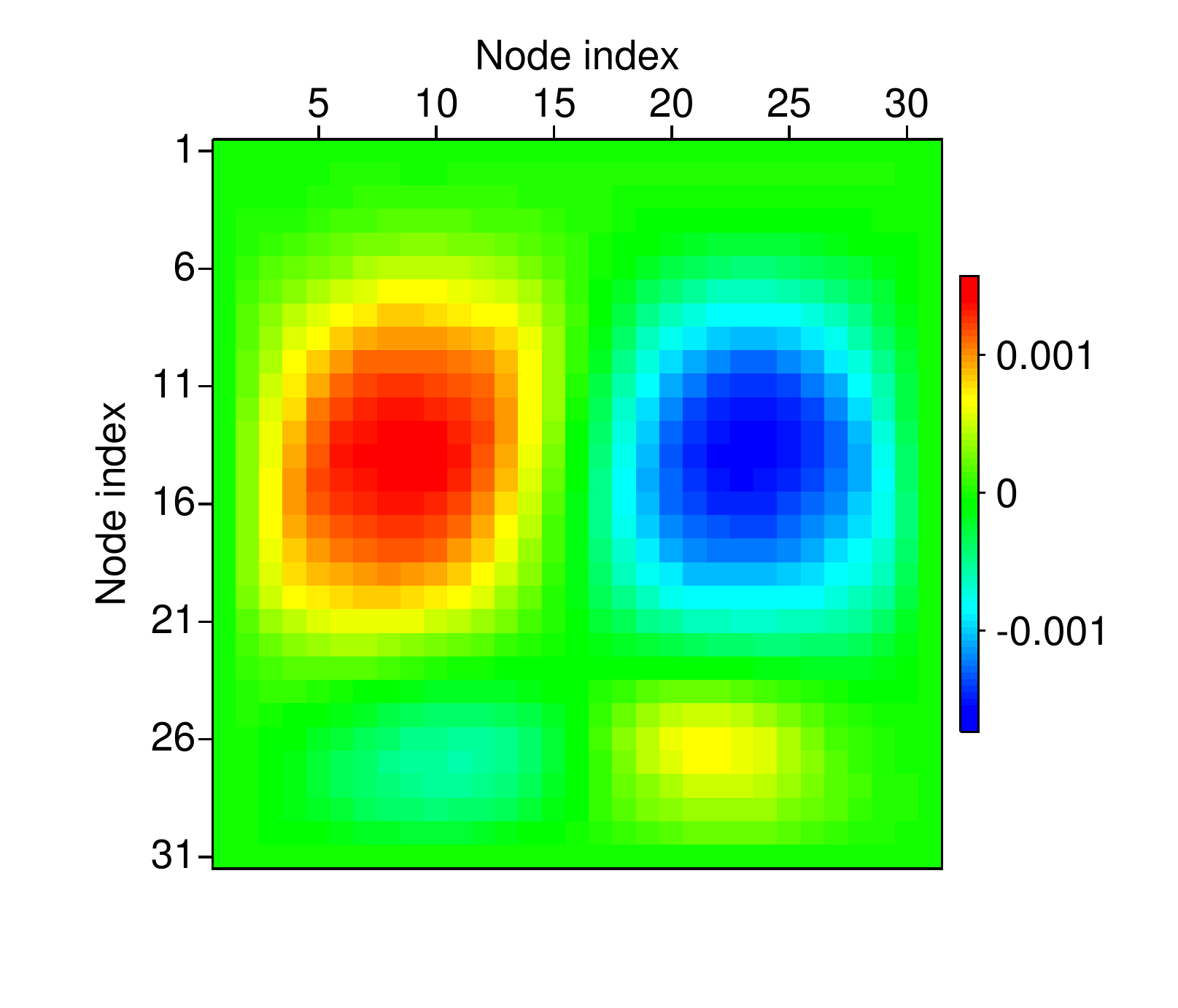}
}
\subfigure[]{
\label{fig:typeII_isohetero_u1_interior_6}
\includegraphics[trim=50 50 30 0,width=0.415\textwidth]{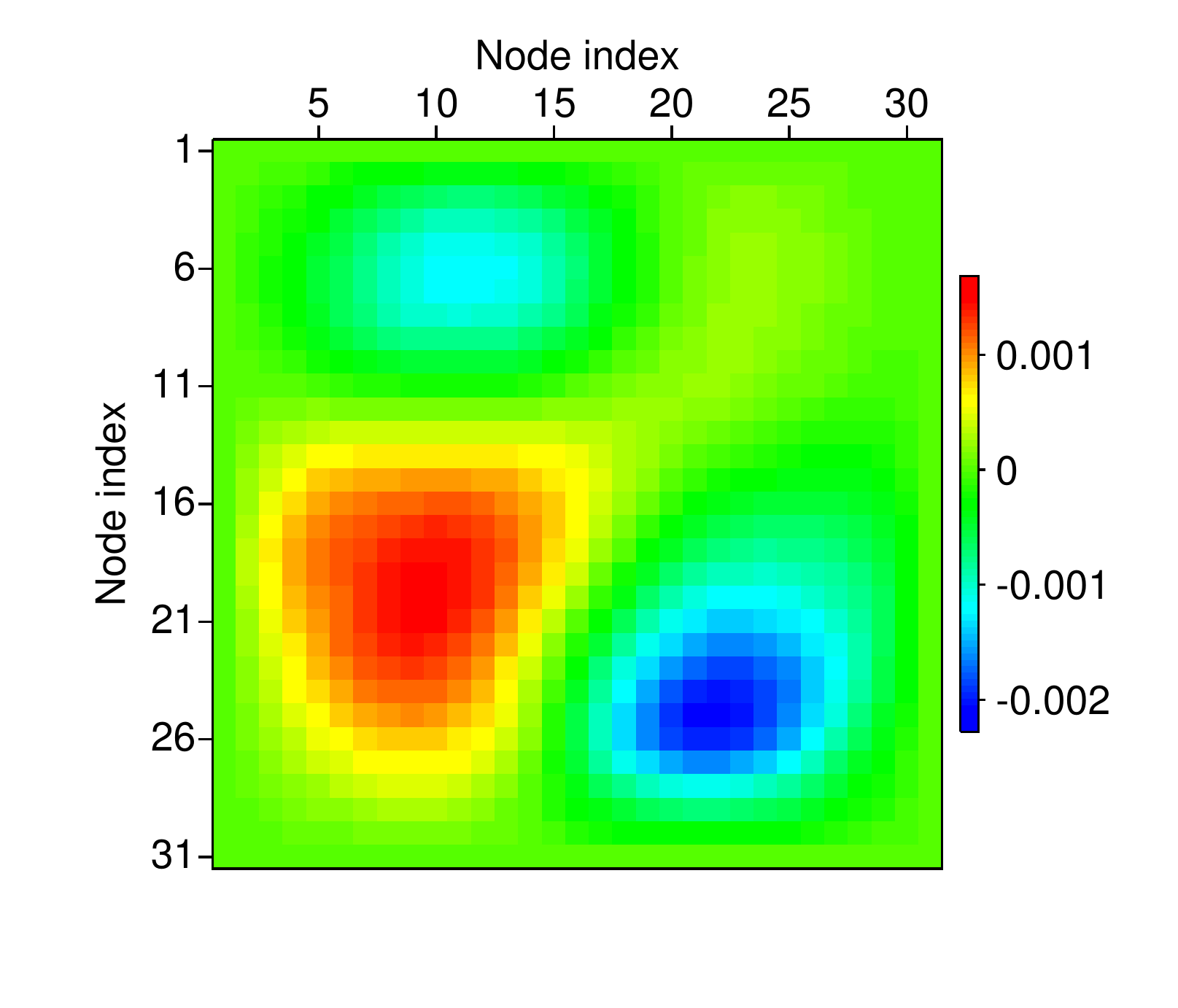}
}
\caption{(a)--(f) represent the $u_1$ component of the first 6 \revision{type II} interior basis functions for the isotropic heterogeneous subgrid model in Figures \ref{fig:hetero_c11} and \ref{fig:hetero_c55}.}
\end{figure}

The space $W_H^1$ is defined to capture the interior eigenmodes for $K$. Consider the local eigenvalue problem in $K$: find the pair $(\mathbf{u},\zeta)$ such that 
\begin{equation} \label{eq:typeii_2}
\begin{aligned}
-\nabla\cdot\boldsymbol{\sigma} & = \zeta \rho \mathbf{u}, \\
\boldsymbol{\sigma} & = \mathbf{c}: \boldsymbol{\varepsilon}, \\
\boldsymbol{\varepsilon} &=\frac{1}{2}[\nabla \mathbf{u}+(\nabla \mathbf{u})^{\mathrm{T}}], 
\end{aligned}
\end{equation}
where we set zero Dirichlet boundary condition, i.e., $\mathbf{u}=\mathbf{0}$ on $\partial K$. The above local problem corresponds with the following system: 
\begin{equation}
\mathbf{A}_{\text{interior}}\mathbf{U}=\zeta \mathbf{M}_{\text{interior}}\mathbf{U}, \label{eq:local_eig_interior}
\end{equation}
with $\mathbf{A}$ and $\mathbf{M}$ defined in equations \ref{eq:local_A} and \ref{eq:local_M}, respectively, and the subscript ``interior'' represents the nodes that are not on $\partial K$.  This local problem is quite similar in form with that defined in equation \ref{eq:local_eig}, 
but the solutions will be fundamentally different due to different boundary conditions in these two problems. 
In a similar way \RLG{to} previous local problems, we will select the first $m_1$ eigenfunctions $\boldsymbol{\phi}_1,\boldsymbol{\phi}_2,\cdots,\boldsymbol{\phi}_{m_1}$ corresponding to the first $m_1$ smallest eigenvalues $0\leq \zeta_1\leq \zeta_2 \leq \cdots \leq \zeta_{m_1}$ of the above problem, and then the space $W_H^1$ is defined as  
\begin{equation}
W_H^1(K)=\mathrm{span}\{\boldsymbol{\phi}_1,\boldsymbol{\phi}_2, \cdots, \boldsymbol{\phi}_{m_1}\}. \label{eq:type2_space_1}
\end{equation}
The multiscale basis functions from $W_H^1$ are called interior basis functions. 
\revision{Figures \ref{fig:typeII_isohetero_u1_interior_1}--\ref{fig:typeII_isohetero_u1_interior_6} show the corresponding first 6 interior basis functions solved from local spectral problem \ref{eq:local_eig_interior} for the isotropic heterogeneous model mentioned in the example for type I basis function, and it should be noted that the interior basis functions are different from those defined through local spectral problem in equation \ref{eq:local_eig} (Figure \ref{fig:typeI_isohetero_u1_1}--\ref{fig:typeI_isohetero_u1_6}). Clearly, the interior basis functions can also capture the fine-scale variations of medium properties as type I basis functions}

In the above definition for interior basis functions, we have set $\mathbf{u}=\mathbf{0}$ on $\partial K$. Consequently, the solution cannot represent a wavefield propagating across grid cells and their boundaries $\partial K$. We therefore define the space $W_H^2$ which takes care of the contribution of the boundaries of $K$.  For a domain $K$, we first solve the local linear elasticity problem 
\begin{subequations}\label{eq:typeii_1}
\begin{align}
-\nabla\cdot\boldsymbol{\sigma} & = \mathbf{0}, \\
\boldsymbol{\sigma} & = \mathbf{c}: \boldsymbol{\varepsilon}, \\
\boldsymbol{\varepsilon} &=\frac{1}{2}[\nabla \mathbf{u}+(\nabla \mathbf{u})^{\mathrm{T}}], 
\end{align}
\end{subequations}
with Dirichlet boundary conditions $\mathbf{u}=\boldsymbol{\delta}_j$, where $j$ indexes
boundary nodes on $\partial K$. For example, in 2D, we can set $\mathbf{u}=(\delta_j,0)$ or $\mathbf{u}=(0,\delta_j)$ at the $j$-th boundary node of $K$, where $\delta_j$ is the delta function and $j=1,2,\cdots,p$, with $p$ being the total number of boundary nodes. We denote the solutions as $\mathbf{u}_1,\mathbf{u}_2, \cdots, \mathbf{u}_{dp}$, 
where $d=1$, 2, or 3 is the number of 
spatial dimensions, and then a trial basis function space $\tilde{W}_H^2$ is defined as
\begin{equation}
\tilde{W}_H^2(K)=\mathrm{span}\{\mathbf{u}_1,\mathbf{u}_2, \cdots, \mathbf{u}_{dp}\},
\end{equation}
For a rectangular $K$ that is composed of $30\times 30$ finer elements, for instance, a total of 240 solutions will be calculated with the two Dirichlet boundary conditions on each boundary node, and these solutions can effectively reflect the medium property variations within $K$ associated with varying values on the boundaries. \oldrevision{At  first glance, such a choice of boundary conditions could make the dimension of \RLG{the} local problem large. However, we illustrate in Appendix A that the solution of this series of local problems can be conveniently achieved without heavy computations}. In practice, we select only a few important modes from $\tilde{W}_H^2$ to form a basis function space $W_H^2$, and the important modes are obtained from the following local spectral problem defined in the 
trial basis function 
space $\tilde{W}_H^2$: 
\begin{equation}
\mathbf{A}\tilde{\mathbf{U}}=\xi \mathbf{N} \tilde{\mathbf{U}}, \label{eq:snapshot}
\end{equation}
where 
\begin{align}
\mathbf{A}=\int_{K}\boldsymbol{\sigma}(\tilde{\boldsymbol{\gamma}}): \boldsymbol{\varepsilon}(\tilde{\boldsymbol{\eta}}) d\mathbf{x}, \\
\mathbf{N}=\int_{\partial K} \rho \tilde{\boldsymbol{\gamma}} \cdot \tilde{\boldsymbol{\eta}} ds,
\end{align}
with $\tilde{\boldsymbol{\gamma}},\tilde{\boldsymbol{\eta}}\in \tilde{W}_H^2$. Note that $\mathbf{N}$ is a mass matrix that is related to the edge of $K$, distinct from the mass matrix $\mathbf{M}$ in equation \ref{eq:local_M}.

\begin{figure}
\centering
\subfigure[]{
\label{fig:typeII_isohetero_u1_boundary_1}
\includegraphics[trim=50 50 30 0,width=0.415\textwidth]{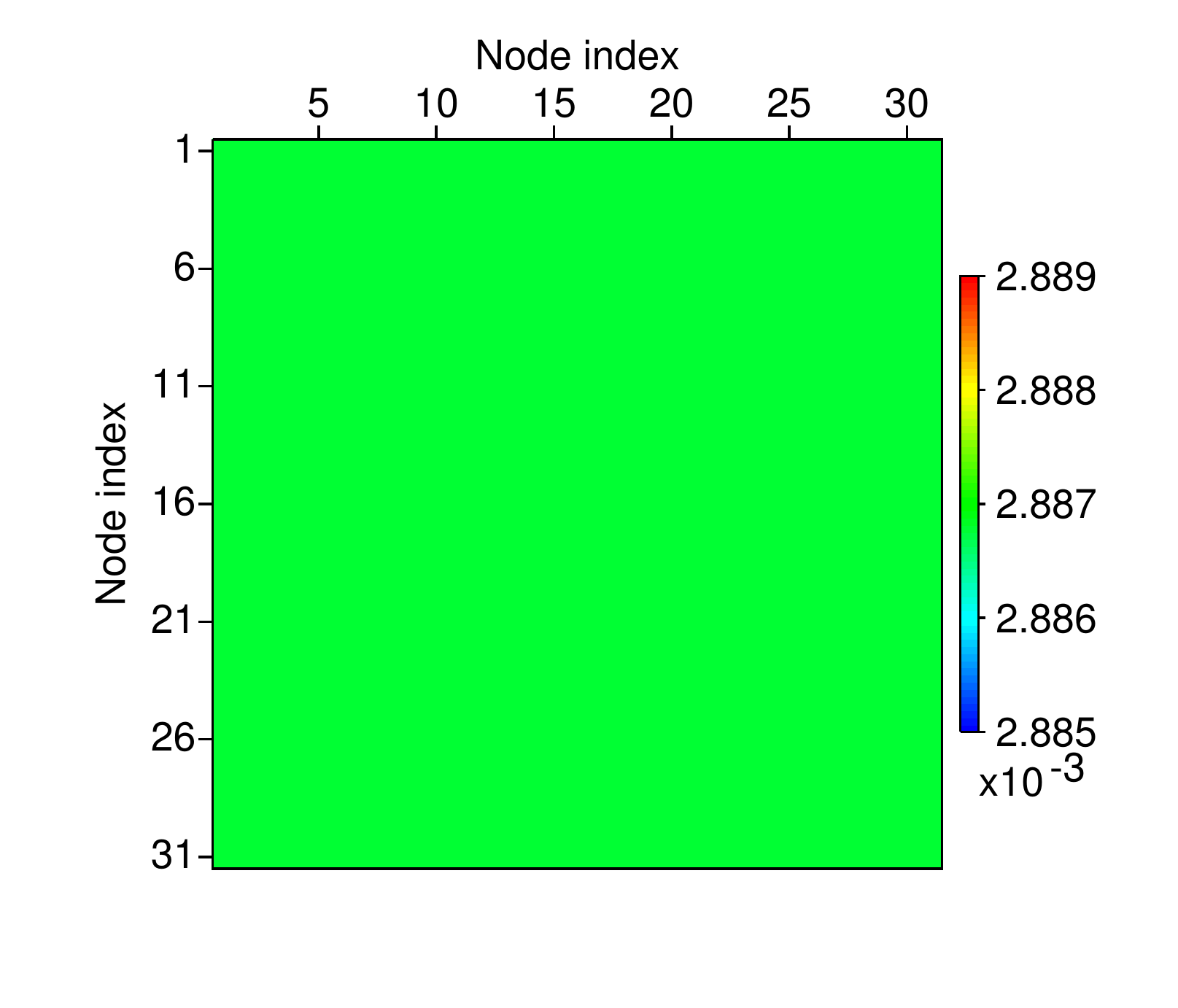}
}
\subfigure[]{
\label{fig:typeII_isohetero_u1_boundary_2}
\includegraphics[trim=50 50 30 0,width=0.415\textwidth]{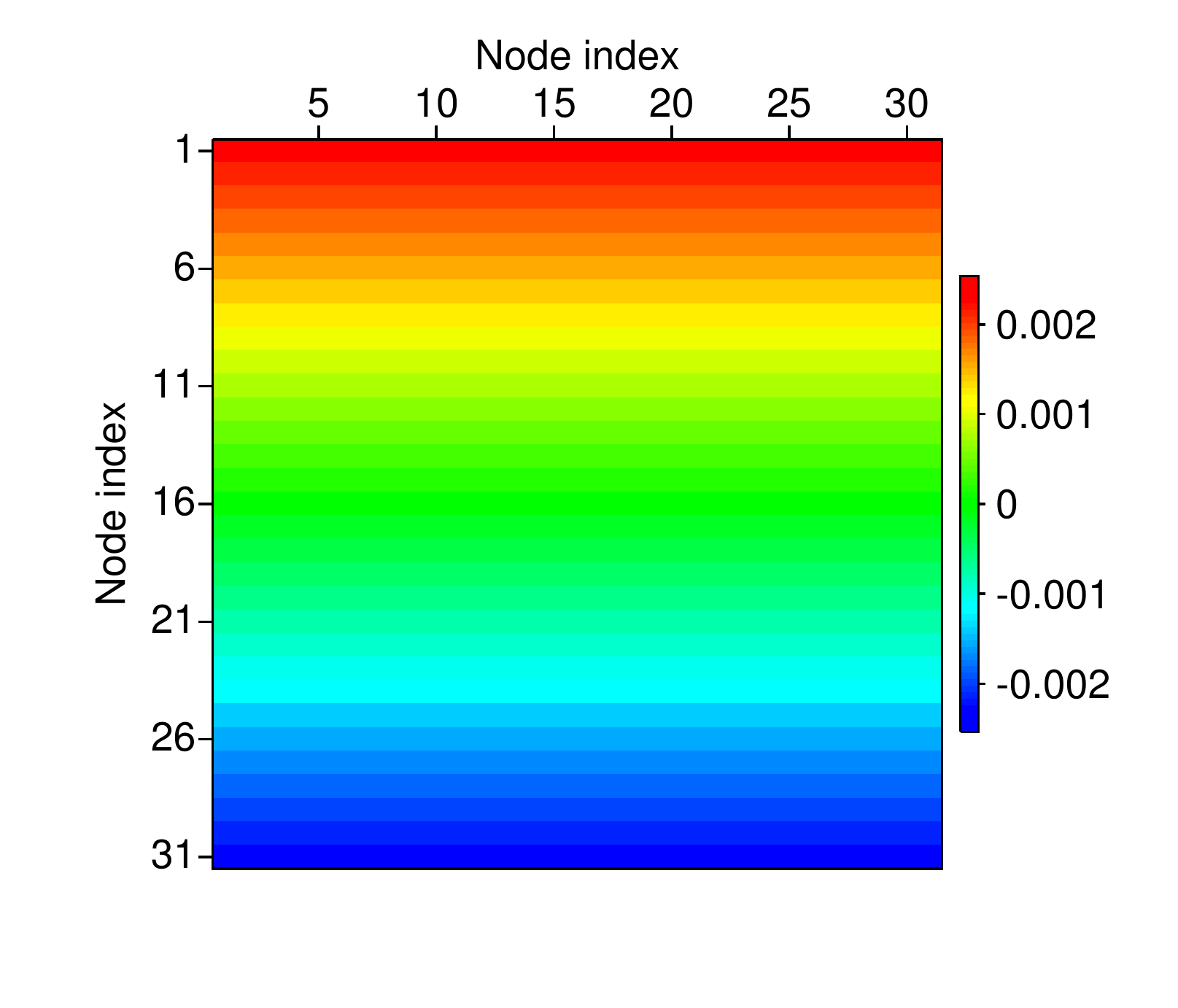}
}
\subfigure[]{
\label{fig:typeII_isohetero_u1_boundary_3}
\includegraphics[trim=50 50 30 0,width=0.415\textwidth]{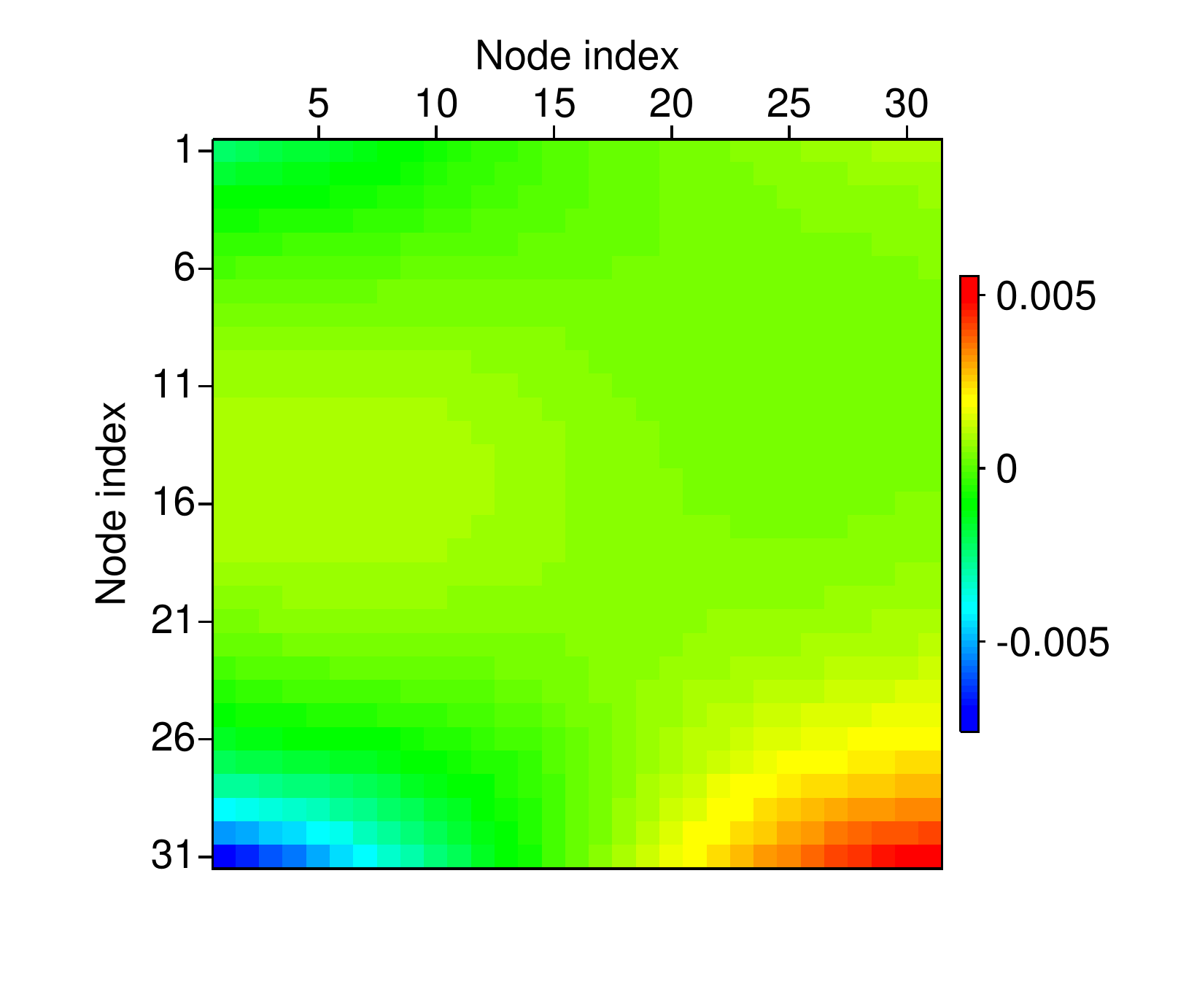}
}
\subfigure[]{
\label{fig:typeII_isohetero_u1_boundary_4}
\includegraphics[trim=50 50 30 0,width=0.415\textwidth]{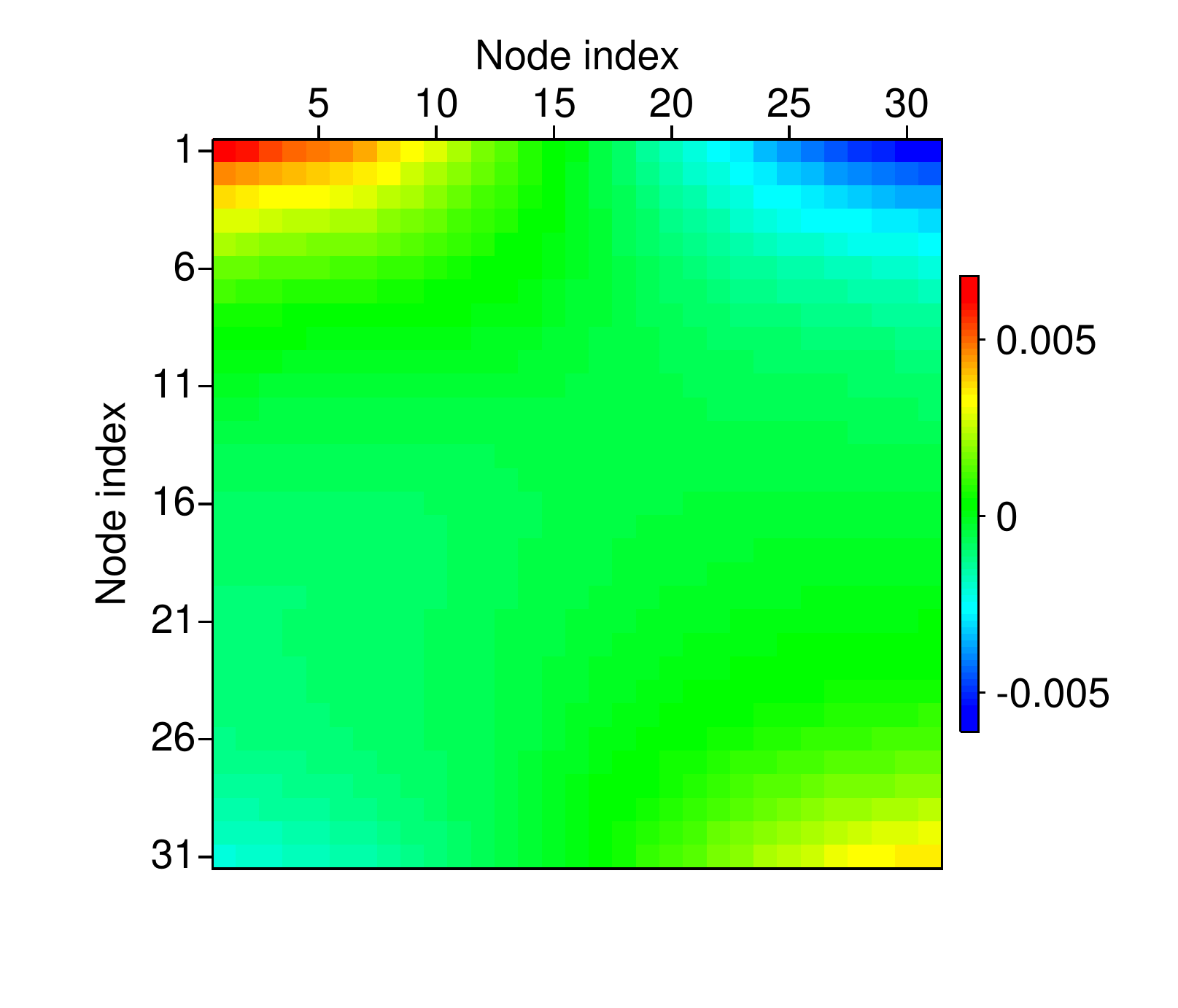}
}
\subfigure[]{
\label{fig:typeII_isohetero_u1_boundary_5}
\includegraphics[trim=50 50 30 0,width=0.415\textwidth]{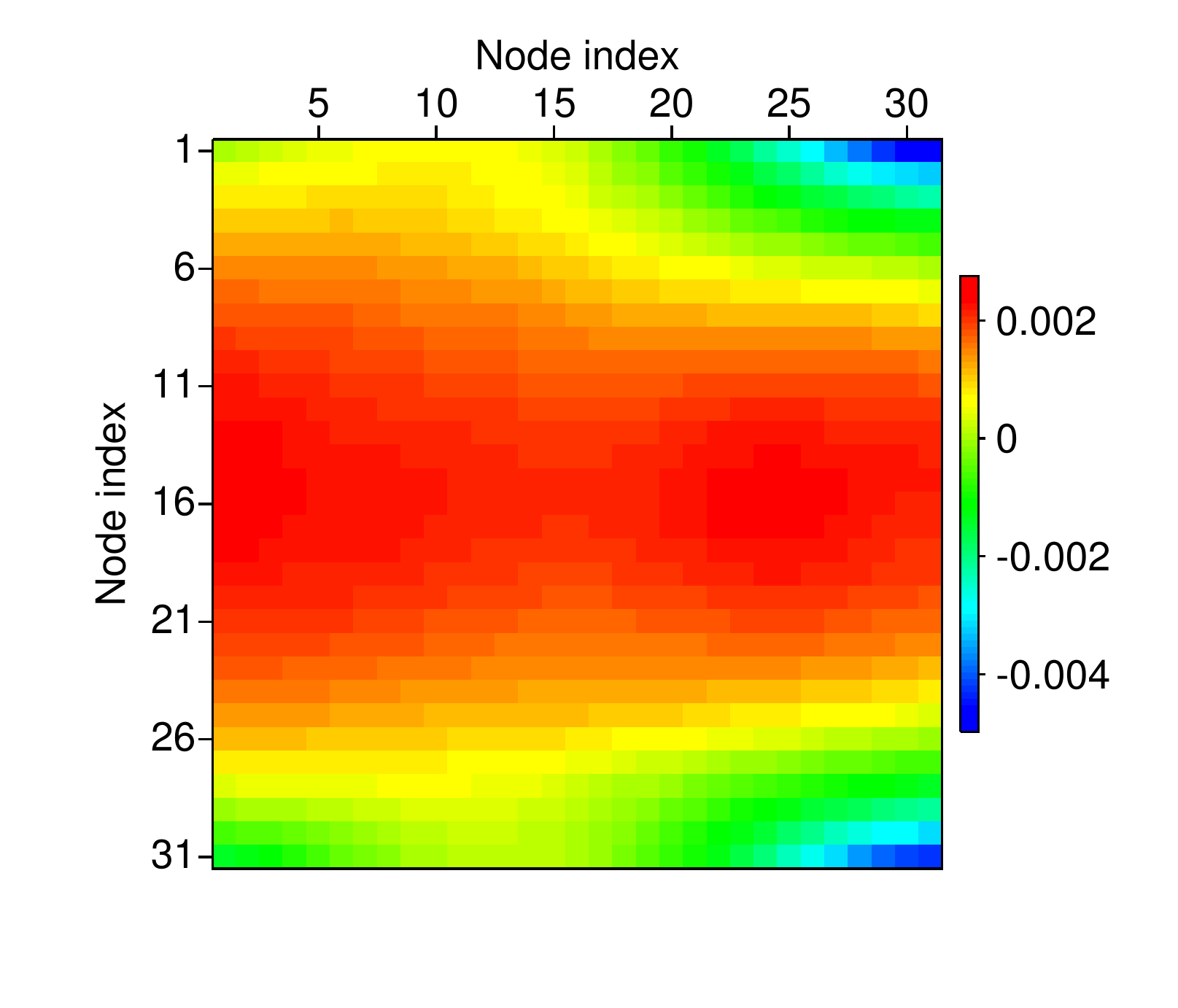}
}
\subfigure[]{
\label{fig:typeII_isohetero_u1_boundary_6}
\includegraphics[trim=50 50 30 0,width=0.415\textwidth]{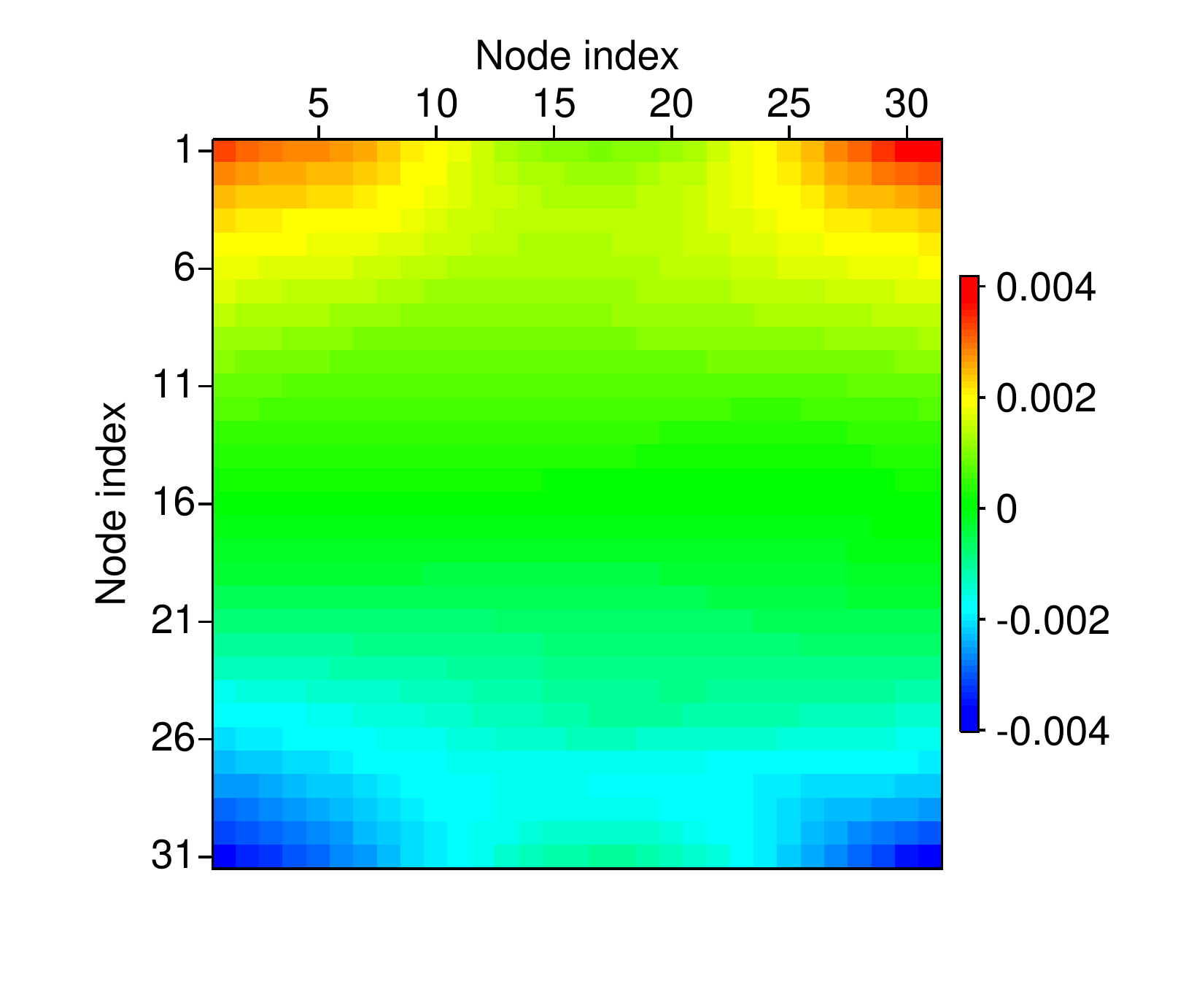}
}
\caption{(a)--(f) represent the $u_1$ component of the first 6 \revision{type II} boundary basis functions for the isotropic heterogeneous subgrid model in Figures \ref{fig:hetero_c11} and \ref{fig:hetero_c55}.}
\end{figure}

The space $\tilde{W}_H^2(K)$ contains a large number of eigenvector solution\RLG{s} when the dimension of $K$ is large, and to construct a reduced space $W_H^2(K)$, we select the first $m_2$ eigenvectors $\tilde{\mathbf{u}}_1, \tilde{\mathbf{u}}_2, \cdots, \tilde{\mathbf{u}}_{m_2}$ corresponding to the first $m_2$ smallest eigenvalues,  $0\leq \xi_1 \leq \xi_2 \leq \cdots \xi_{m_2}$, and define the space $W_H^2$ by 
\begin{equation}
W_H^2(K)=\mathrm{span}\{\boldsymbol{\varphi}_1,\boldsymbol{\varphi}_2, \cdots, \boldsymbol{\varphi}_{m_2}\}, \label{eq:type2_space_2}
\end{equation}
with the basis
\begin{equation}
\boldsymbol{\varphi}_{i,l}=\sum_{j=1}^{dp}(\tilde{\mathbf{u}}^{\mathrm{T}})_{i,j}  \mathbf{u}_{j,l},
\end{equation}
where in each terms of the above equation, $(i,j)$ represents the $j$-th node in the $i$-th vector.  These multiscale basis functions from $W_H^2$ are called boundary basis functions. Figures \ref{fig:typeII_isohetero_u1_boundary_1}--\ref{fig:typeII_isohetero_u1_boundary_6} show the first 6 boundary basis functions solved from local spectral problem \ref{eq:snapshot}, with snapshot solutions solved with local linear elasticity problem \ref{eq:typeii_1} for the isotropic heterogeneous model. \revision{We can see that like the interior basis functions, the boundary basis functions are also affected by the fine-scale variations of medium parameters. }

The above discussions are valid for the DG formulation. For CG formulation, the type II basis functions can be constructed in exactly the same way, except that the calculated eigenfunctions should be multiplied with partition of unity $\chi_K$, as is in equation \ref{eq:type1_space_cg}, i.e., 
\begin{align}
V_H^1(K)&=\mathrm{span}\{\boldsymbol{\chi}_K \boldsymbol{\phi}_1,\boldsymbol{\chi}_K \boldsymbol{\phi}_2, \cdots,\boldsymbol{\chi}_K  \boldsymbol{\phi}_{m_1}\}, \label{eq:type2_space_cg_2} \\
V_H^2(K)&=\mathrm{span}\{\boldsymbol{\chi}_K \boldsymbol{\varphi}_1,\boldsymbol{\chi}_K \boldsymbol{\varphi}_2, \cdots, \boldsymbol{\chi}_K \boldsymbol{\varphi}_{m_2}\}, \label{eq:type2_space_cg_1}
\end{align}
with eigenfunctions $\boldsymbol{\phi}_i$ and $\boldsymbol{\varphi}_i$ same with those in equations \ref{eq:type2_space_1} and \ref{eq:type2_space_2}, respectively. 

\subsubsection{Oversampling}
\begin{figure}
\centering
\includegraphics[width=0.4\textwidth]{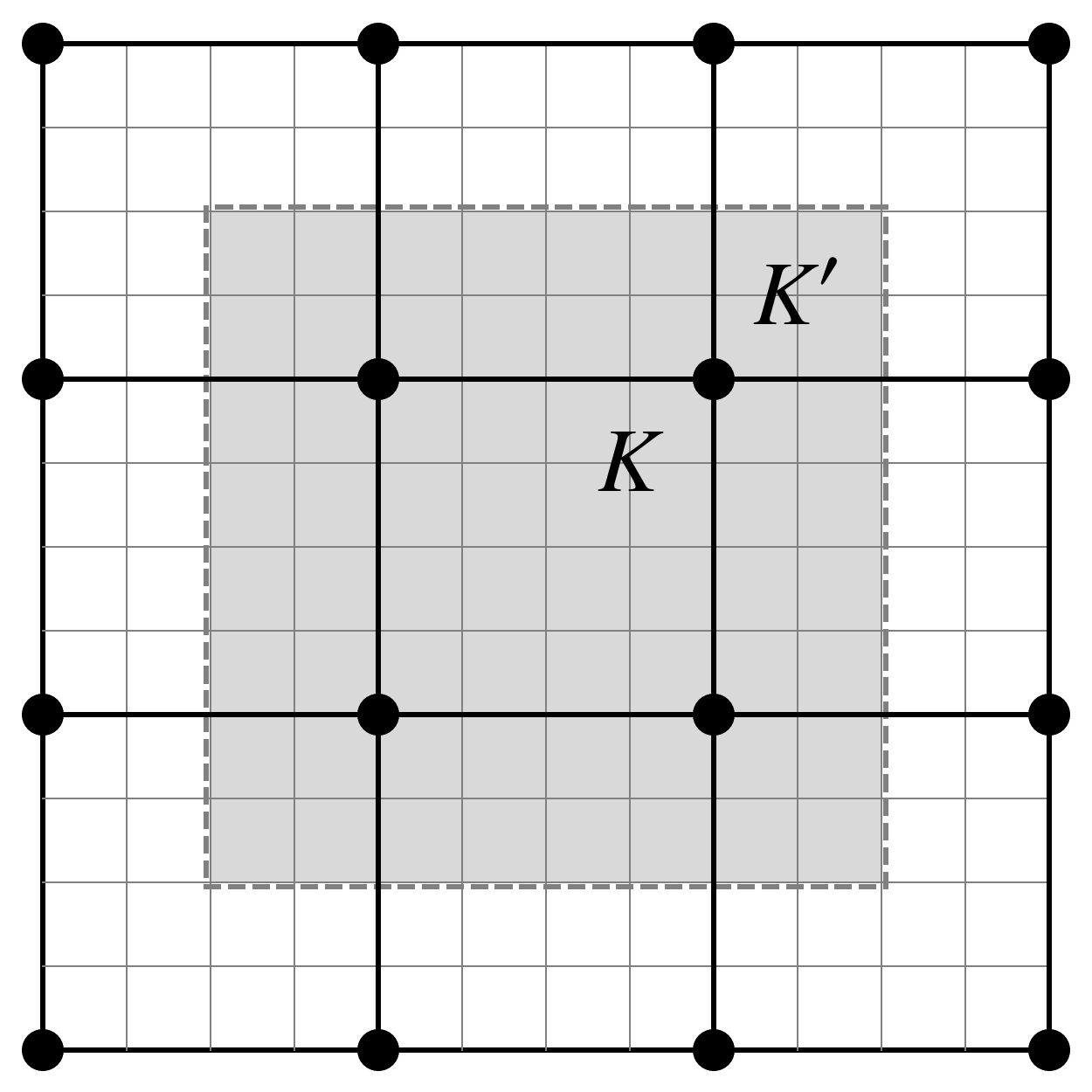}
\caption{A sketch of oversampling for DG formulation. $K$ is the coarse element where the corresponding problems needed to be solved, while $K'$ represented by gray dashed rectangle is the oversampled coarse element. After solving local problems in $K'$, we take the solutions corresponding with the nodes in $K$ as basis functions. For CG, a similar sketch \RLG{applies}. }
\label{fig:mesh_oversampling}
\end{figure}

The oversampling technique is a way to reduce the influence of fixed boundary conditions that are prescribed on $K$ when solving local problems \cite{Hou-Wu_1997,Efendiev-etal_2013b}. The concept of oversampling is shown  by Figure \ref{fig:mesh_oversampling}. When solving for the two types of basis functions, we solve the local problems on a larger region $K'$ that includes a region outside $K$, as indicated by the dashed black rectangle in Figure \ref{fig:mesh_oversampling}. We still apply the boundary conditions and local problems that are defined in equations \ref{eq:local_eig}, \RLG{\ref{eq:local_eig_interior} and \ref{eq:snapshot}}, where the boundary conditions are prescribed on $\partial K'$, rather than $\partial K$. After we obtains the solutions on $K'$, we select values on the interior region corresponding to $K$ and take them as the oversampling multiscale basis functions. In this way, the boundary nodes on $\partial K$, which are the interior nodes of $K'$, are less affected by the prescribed boundary conditions of various local problem and therefore can better represent the local properties of the elastic wave equation. \oldrevision{We must mention that the oversampling technique is quite important to reduce error of GMsFEM, and is especially important for DG-GMsFEM, since there  \RLG{is no overlap} between the coarse elements in DG-GMsFEM. }

\subsection{Stability condition and dispersion relation}
A rigorous proof of the stability condition as well as the dispersion relation of the multiscale method would be beyond the scope of this paper. 
Chung~et.~al. \cite{Chung-etal_2013b} present a complete and rigorous proof of the stability and convergence of the similar multiscale method for acoustic wave equation case, and similar behavior should result for the current algorithm. \oldrevision{Chung~et.~al. \cite{Chung_etal_2014} provide the theory and error analysis of DG-GMsFEM for the isotropic linear elasticity problem, which is exactly the static correspondence, or spatial part, of the elastic wave equation}. Also, we suggest that application of some standard results of stability condition for conventional continuous and discontinuous Galerkin finite-element method \cite{DeBasabe-Sen_2007,DeBasabe-etal_2008,Cockburn_2003} in our multiscale method at present. In the Numerical Results section below, where we present comparisons between  conventional CG/DG and the new multiscale CG/DG method, we apply a $\Delta t$ value selected by dispersion analysis for conventional CG or DG, and the results suggest this value is also adequate for stability of the multiscale methods.
\oldrevision{We aim to provide analysis for the stability condition and the dispersion relation of our CG- and DG-GMsFEM for anisotropic elastic wave equation in the future. }

\subsection{Implementation}
\subsubsection{Semi-discrete form of the GMsFEM}
With the basis functions we have introduced above, the semi-discrete system of the GMsFEM can be expressed as 
\begin{equation}
\mathbf{M} \ddot{\mathbf{d}}_H + \mathbf{K} \mathbf{d}_H = \mathbf{F}, \label{eq:semi-discrete-no-damping}
\end{equation}
where $\mathbf{M}$, $\mathbf{K}$ and $\mathbf{F}$ are the global mass matrix, stiffness matrix and force vector, respectively. For example, for CG-GMsFEM,
\begin{align}
M_{ij}&=\int_{\Omega} \boldsymbol{\Phi}_i^{\mathrm{T}} \cdot \boldsymbol{\Phi}_j d\mathbf{x}, \label{eq:global_M}\\
K_{ij}&=a_{\text{CG}}( \boldsymbol{\Phi}_i,\boldsymbol{\Phi}_j), \label{eq:global_K}\\
F_{i}&= \int_{\Omega} \mathbf{f} \cdot \boldsymbol{\Phi}_i d\mathbf{x}, \label{eq:global_F}
\end{align}
which can be calculated by matrix multiplication  \cite{Kaufmann-etal_2008,Bengzon-Larson_2013}. \oldrevision{For DG formulation}, all the expressions are the same, except that the basis functions are $\boldsymbol{\Psi}_i$, and $a_{\text{CG}}(\boldsymbol{\Phi}_i,\boldsymbol{\Phi}_j)$ is replaced with $a_{\text{DG}}(\boldsymbol{\Psi}_i,\boldsymbol{\Psi}_j)$. 

\oldrevision{For DG formulation, it is possible to adopt an coarse element based implementation rather than the global approach in equation \ref{eq:semi-discrete-no-damping}, i.e., the mass, stiffness, damping matrices and multiscale basis functions are formulated for each coarse element rather than all the coarse elements, which favors convenient parallel implementation with OpenMP or MPI. Nevertheless, to ensure a fair comparison between different schemes, we still use global approach as described by equation \ref{eq:semi-discrete-no-damping} for both CG and DG formulation in the Numerical Results section. }

\subsubsection{Absorbing boundary conditions}
In any practical applications of wave equation modeling, it is necessary to set appropriate boundary conditions  at the computation domain boundaries, including a free surface boundary condition and an absorbing boundary condition (ABC). Since the free surface boundary conditions can be naturally satisfied by setting $\boldsymbol{\sigma}\cdot \mathbf{n}=\mathbf{0}$ \cite{Bengzon-Larson_2013,Komatitsch-etal_1999}, we focus on choosing appropriate boundary conditions that can damp or absorb outgoing waves at the boundaries. 

There have been many different approaches that can achieve this goal, e.g., one-way wave equation based ABC \cite{Engquist-Majda_1977,Higdon_1991,Givoli-etal_2006,Hagstrom-etal_2008,Liu-Sen_2010,Liu-Sen_2012}, attenuation-based approach \cite{Cerjan-etal_1985,Kosloff-Kosloff_1986,Sarma-etal_1998}, and perfectly matched layers \cite{Berenger_1994,Collino-Tsogka_2001,Gao-Zhang_2008,Komatitsch-Martin_2007,Fajardo-Papageorgiou_2008,Ping-etal_2014}. Here, we adopt the Rayleigh damping \cite[]{Sarma-etal_1998}, or so-called proportional damping, to reduce the amplitude of outgoing waves at the boundaries. We also set a non-constant damping zone for Rayleigh damping by changing the spatial weight from the inner to the outer nodes, and the weight profile in the $i$-th axis direction we have chosen is a power-law curve, i.e.,
\begin{equation}
w_{i,j}(x_i)=\left(\frac{j-1}{L_i}\right)^{b_i},
\label{eq:weight}
\end{equation}
where $j$ is the $j$-th node counting from the boundary between computation domain and the attenuating zone, $L_i$ is the total number of nodes in the attenuating zone in the $i$-th direction, and $b_i$ is the power-law exponent for the damping zone. The reason for choosing such a varying weight is to avoid rapid changes in medium properties, since by adding Rayleigh damping the medium has changed to viscous medium, which will cause reflections at the boundary of damping zone and central computational domain. The weight in equation \ref{eq:weight} is similar with the idea \oldrevision{by Liu and Sen} \cite{Liu-Sen_2010}, yet they applied a linear weight where $b_i=1$. 
Combining
the weights in all directions, we get
\begin{equation}
w(\mathbf{x})=\sum_{i=1}^{3} w_i (x_i).
\end{equation}

By introducing the proportional damping boundary condition, the modeling system \ref{eq:semi-discrete-no-damping} will become
\begin{equation}
\mathbf{M} \ddot{\mathbf{d}}_H + \mathbf{E} \dot{\mathbf{d}}_H + \mathbf{K} \mathbf{d}_H = \mathbf{F}, \label{eq:semi-discrete}
\end{equation}
where $\mathbf{E}$ is the global damping matrix with elements that are only non-zero on the damping boundary zone. For each element $K$ in the damping boundary zone, the damping matrix can be written as the sum of mass matrix and stiffness matrix with some coefficients as 
\begin{equation}
\mathbf{E}_K=\alpha_1 \mathbf{M}_K +\alpha_2 \mathbf{K}_K,
\end{equation}
where the damping coefficients satisfy
\begin{equation}
2\omega_i \xi_i =\alpha_1 + \alpha_2 \omega_i^2, \label{eq:charac_damping}
\end{equation}
 the parameters $\omega_i$ are related the frequencies of the source wavelet \cite{Sarma-etal_1998}, and the $\xi_i$ are the damping ratio with respect to the critical damping ratio related to the medium properties and to the width of the damping zone around the computation zone. The coefficients can be solved directly from equation \ref{eq:charac_damping} by choosing two distinct frequencies $\omega_1$ and $\omega_2$, and two different damping ratios $\xi_1$ and $\xi_2$:
\begin{align}
\alpha_1&=\frac{2\omega_1 \omega_2 (\xi_2 \omega_1 - \xi_1 \omega_2)}{\omega_1^2- \omega_2^2}, \\
\alpha_2&=\frac{2(\xi_1 \omega_1 - \xi_2 \omega_2)}{\omega_1^2- \omega_2^2}.
\end{align}
We remark that the choice of two different $\xi_i$ is different from that \oldrevision{adopted by Sarma~et.~al.} \cite{Sarma-etal_1998}, where the two damping ratios are set to be the same, i.e., $\xi_1=\xi_2$.

\subsubsection{Adaptability in choosing the number of basis functions}
The accuracy of the multiscale solution is closely related to the number of basis functions in the coarse elements. In principle, 
for a fixed ratio of coarse to fine element dimensions, 
the shorter the wavelength of the wavefield traveling through the coarse element, the more basis functions are required to represent the wavefield in this coarse element. This is a natural conclusion from the physical meaning of the multiscale basis functions, since in the last section, we have known that the multiscale basis functions are solved from local spectral problems, and the selection of first eigenfunctions corresponds with selecting the eigenmodes with lowest frequencies. Therefore, to represent the shorter wavelength portion of a wavefield, more eigenfunctions, i.e., more multiscale basis functions are required. 

However, in a certain model, the elasticity parameters and density may be spatially heterogeneous, and in some circumstances we may encounter highly heterogeneous media. When we solve the wave equation with GMsFEM on the coarse mesh, in some coarse elements we may need greater number of basis functions than the others. 
Low velocity portions of the model with small wavelength will require more basis functions, but this will be too many
for regions with larger velocities.
We therefore propose an adaptive way to quantify and set the number of basis functions in each coarse element. 

In a particular model,
for each coarse element, say, $K_j$, we calculate the harmonic average of S-wave velocity, i.e.,
\begin{equation} \label{eq:vs_harmonic}
v_{\text{S,Harmonic}}=n_1 n_3 \left(\sum_{i_1=1}^{n_1}\sum_{i_3=1}^{n_3}\sqrt{\frac{\rho(i_1,i_3)}{C_{55}(i_1,i_3)}}\right)^{-1},
\end{equation}
and then a time duration $\delta t_j$, which characterizes the average time for a plane wave propagating through the coarse element $K_j$, can be calculated as
\begin{equation}
\delta t_j = \frac{1}{v_{\text{S,Harmonic}}},
\end{equation}
from which we can know the maximum and minimum time differences in the model:
\begin{align}
\delta t_{\max} & = \max_{\cup K_j} \delta t_j, \\
\delta t_{\min} & = \min_{\cup K_j} \delta t_j,
\end{align}
where $K_j$ denotes in the coarse block $K_j$, and $\cup K_j$ means the set of all coarse blocks. \revision{Note that the calculation with equation \ref{eq:vs_harmonic} is not perfectly accurate in anisotropic media, since the phase velocities of qS-wave along different propagation directions might be different in anisotropic media. Therefore, equation \ref{eq:vs_harmonic} can only serve as an approximation in such situations. }

Assume the maximum and minimum number of basis functions we assign to the coarse element are $n_{\max}$ and $n_{\min}$, respectively, then for some coarse element $K_j$ the number of basis functions we assign satisfies
\begin{equation}
\frac{n_{\max}-n_j}{n_{\max}-n_{\min}}=\frac{\delta t_{\max}-\delta t_j}{\delta t_{\max}-\delta t_{\min}},
\end{equation}
where we take the integer part of $n_j$, if necessary. In this way, the coarse elements where the wave velocity is slower, i.e., the wavelength is shorter, will be assigned with greater number of basis functions, and vice versa. 

It should be noted that this method determining the number of basis functions can only give a relative indication of which cells need more or fewer bases. We still need to set minimum and maximum numbers of basis functions $n_{\min}$ and $n_{\min}$ beforehand, which requires test evaluations.

\subsubsection{A global projection approach}
The global matrices can be calculated by projecting the global matrices of the corresponding fine mesh problem onto the coarse mesh with a global projection matrix assembled from the calculated multiscale basis functions. Assume we can first assemble the global matrices $\mathbf{M}_h$, $\mathbf{K}_h$ and $\mathbf{F}_h$ on the fine mesh with traditional finite-element assembly methods \cite[]{Riviere_2008,Kaufmann-etal_2008, Bengzon-Larson_2013}, then for CG formulation, we form a global projection matrix $\mathbf{R}$ with the multiscale basis functions as 
\begin{equation}
\mathbf{R}=(R_1,R_2,\cdots,R_N)^{\mathrm{T}}, \label{eq:R}
\end{equation}
where
\begin{equation}
R_i=\left[\boldsymbol{\Phi}_{i,1}, \boldsymbol{\Phi}_{i,2}, \cdots, \boldsymbol{\Phi}_{i,m_i} \right], \label{eq:Ri}
\end{equation}
with $\boldsymbol{\Phi}_{i,j}$ being the $j$-th multiscale basis function of the $i$-th coarse node, which follows the definition \ref{eq:type1_space_cg} of type I basis function, or \ref{eq:type2_space_1} and \ref{eq:type2_space_2} of type II basis function, $m_i$ is the total number of basis functions of the $i$-th coarse node. For DG formulation, $\mathbf{R}$ can be constructed in the same way. 

The global projection matrix $\mathbf{R}$ therefore has the dimension $(\sum_{i=1}^{N} m_i) \times n$, where $N$ is the number of coarse nodes in CG formulation, and coarse elements in DG formulation, and $m_i$ is the number of basis functions in $K_i$, and $n$ is the number of degrees of freedom of fine mesh $\mathcal{T}_h$ or $\mathcal{P}_h$. With $\mathbf{R}$, the semi-discrete system \ref{eq:semi-discrete} can be written as
\begin{equation}
\mathbf{R}\mathbf{M}_h\mathbf{R}^{\mathrm{T}}\ddot{\mathbf{d}}_H + \mathbf{R}\mathbf{E}_h\mathbf{R}^{\mathrm{T}} \dot{\mathbf{d}}_H+ \mathbf{R}\mathbf{K}_h\mathbf{R}^{\mathrm{T}} \mathbf{d}_H = \mathbf{R}\mathbf{F}_h. \label{eq:multi_sys}
\end{equation}
Clearly, $\mathbf{d}_H$ has the length of $\sum_{i=1}^{N} m_i$, compared with $n$ of $\mathbf{d}_h$ in the corresponding fine mesh problem. Importantly, the expected wavefield on the fine mesh can be recovered through 
\begin{equation}
\mathbf{d}_h=\mathbf{R} \mathbf{d}_H,
\end{equation}
which means that the degrees of freedom that are required to save and recover the complete wavefield can be greatly reduced, given that normally the ratio between $n$ and $\sum_{i=1}^{N} m_i$ is large. For example, assume there is an equal number of basis functions in all $K_i$, say, $m$, then for a rectangular domain $\Omega$ with rectangular elements $K$ in 2D, this ratio is $2(n_1+1)(n_2+1)/[(n_1/r_1+1)(n_2/r_2+1)m]$ for the CG formulation, and \revision{$8 r_1 r_2/m$} for the DG formulation. Here $n_i$ is the number of element in $i$-th direction on fine mesh and $r_i$ is the number of element contained in $i$-th direction in $K$. This ratio can be large if $r_i$ is large. 

\subsubsection{Time stepping}
For temporal discretization, we simply use central finite difference, i.e.,  
\begin{align}
\ddot{\mathbf{d}}_H&=\frac{\mathbf{d}_H^{t+\Delta t}-2\mathbf{d}_H^t+\mathbf{d}_H^{t+\Delta t}}{\Delta t^2}, \\
\dot{\mathbf{d}}_H&=\frac{\mathbf{d}_H^{t+\Delta t}-\mathbf{d}_H^{t-\Delta t}}{2\Delta t}
\end{align}
which has second-order accuracy. More complicated time stepping schemes, e.g., the Newmark scheme \cite[e.g.,][]{Marfurt_1984,Hughes_1987} could be adopted for the temporal discretization.

\subsubsection{Source term}
We have set the source term as a force vector, \revision{which is}
\begin{equation}
\mathbf{f}(\mathbf{x},\theta,t)=G(\mathbf{x})\mathbf{P}(\theta)R(t),
\end{equation}
where \revision{$\mathbf{P}(\theta)=(\cos \theta,\sin\theta)$}, with $\theta$ being the polar angle of the source force vector, and $\theta=0$ being the force points along $x_1$ axis. The temporal signature $R(t)$ is a Ricker wavelet, which can be written as
\begin{equation}
R(t)=[1-2\pi^2f_0^2(t-t_0)^2]\exp[-\pi^2f_0^2(t-t_0)^2],
\end{equation}
where $f_0$ is the central frequency of the wavelet, $t_0=1/f_0$. \revision{The spatial function $G(\mathbf{x})$ is determined by the half source width parameter $\eta$ in terms of node number, which is
\begin{equation}
G(\mathbf{x})=\begin{cases}
\delta(\mathbf{x}-\mathbf{x}_0) & \qquad \text{if } \eta=0 \text{ (point source)}, \\
\exp\left[-\dfrac{(\mathbf{x}-\mathbf{x}_0)^2}{2\beta^2}\right]  &\qquad \text{if } \eta\neq 0 \text{ (Gaussian-correlated source)},
\end{cases}
\end{equation}
where we set $\beta=\eta/(2.5\sqrt{2\ln 2})$, and $\mathbf{x}_0$ is the source position. For example, when the source width in terms of node number is 15 (corresponding to $\eta=7$), $\beta\approx 2.38$.}

\section{Numerical Results}
We present three sets of results to demonstrate that the GMsFEM approach provides accurate solutions for
various types of earth models for the various choices of implementations described above. The first
test applies a \revision{three-layer} model combining isotropic and anisotropic media using the
CG- \revision{and DG-}GMsFEM \revision{formulations}, and the second case considers a complex,  2D, heterogeneous anisotropic
medium and the DG-GMsFEM solution.  In both cases, \revision{we use conventional CG- or DG-FEM with either \RLG{a} very finely discretized mesh or \RLG{a} very small time step to obtain the reference solutions, and compare the DOF, accuracy and computational time of the simulations with those of our CG- or DG-GMsFEM.}
The third example utilizes a subset of 
a common test model, the elastic Marmousi 2 model, and demonstrates the application of the 
spatial adaptivity approach described above. 

\oldrevision{Our implementation of the GMsFEM \revision{codes}, as well as the conventional FEM codes, are \RLG{prototypes} 
at the current stage, and we implement all the following numerical tests in MATLAB. Due to the limitations of the intrinsic functions of MATLAB, the time stepping part of the wave equation simulation is calculated with only one core on the Intel 4770K 3.5~GHz CPU. However, we remark that both the conventional FEM and our GMsFEM can be appropriately parallelized with OpenMP or MPI, or even GPU technique\RLG{s}, the details of which are beyond the scope of this paper.  }

\subsection{\revision{VTI-TTI-isotropic three-layer heterogeneous model}}
\revision{
In the first example, we use a model composed of three layers to verify the effectiveness of our multiscale method for numerical modeling on the coarse mesh. The model is 6000~m in the horizontal direction and 6000~m in the vertical direction. The first layer is a homogeneous  VTI (transversely isotropic with vertical axis) medium ranging from 0~m to 2100~m, with elasticity constants $C_{11}=20$~GPa, $C_{13}=8$~GPa, $C_{33}=16$~GPa, $C_{55}=4$~GPa, and $C_{15} =C_{35}=0$. The second layer is a homogeneous TTI (transversely isotropic with tilted axis) medium ranging from 2100~m to 3900~m, with elasticity constants $C_{11}=10.8125$~GPa, $C_{13}=4.1875$~GPa, $C_{15}=-1.1908$~GPa, $C_{33}=15.8125$~GPa, $C_{35}=-3.1393$~GPa, $C_{55}=5.6875$~GPa. The third layer is a homogeneous isotropic medium ranging from 3900~m to 6000~m, with elasticity constants  $C_{11}=C_{33}=24$~GPa, $C_{13}=8$~GPa, $C_{55}=8$~GPa, and $C_{15} =C_{35}=0$. We assume constant mass density with a value of 1000~kg/m$^3$ for the three media for convenience. The point source (i.e., $\eta=0$) is placed at the center of the model, i.e., $(3000,3000)$~m. The direction of the force vector is set to be $\theta=\pi/3$, and the central frequency of the Ricker wavelet is 10~Hz. The time step size is 0.5~ms for all the wavefield simulations. 
}

\revision{
To quantify the accuracy of the FEM or GMsFEM solutions, we define the $L^2$-norm error of the wavefield as
\begin{equation}
e(\mathbf{u})=\frac{\|\mathbf{u}-\mathbf{u}_{\text{ref}}\|}{\|\mathbf{u}_{\text{ref}}\|},
\end{equation}
where $\mathbf{u}$ is the FEM or GMsFEM solution, $\mathbf{u}_{\text{ref}}$ is the reference solution which is usually solved on \RLG{a} very finely discretized mesh with FEM, and $||\cdot ||$ represents $L^2$-norm.
}

\begin{figure} \label{fig:three_layer_cg}
\centering
\subfigure[]{
\label{fig:three_layer_CG_reference}
\includegraphics[width=0.4\textwidth,trim=45 55 45 0]{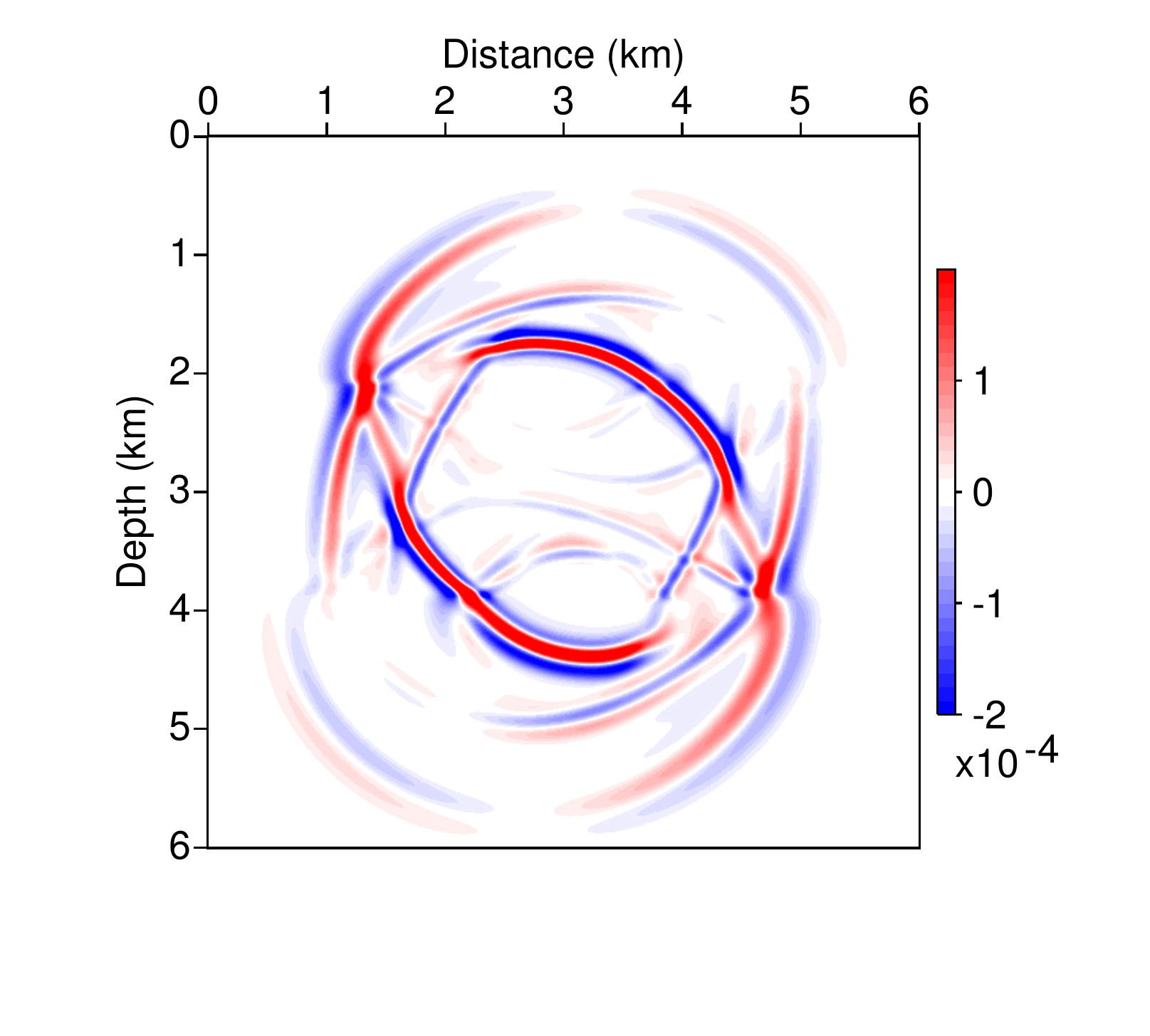}
}
\subfigure[]{
\label{fig:three_layer_CG_200x200}
\includegraphics[width=0.4\textwidth,trim=45 55 45 0]{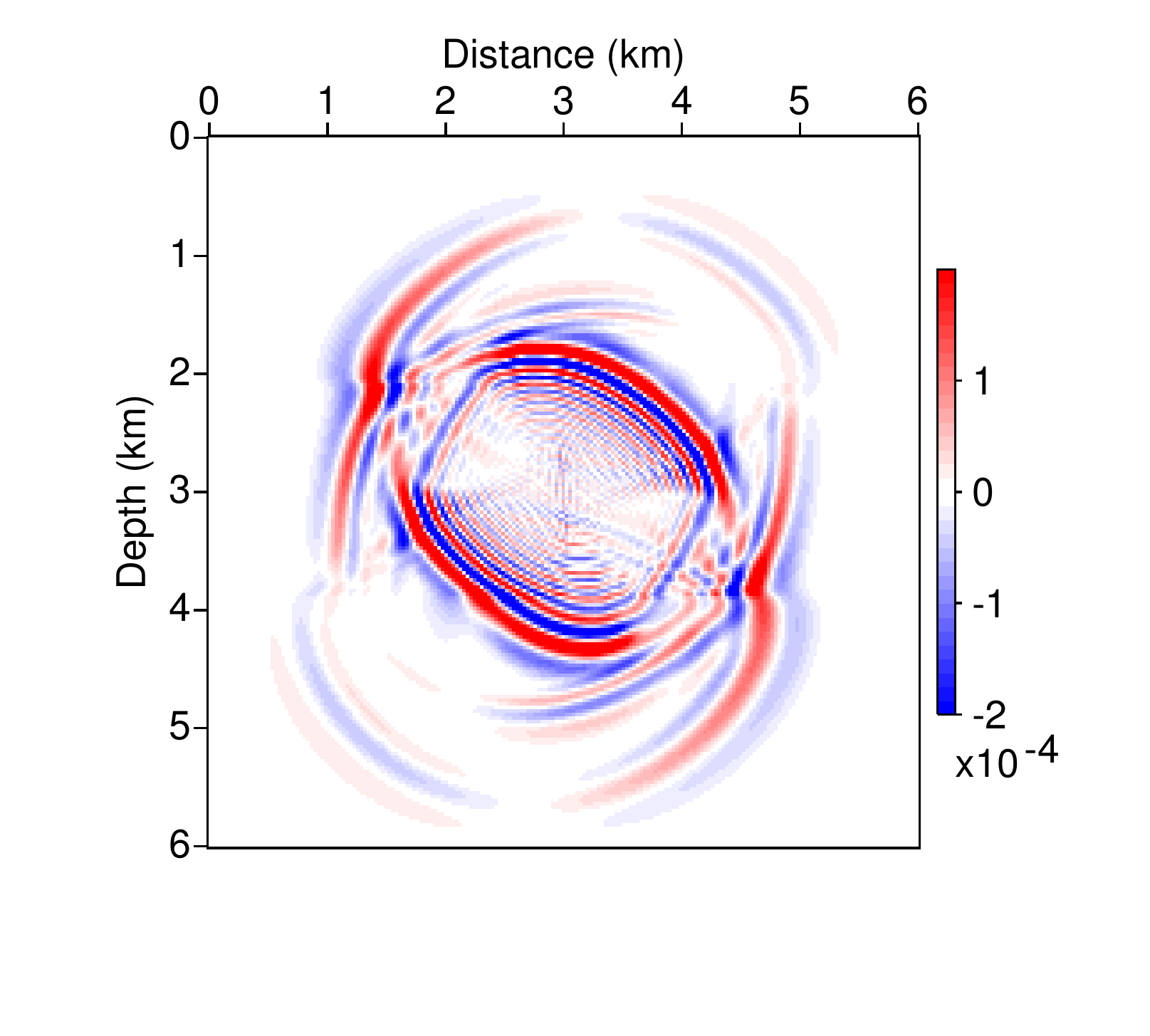}
}
\subfigure[]{
\label{fig:three_layer_CG_300x300}
\includegraphics[width=0.4\textwidth,trim=45 55 45 0]{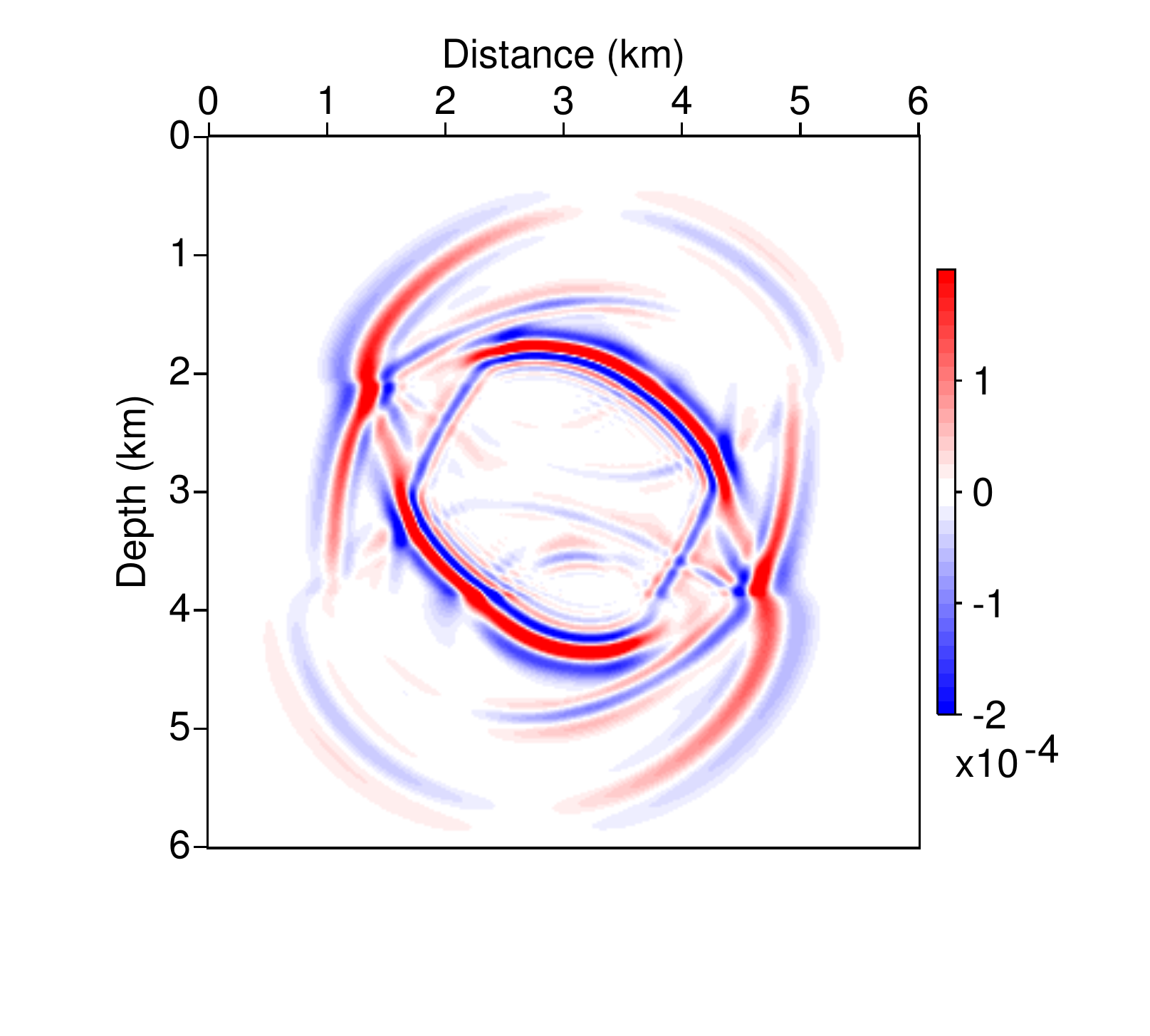}
}
\subfigure[]{
\label{fig:three_layer_CG_ms_23}
\includegraphics[width=0.4\textwidth,trim=45 55 45 0]{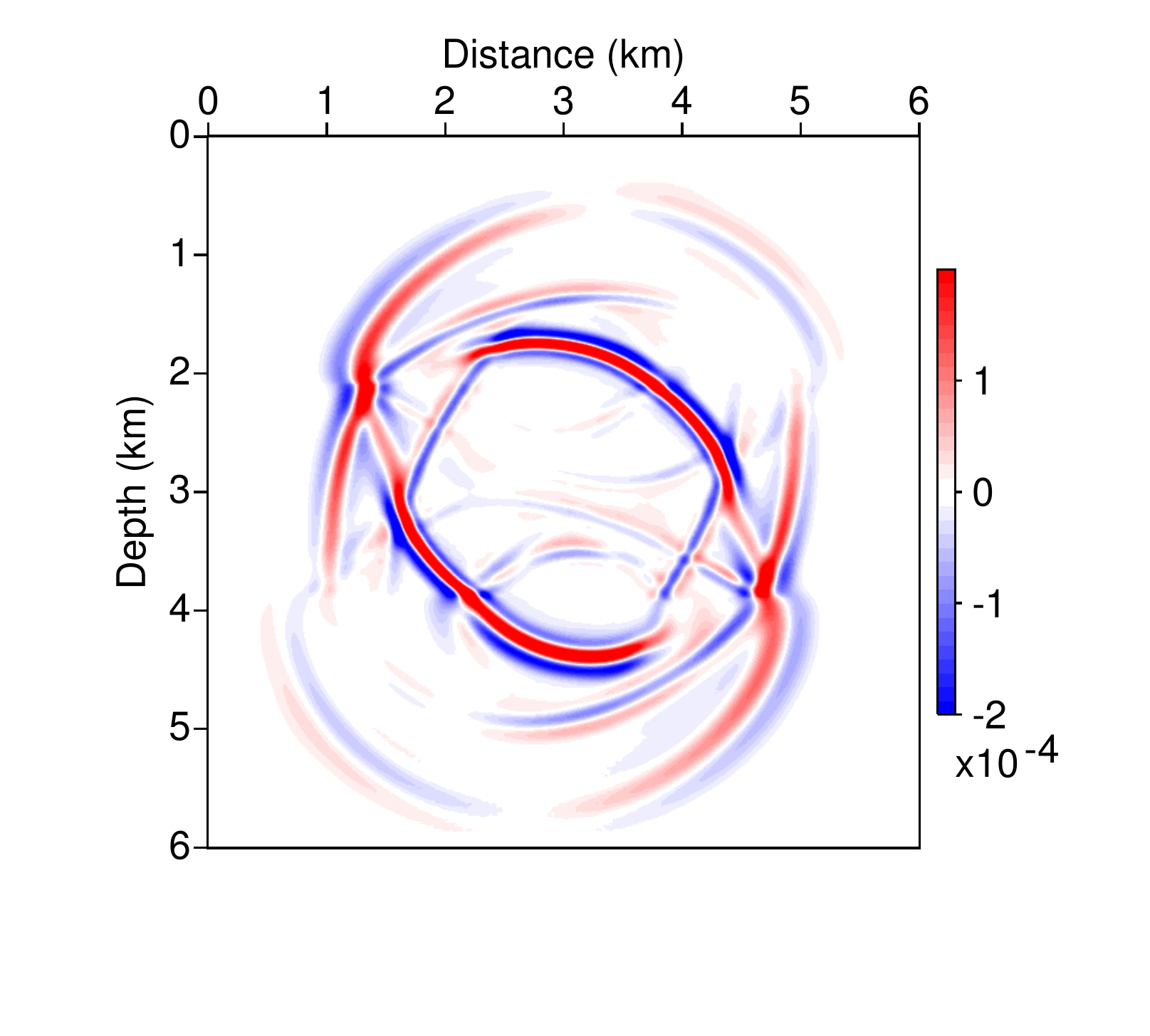}
}
\subfigure[]{
\label{fig:three_layer_CG_ms_35}
\includegraphics[width=0.4\textwidth,trim=45 55 45 0]{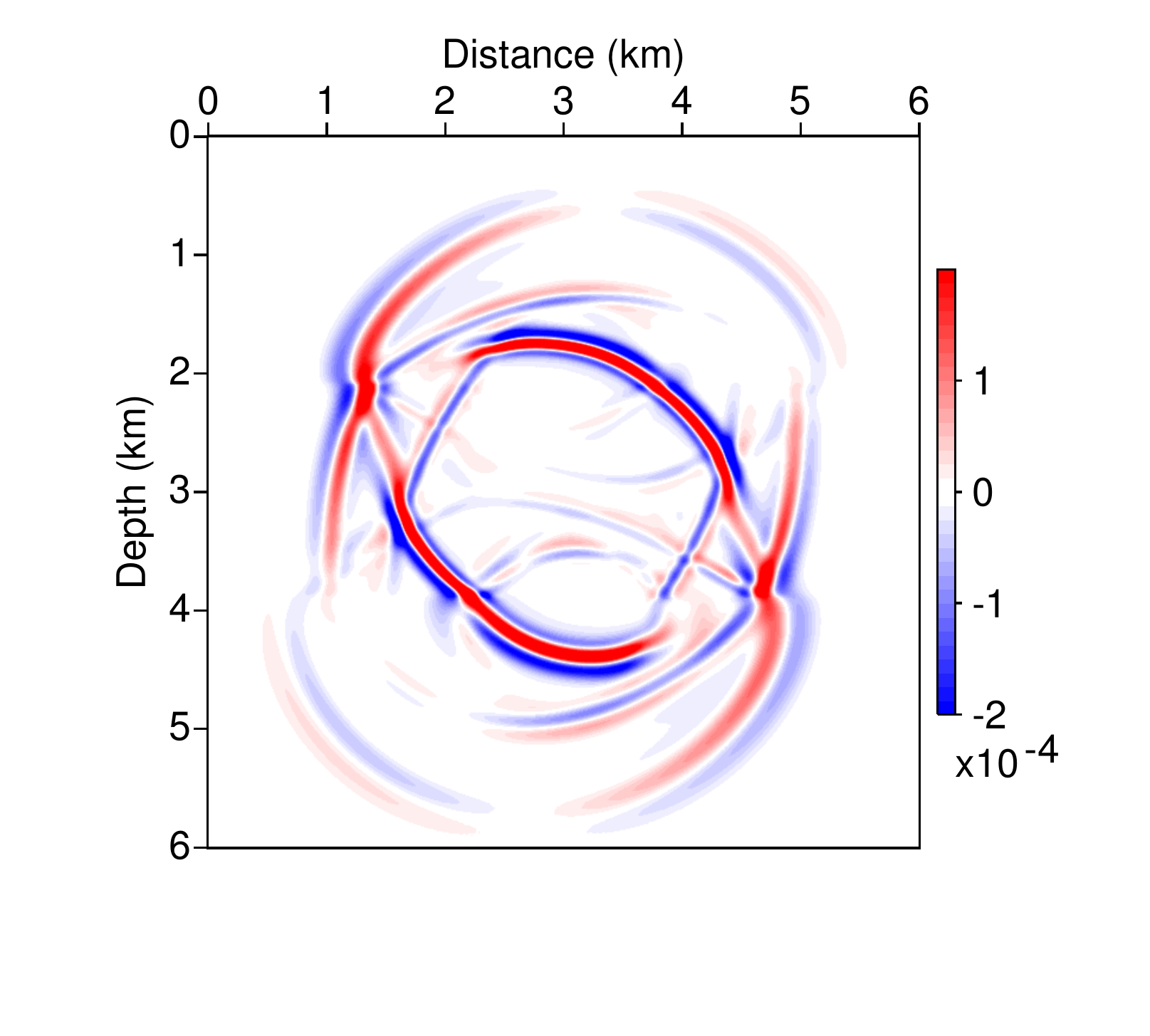}
}
\subfigure[]{
\label{fig:three_layer_CG_ms_typeII_35}
\includegraphics[width=0.4\textwidth,trim=45 55 45 0]{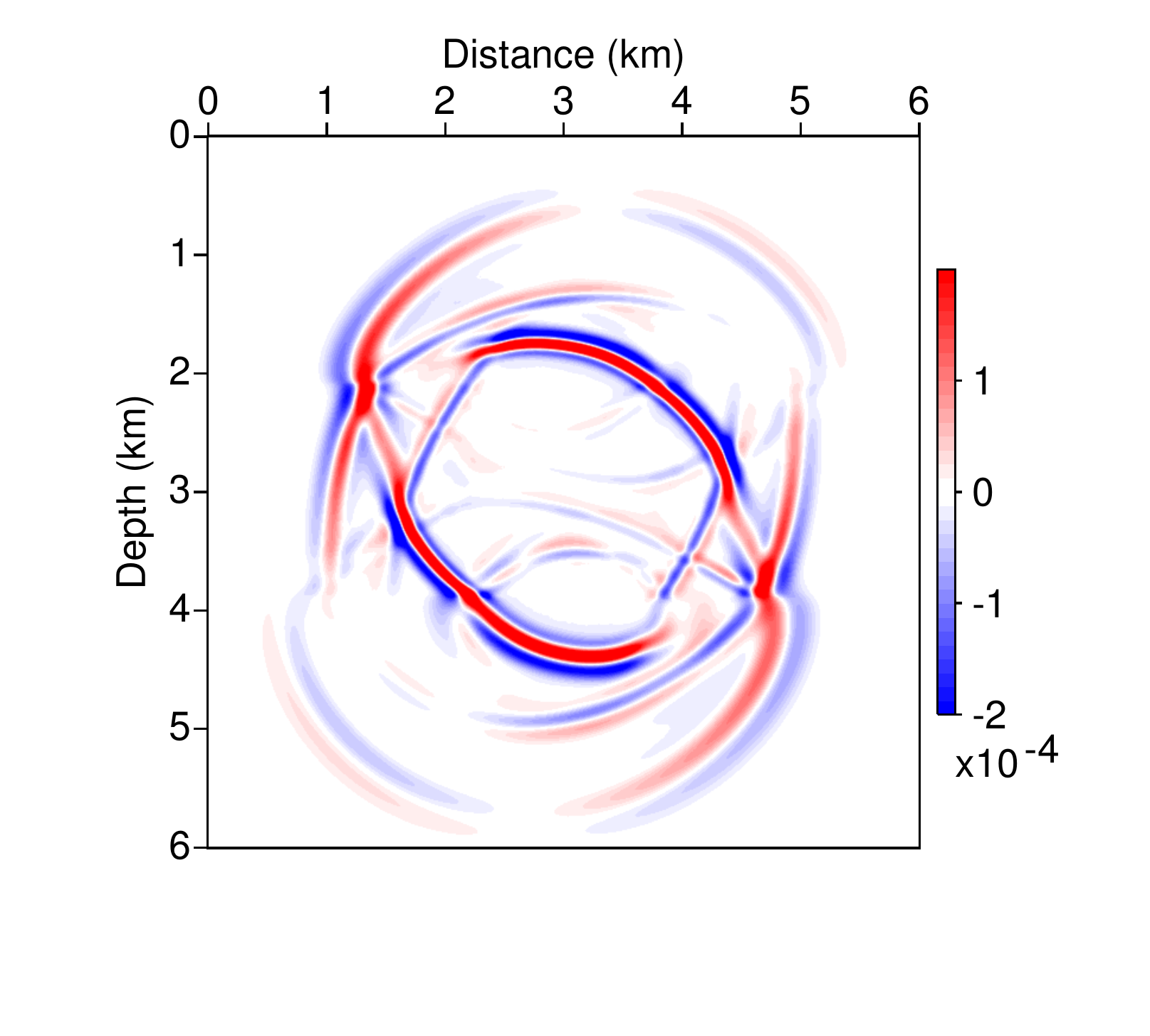}
}
\caption{$u_1$ wavefield snapshots at 0.7~s after source excitation: (a) reference solution and (b) CG-FEM solution on $200\times 200$-element mesh, (c) CG-FEM solution on $300\times 300$-element mesh, (d) CG-GMsFEM on $60\times 60$-element mesh with 23 type I basis functions, (e) CG-GMsFEM solution on $60\times 60$-element mesh with 35 type I basis functions, and (f) CG-GMsFEM solution on $60\times 60$-element mesh with 20 type II interior basis functions and 15 type II boundary basis functions.}
\end{figure}

\revision{
We first compare the results of our CG-GMsFEM and conventional CG-FEM. Figure \ref{fig:three_layer_CG_reference} shows the reference solution of $u_1$ component at 0.7~s after source excitation. The anisotropy and the heterogeneities result in complicated wavefronts as well as reflections. 
We solve the elastic wave equation with CG-FEM by discretizing the three-layer model with \RLG{a} conformal mesh which is composed of $200\times 200$ elements, and the DOF of the system is $8.08\times 10^4$. The $u_1$ wavefield snapshot is shown in Figure \ref{fig:three_layer_CG_200x200}. Obviously there is strong numerical dispersion associated with the qS-wave (the slower wave), and the qP-wave is not accurate compared with the reference solution as well. In fact, the $L^2$-norm error of this CG-FEM solution with respect to the reference solution is about 1.03  (see Table \ref{tab:three_layer_cg}). 
Figure \ref{fig:three_layer_CG_300x300} shows the $u_1$ wavefield snapshot solved with CG-FEM in the same model, but now discretized with $300\times 300$ elements. The DOF of the system is $1.81\times 10^5$. Compared with the solutions in Figure \ref{fig:three_layer_CG_200x200}, the numerical dispersion is much reduced, yet there is still visible numerical dispersion  with the qS-wave. The error for for this CG-FEM solution is about $5.77\times 10^{-1}$.
On the other hand, we calculate the wavefield with CG-GMsFEM in Figure \ref{fig:three_layer_CG_ms_23} for the same model, but with only $60\times 60$ elements, i.e., the element size is 100~m by 100~m, which is much coarser than those in Figures \ref{fig:three_layer_CG_200x200} and \ref{fig:three_layer_CG_300x300}. We use 23 type I basis functions in our CG-GMsFEM, making the DOF of the system $8.56\times 10^4$, which is almost the same as that of CG-FEM on the $200\times 200$-element mesh in Figure \ref{fig:three_layer_CG_200x200}. Obviously, our CG-GMsFEM solution is much better than CG-FEM solution, since there is no obvious numerical dispersion with either qS- or qP-wave. In Table \ref{tab:three_layer_cg}, we can find that the relative error for the CG-GMsFEM solution with 23 type I basis functions is about  $1.08\times 10^{-1}$. 
We also calculate the wavefield with our CG-MsFEM for the model discretized with $60\times 60$ elements, but now with 35 type I multiscale basis functions (Figure \ref{fig:three_layer_CG_ms_35}). The DOF of this system is $1.30\times 10^5$, which is on the same order of that in CG-FEM on $300\times 300$-element mesh (about 70\%). Again, our CG-GMsFEM solution is with much smaller error, which is only about $3.20\times 10^{-2}$, compared with about $5.77\times 10^{-1}$ for the CG-FEM solution. By using type II basis functions (20 interior and 15 boundary basis functions), our CG-GMsFEM solution (Figure \ref{fig:three_layer_CG_ms_typeII_35}) is also accurate, with about $3.17\times 10^{-2}$ relative error. It can be reasonably expected that with 49 basis functions where the DOF of our CG-GMsFEM will be approximately the same as that of CG-FEM on $300\times 300$-element mesh, our CG-GMsFEM solution will be even more accurate, since more basis functions will provide more information about subgrid medium properties. These comparisons show that our CG-GMsFEM can achieve higher accuracy with similar DOF compared with CG-FEM in anisotropic, heterogeneous models. 
}

\begin{figure} \label{fig:three_layer_dg}
\centering
\subfigure[]{
\label{fig:three_layer_DG_reference}
\includegraphics[width=0.4\textwidth,trim=45 55 45 0]{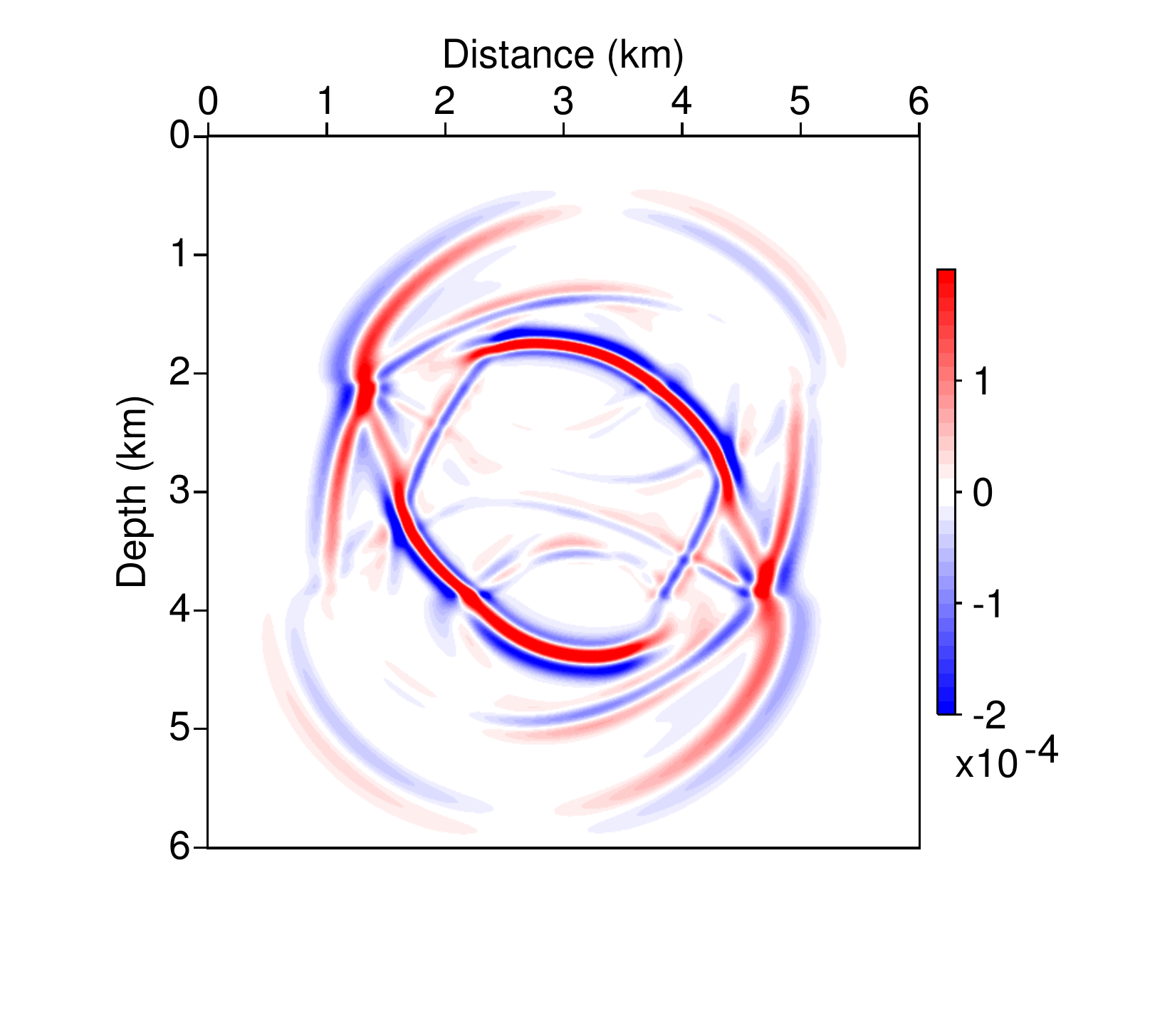}
}
\subfigure[]{
\label{fig:three_layer_DG_200x200}
\includegraphics[width=0.4\textwidth,trim=45 55 45 0]{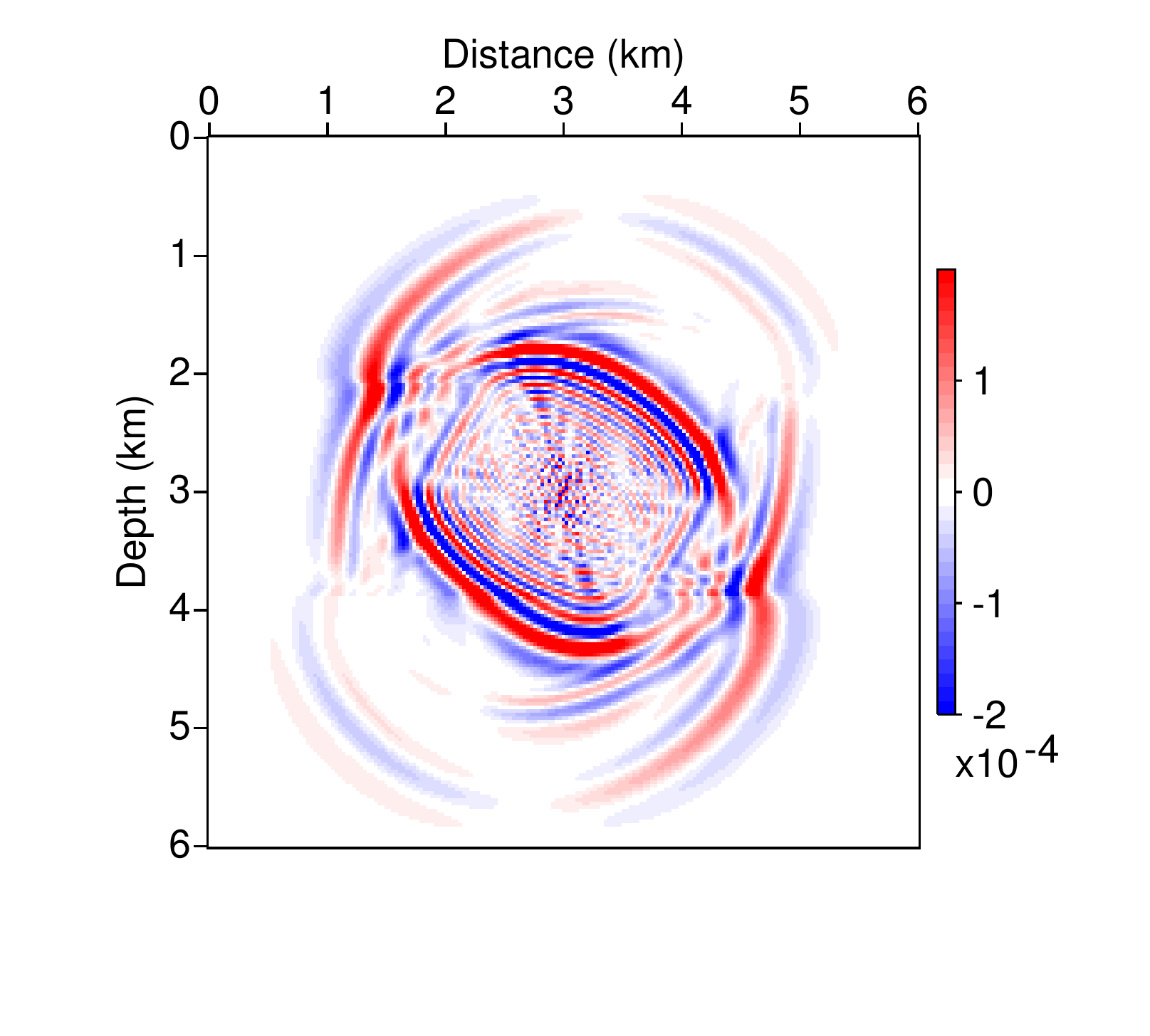}
}
\subfigure[]{
\label{fig:three_layer_DG_300x300}
\includegraphics[width=0.4\textwidth,trim=45 55 45 0]{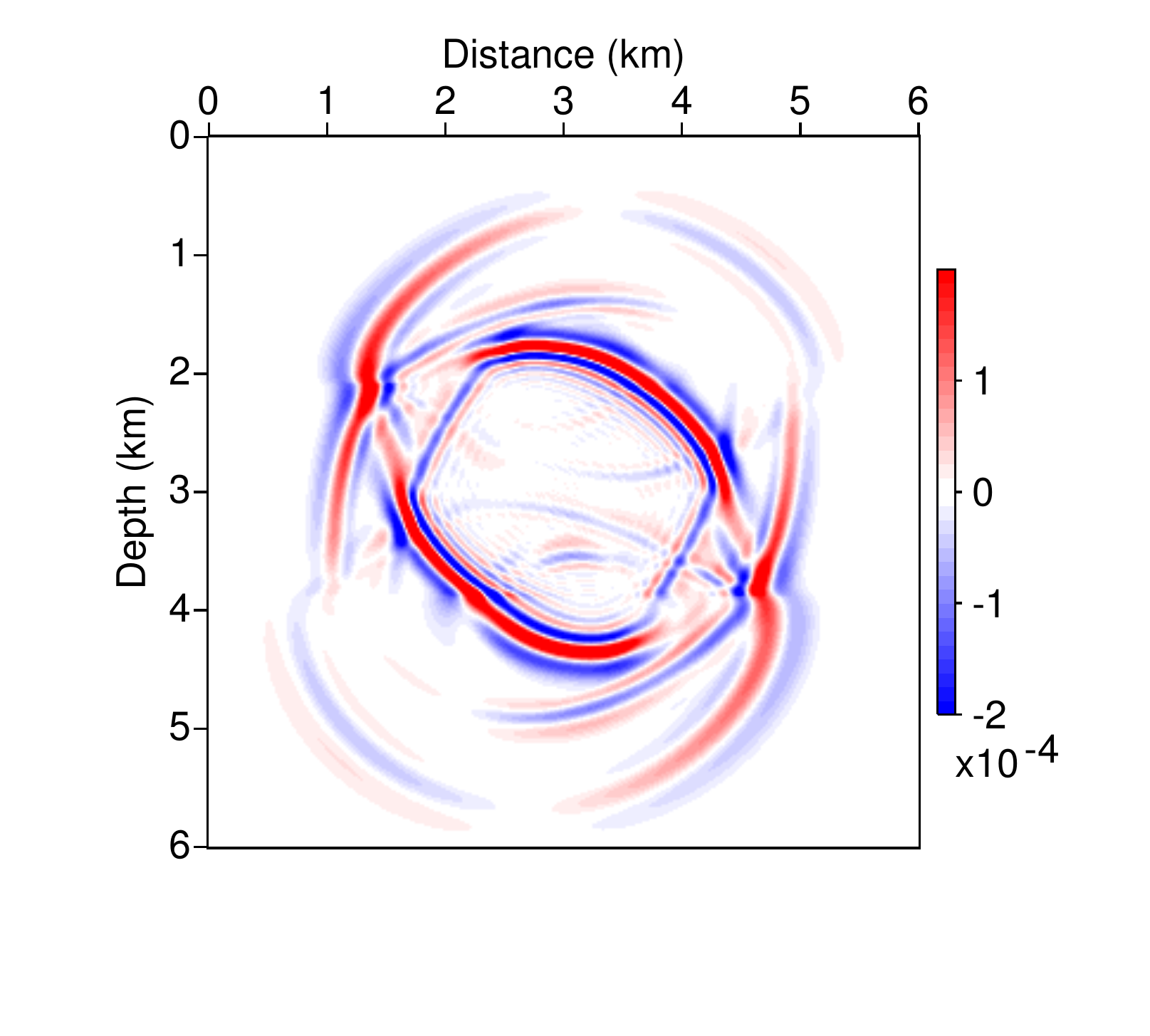}
}
\subfigure[]{
\label{fig:three_layer_DG_ms_i50b30}
\includegraphics[width=0.4\textwidth,trim=45 55 45 0]{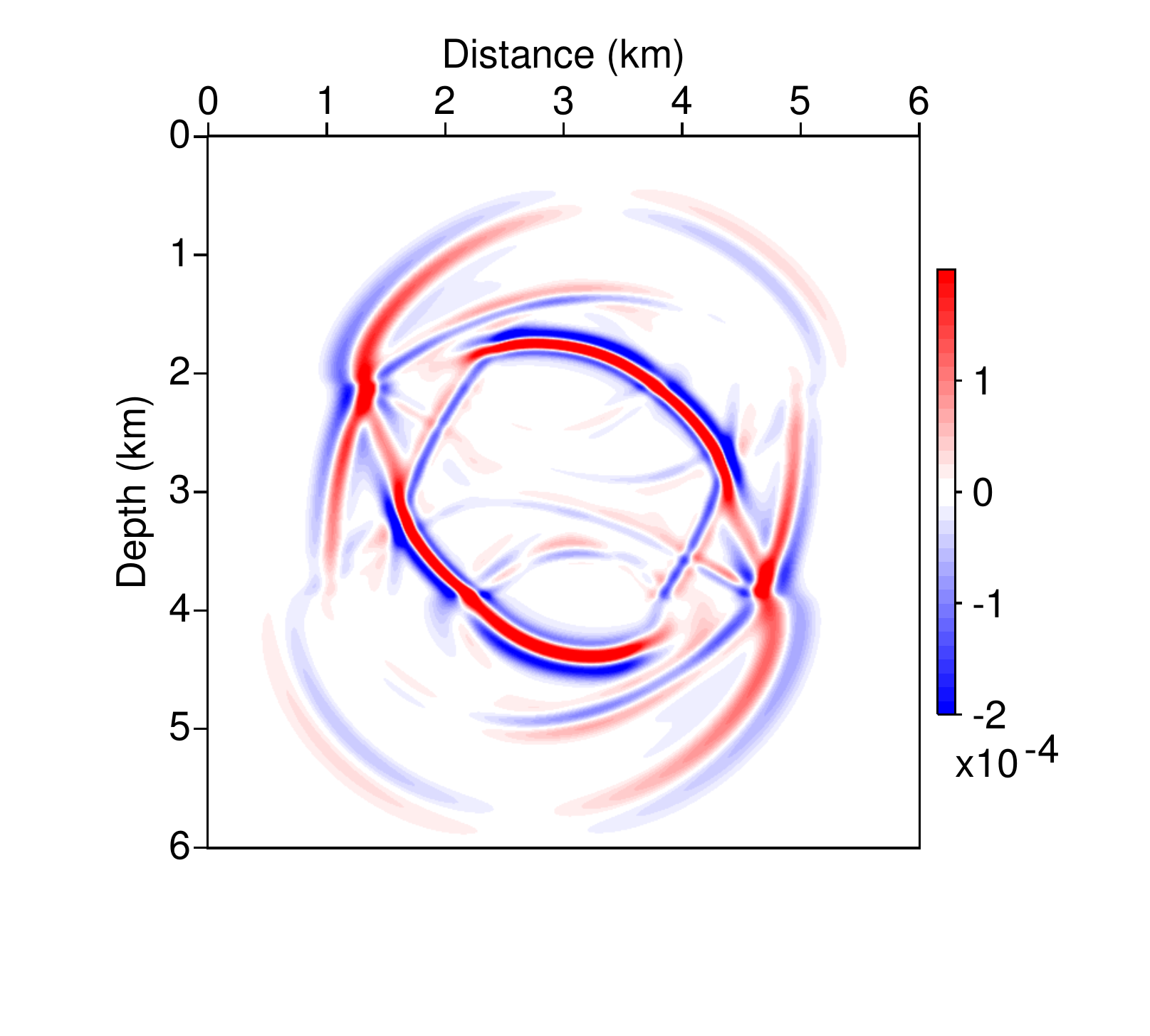}
}
\subfigure[]{
\label{fig:three_layer_DG_ms_i150b150}
\includegraphics[width=0.4\textwidth,trim=45 55 45 0]{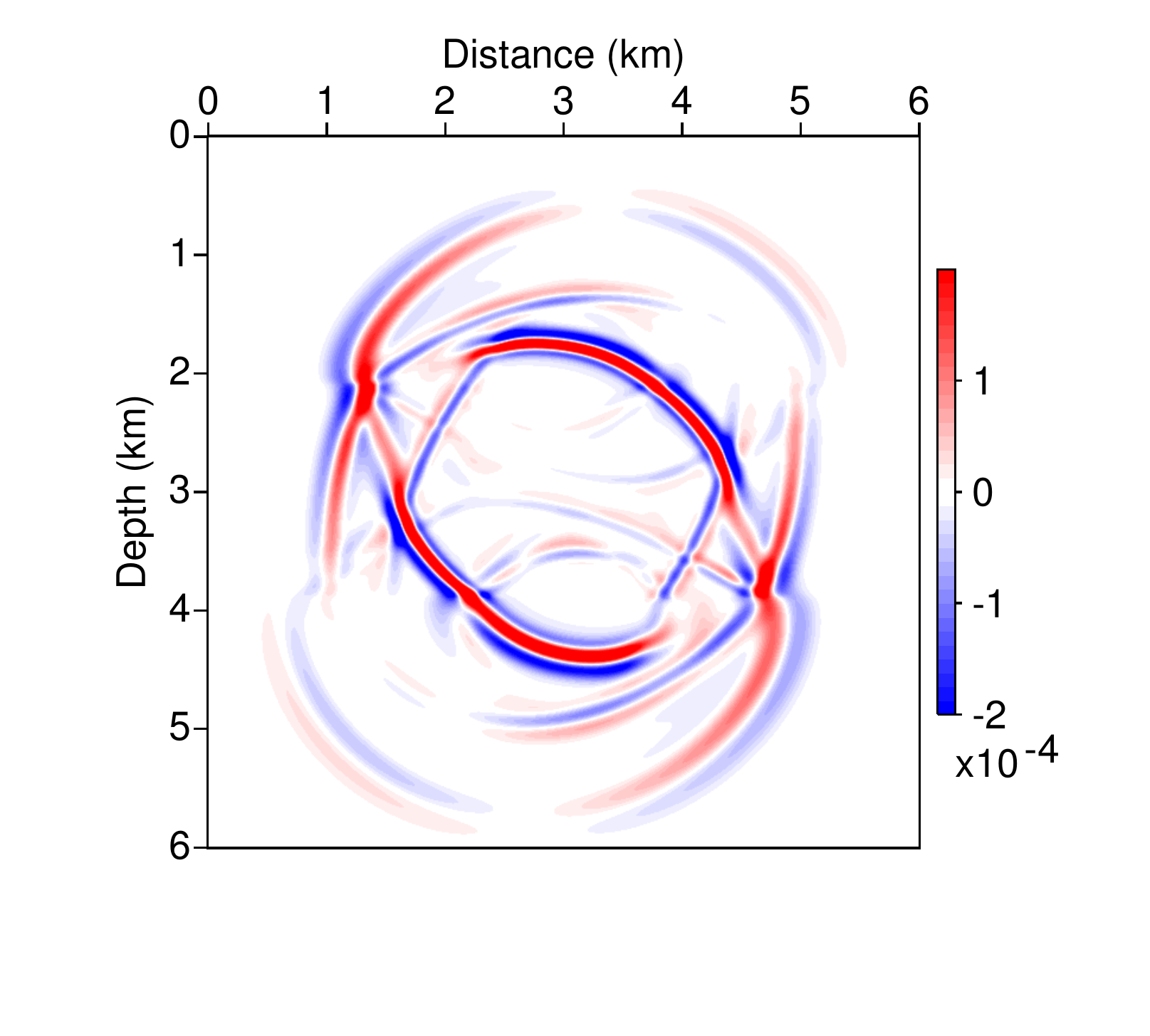}
}
\subfigure[]{
\label{fig:three_layer_DG_ms_typeI_300}
\includegraphics[width=0.4\textwidth,trim=45 55 45 0]{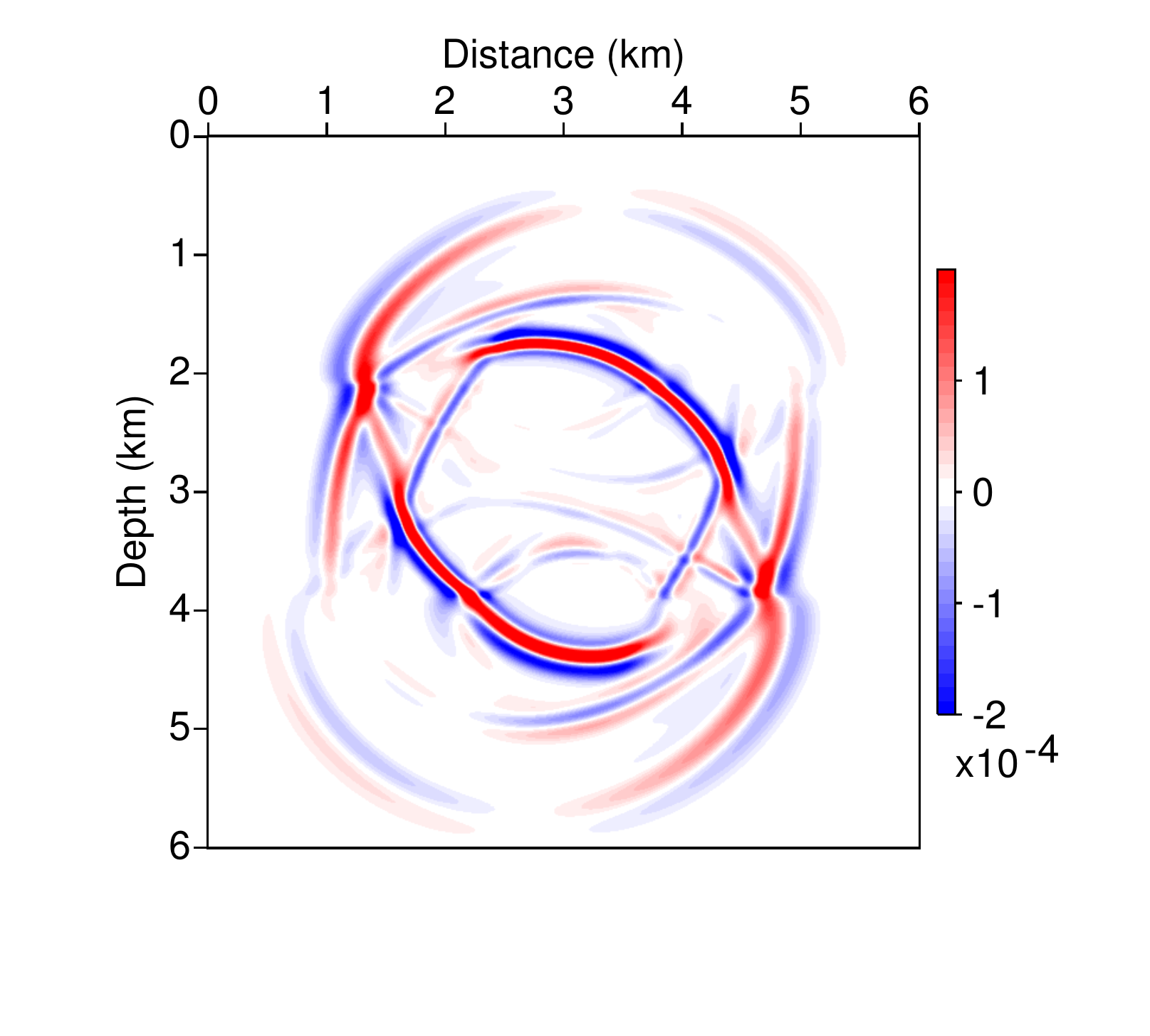}
}
\caption{$u_1$ wavefield snapshots at 0.7~s after source excitation: (a) reference solution and (b) DG-FEM solution on $200\times 200$-element mesh, (c) DG-FEM solution on $300\times 300$-element mesh, (d) DG-GMsFEM on $60\times 60$-element mesh with 50 type II interior basis functions and 30 type II boundary basis functions, (e) DG-GMsFEM solution on $20\times 20$-element mesh with 150 type II interior basis functions and 150 type II boundary basis functions, and (f) DG-GMsFEM solution on $20\times 20$-element mesh with 300 type I basis functions. Note the obvious numerical dispersion in DG-FEM solutions.}
\end{figure}

\revision{
In the same model, we also test the effectiveness of our DG-GMsFEM. We use \RLG{the} oversampling technique for our DG-GMsFEM, with 5 element oversampling, i.e., on each of the four boundaries of coarse element $K$, we oversample $K$ with 5 more fine elements. Figure \ref{fig:three_layer_DG_reference} shows the reference $u_1$ component wavefield snapshot at 0.7~s. Again, we calculated wavefields with DG-FEM in the model that is discretized with $200\times 200$ and $300\times 300$ elements, and show the $u_1$ component in Figures \ref{fig:three_layer_DG_200x200} and \ref{fig:three_layer_DG_300x300}. The DOFs of these two systems are $3.2\times 10^5$ and 
$7.2\times 10^5$, respectively. Similar \RLG{to} the results \RLG{of} CG-GMsFEM, both of these two DG-FEM solutions show obvious or visible numerical dispersions, and the relative error is quite large (see Table \ref{tab:three_layer_dg}). Meanwhile, Figure 
\ref{fig:three_layer_DG_ms_i50b30} show\RLG{s} the $u_1$ component snapshot calculated with our DG-GMsFEM in the same model discretized with $60\times 60$ coarse elements, and we use 50 type II interior basis functions and 30 type II boundary basis functions in the simulation. The DOF of such system is $2.88\times 10^5$, which is on the same order as that in DG-FEM on $200\times 200$-element mesh, \RLG{but} the solution of our DG-GMsFEM is much better, with only $3.36\times 10^{-3}$ error compared with 1.06 \RLG{for the} DG-FEM solution. Figure \ref{fig:three_layer_DG_ms_i150b150} show\RLG{s} the solution calculated with our DG-GMsFEM on a\RLG{n} even coarser mesh ($20\times 20$ elements, i.e., the element size is 300~m by 300~m), and we use 150 type II interior multiscale basis functions along with 150 type II boundary basis functions. This makes the DOF of the system $1.2\times 10^5$, which is only $1/6$ of that with DG-GMsFEM on $300\times 300$-element mesh, yet our DG-GMsFEM is again much more accurate than DG-FEM, with $3.98\times 10^{-3}$ error compared with about $5.98\times 10^{-1}$ of DG-FEM solution. If we use a total of 1800 type II basis functions for DG-GMsFEM on this $20\times 20$-element mesh, the solution will much more accurate, which is a reasonable guess as is the case of CG-GMsFEM. However, perhaps there is no such necessity in practice. We can see from Table \ref{tab:three_layer_dg} that by using 300 type I basis functions for DG-GMsFEM on the $20\times 20$-element mesh, the solution (Figure \ref{fig:three_layer_DG_ms_typeI_300}) is also accurate, with about $1.50\times 10^{-2}$ error. These comparisons show that our DG-GMsFEM is also able to achieve higher accuracy under the same DOF compared with conventional DG-FEM. The very small error of our GMsFEM also indicates that the multiscale method can provide very satisfactory approximation to the fine scale reference solution, with much smaller DOF in the system. For example, the DG-GMsFEM uses only no more than 5\% to 10\% of the DOF compared with the reference solution, yet the relative error is only about $3.36\times 10^{-3}$ to $1.50\times 10^{-2}$. 
}

\begin{table}
\centering
\begin{tabular}{|c|c|c|c|c|c|c|}
\hline 
Type & Mesh & DOF & $m_{\text{I}}$ & $m_{\text{II-interior}}$ & $m_{\text{II-boundary}}$ & $e(\mathbf{u}) (\%)$\tabularnewline
\hline 
\hline 
CG-FEM reference & $600\times600$ & 7.22E5 & - & - & - & -\tabularnewline
\hline 
CG-FEM & $200\times200$ & 8.08E4 & - & - & - & 1.0304E0\tabularnewline
\hline 
CG-GMsFEM & $60\times60$ & 8.56E4 & 23 & - & - & 1.0812E-1\tabularnewline
\hline 
CG-FEM & $300\times300$ & 1.81E5 & - & - & - & 5.7678E-1\tabularnewline
\hline 
CG-GMsFEM & $60\times60$ & 1.30E5 & 35 & - & - & 3.1999E-2\tabularnewline
\hline 
CG-GMsFEM & $60\times60$ & 1.30E5 & - & 20 & 15 & 3.1667E-2\tabularnewline
\hline 
\end{tabular}
\caption{A comparison of accuracy and DOF for three-layer VTI-TTI-isotropic
model. $m_{\text{I}}$ is the number of type I basis functions, $m_{\text{II-interior}}$
and $m_{\text{II-boundary}}$ are the numbers of type II interior
and boundary basis functions, respectively. $e(\mathbf{u})$ is the
relative $L^{2}$-norm error of CG-FEM or CG-GMsFEM with respect to
the reference solution. }
\label{tab:three_layer_cg}
\end{table}

\begin{table}
\centering
\begin{tabular}{|c|c|c|c|c|c|c|}
\hline 
Type & Mesh & DOF & $m_{\text{I}}$ & $m_{\text{II-interior}}$ & $m_{\text{II-boundary}}$ & $e(\mathbf{u})$\tabularnewline
\hline 
\hline 
DG-FEM reference & $600\times600$ & 2.88E6 & - & - & - & -\tabularnewline
\hline 
DG-FEM & $200\times200$ & 3.20E5 & - & - & - & 1.0572E0\tabularnewline
\hline 
DG-GMsFEM & $60\times60$ & 2.88E5 & - & 50 & 30 & 3.3565E-3\tabularnewline
\hline 
DG-FEM & $300\times300$ & 7.20E5 & - & - & - & 5.9762E-1\tabularnewline
\hline 
DG-GMsFEM & $20\times20$ & 1.20E5 & - & 150 & 150 & 3.9771E-3\tabularnewline
\hline 
DG-GMsFEM & $20\times20$ & 1.20E5 & 300 & - & - & 1.5013E-2\tabularnewline
\hline 
\end{tabular}
\caption{A comparison of accuracy and DOF for three-layer VTI-TTI-isotropic
model. $m_{\text{I}}$ is the number of type I basis functions, $m_{\text{II-interior}}$
and $m_{\text{II-boundary}}$ are the numbers of type II interior
and boundary basis functions, respectively. $e(\mathbf{u})$ is the
relative $L^{2}$-norm error of DG-FEM or DG-GMsFEM with respect to
the reference solution. }
\label{tab:three_layer_dg}
\end{table}

\subsection{Randomly heterogeneous anisotropic \oldrevision{medium} model with curved boundaries}
\begin{figure}
\centering
\subfigure[]{
\label{fig:random_c11}
\includegraphics[trim=50 50 30 0,width=0.41\textwidth]{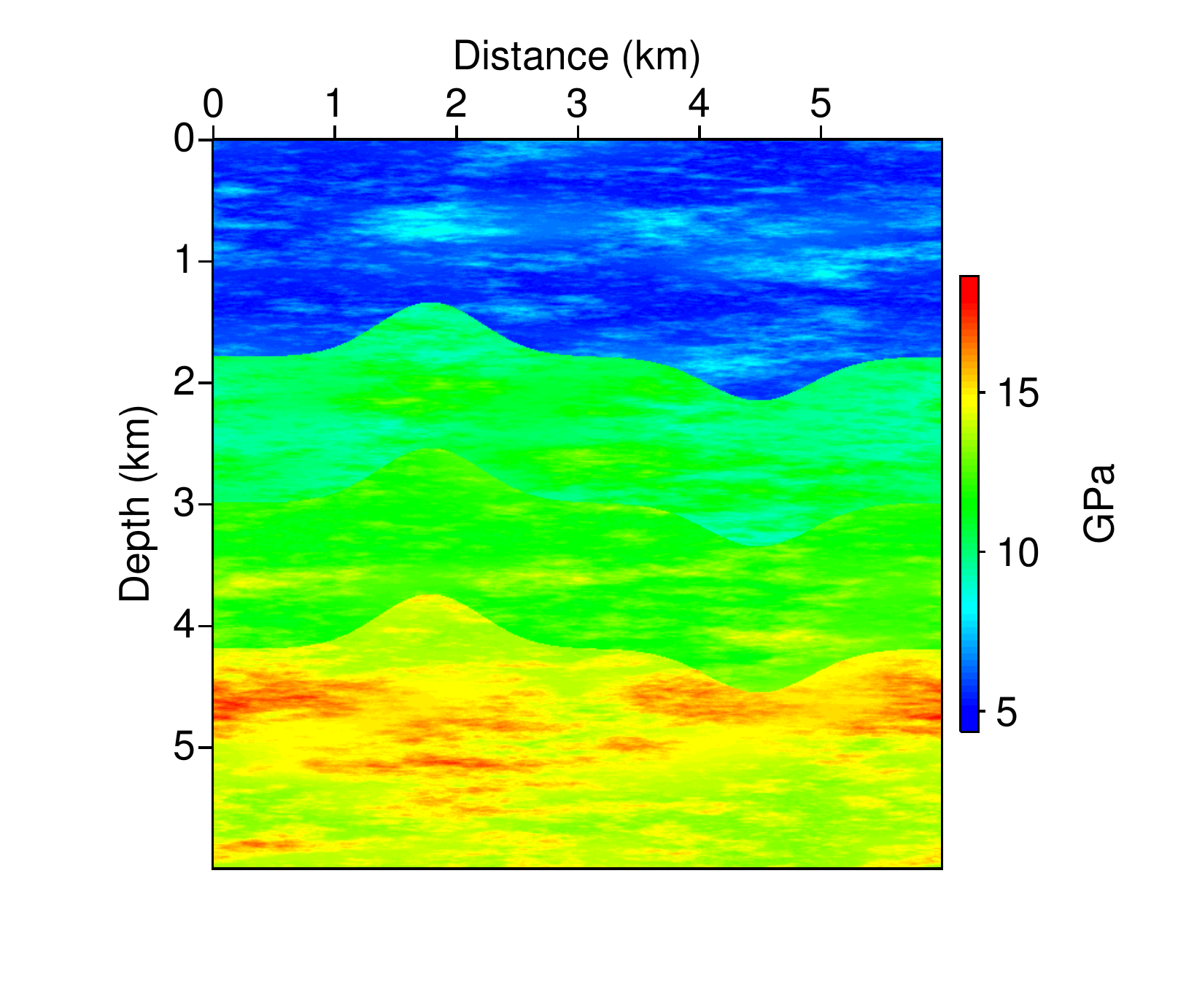}
}
\subfigure[]{
\label{fig:random_c13}
\includegraphics[trim=50 50 30 0,width=0.41\textwidth]{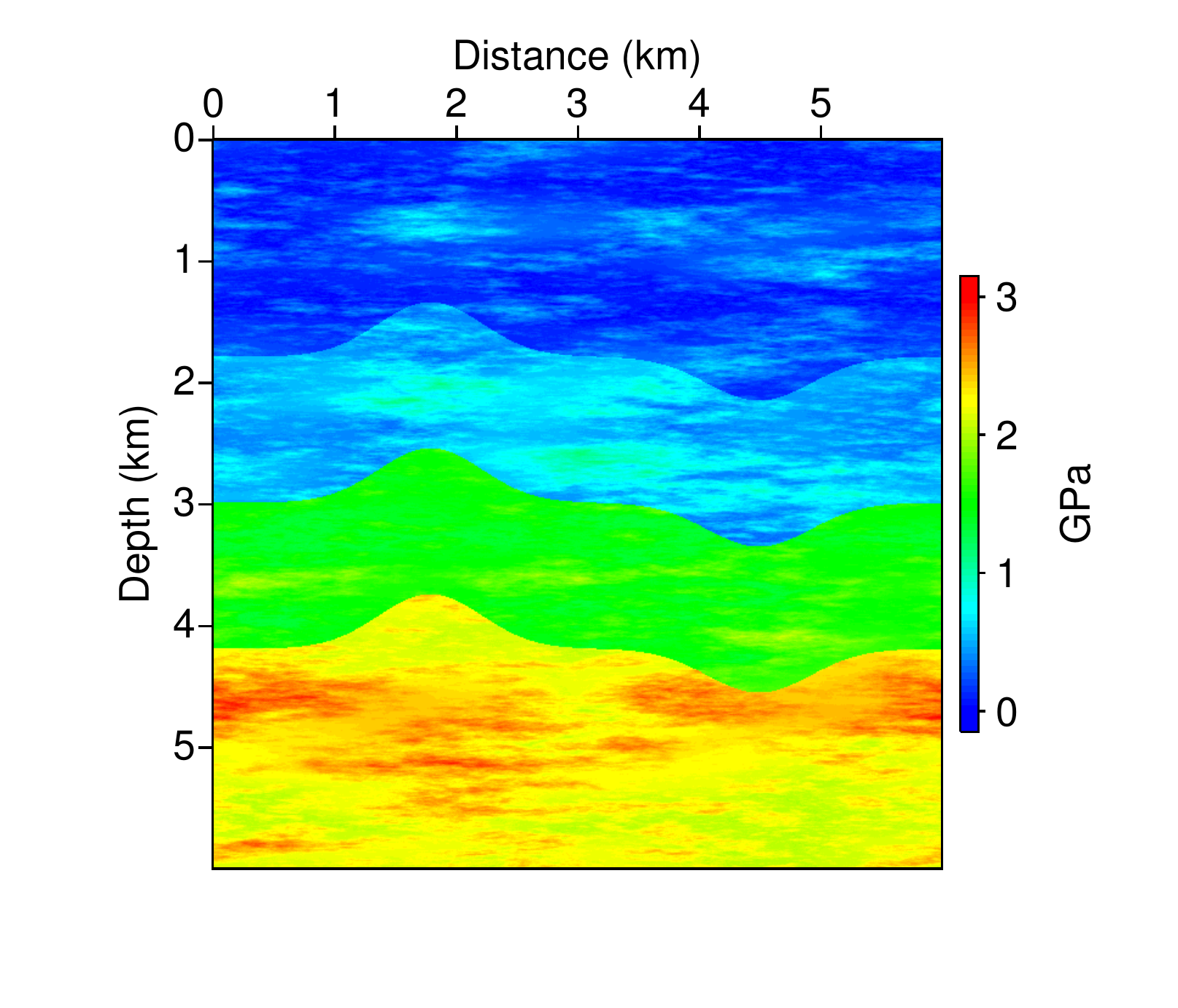}
}
\subfigure[]{
\label{fig:random_c15}
\includegraphics[trim=50 50 30 0,width=0.41\textwidth]{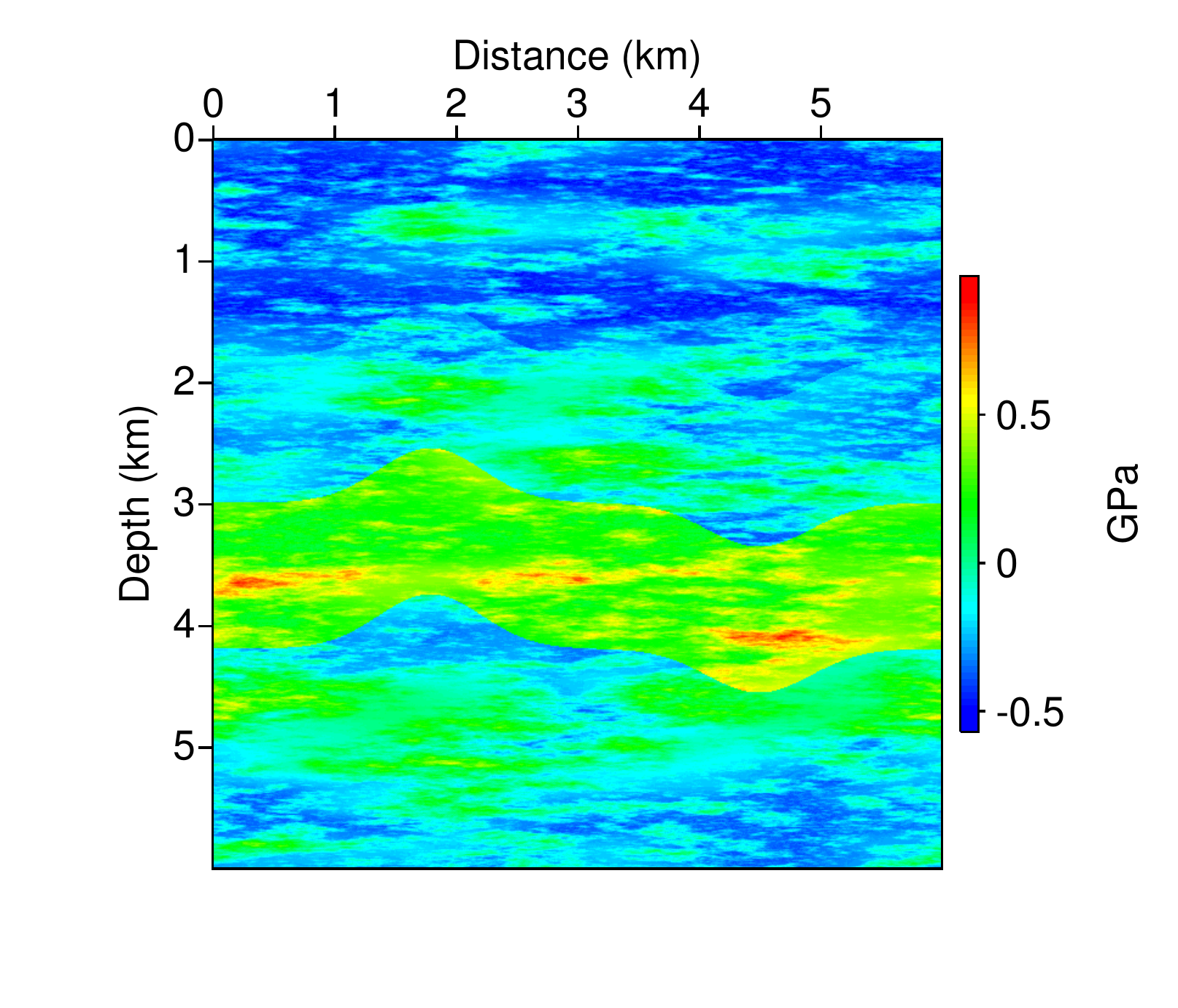}
}
\subfigure[]{
\label{fig:random_c33}
\includegraphics[trim=50 50 30 0,width=0.41\textwidth]{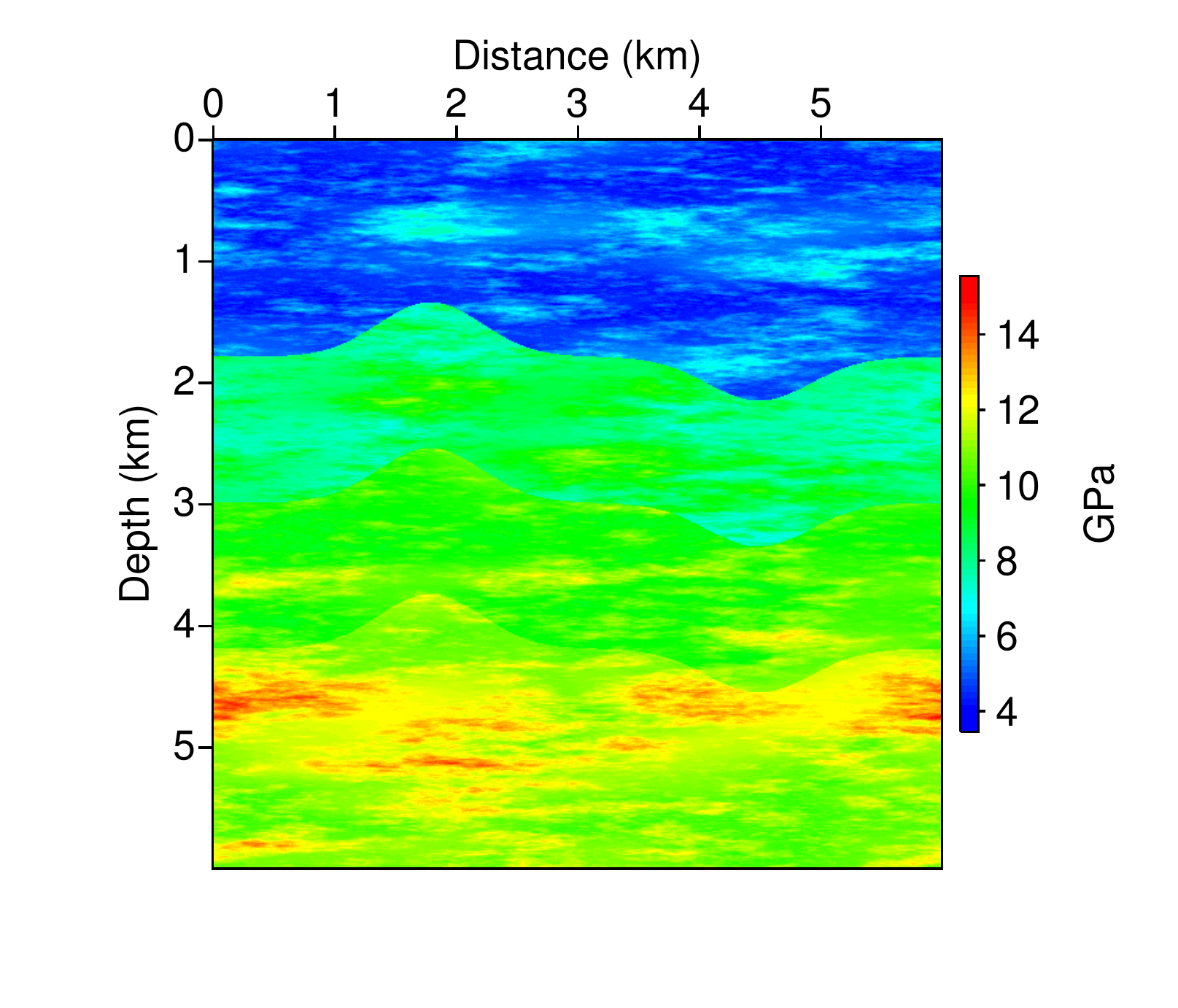}
}
\subfigure[]{
\label{fig:random_c35}
\includegraphics[trim=50 50 30 0,width=0.41\textwidth]{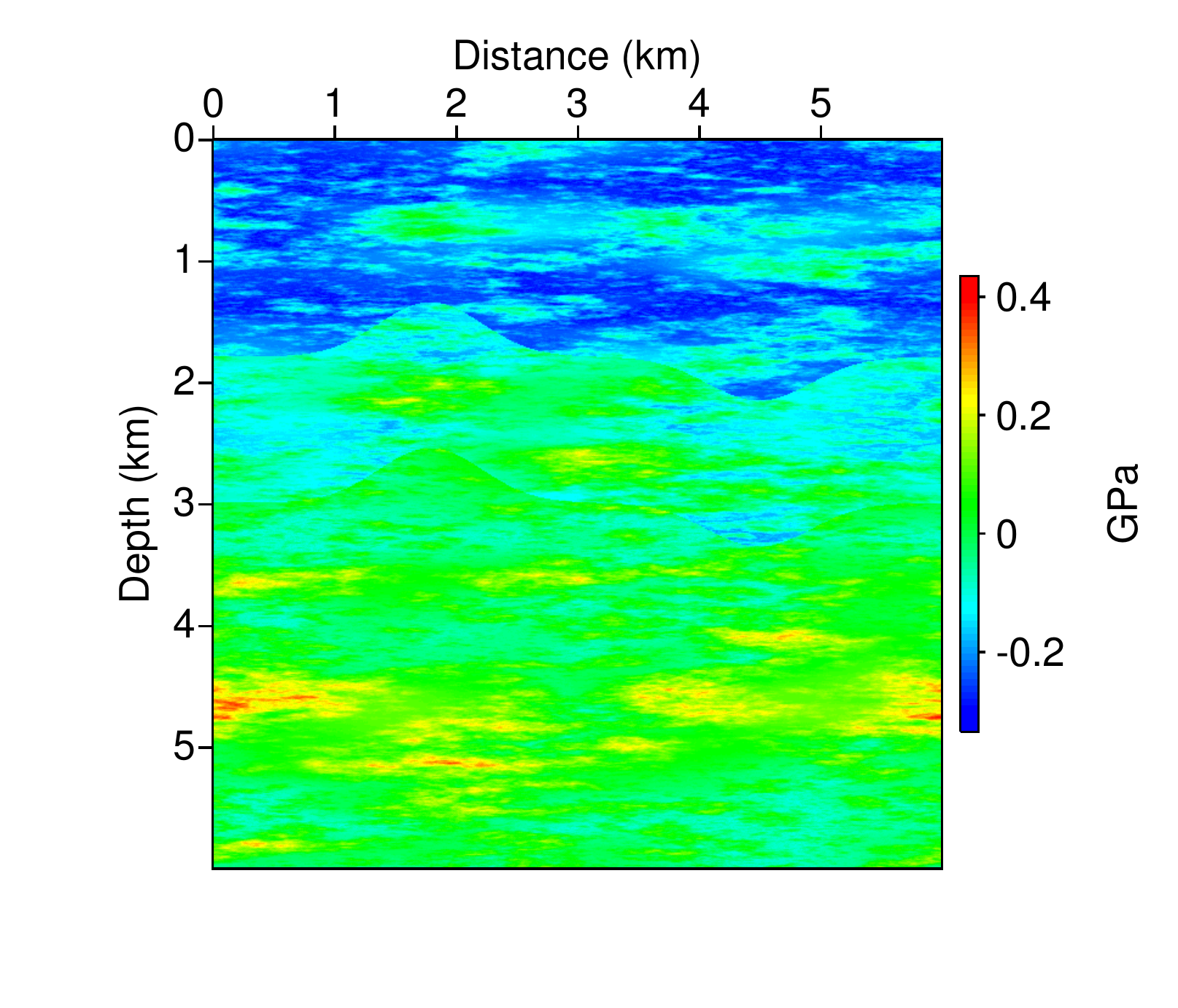}
}
\subfigure[]{
\label{fig:random_c55}
\includegraphics[trim=50 50 30 0,width=0.41\textwidth]{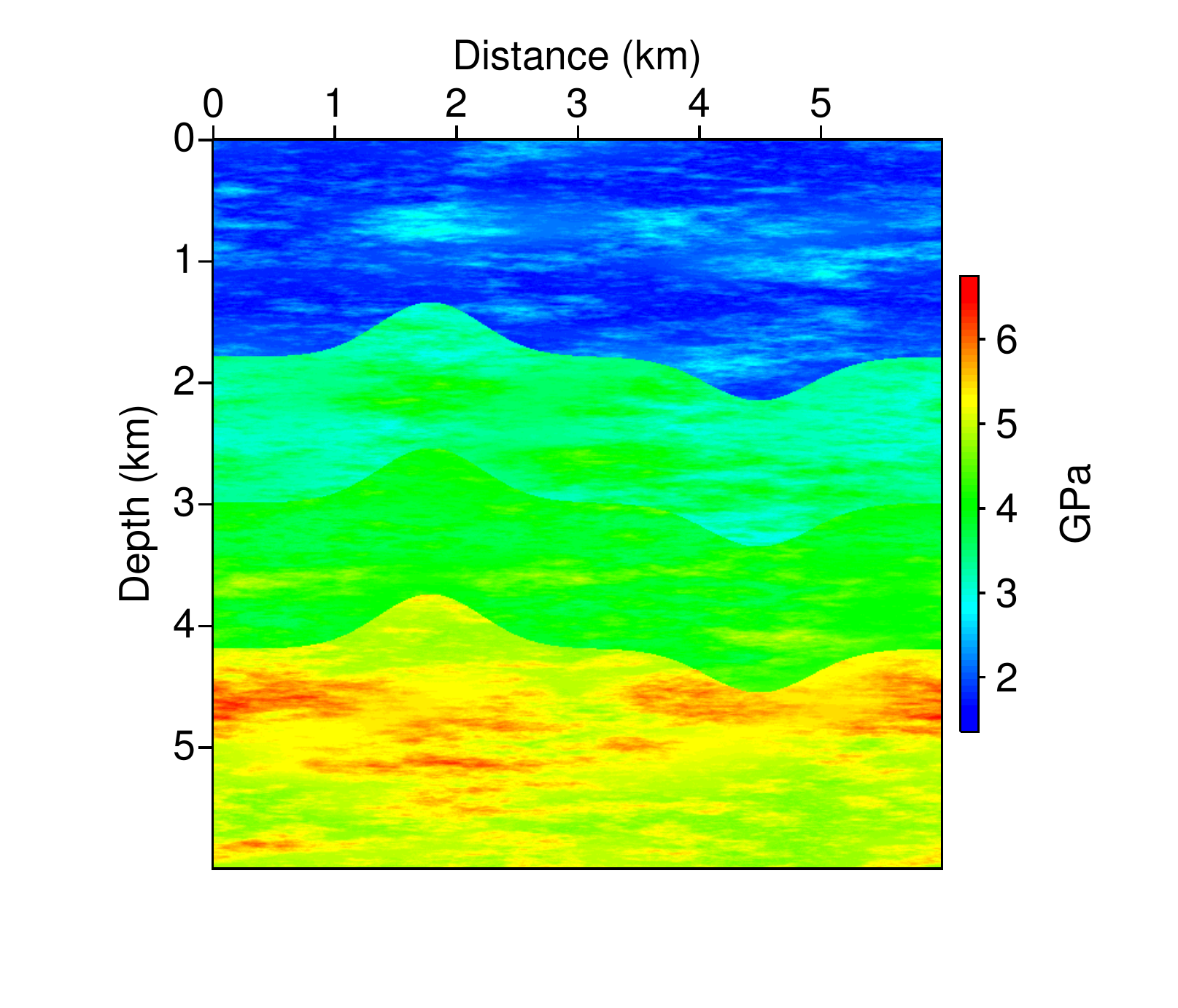}
}
\caption{A heterogeneous, anisotropic elastic model. Parts (a)-(f) show $C_{11}$, $C_{13}$, $C_{15}$, $C_{33}$, $C_{35}$ and $C_{55}$, respectively.}
\end{figure}

We  \oldrevision{further} verify the effectiveness of the DG formulation of our GMsFEM in a heterogeneous anisotropic elastic model, with elasticity parameters shown in Figures \ref{fig:random_c11}-\ref{fig:random_c55}, \oldrevision{by comparing the results from conventional \revision{DG-FEM} and our DG-GMsFEM with the reference solution}. The density is \oldrevision{\RLG{for convenience}} set to be homogeneous with the value 1000~kg/m$^3$. This heterogeneous model is 6000~m in depth and 6000~m in horizontal distance, and consists of 600$\times$600 fine elements. The source is placed at \revision{$(3.0,3.0)$~km}, and we apply a Ricker wavelet with 10~Hz central frequency. \revision{We also set $\eta=5$.} The time sampling interval is $\Delta t=0.5$~ms, \oldrevision{and we implement 1650 time steps, i.e., totally 0.825~s}.

\begin{figure}
\centering
\subfigure[]{
\label{fig:random_curved_u1_reference}
\includegraphics[trim=45 55 50 0,width=0.4\textwidth]{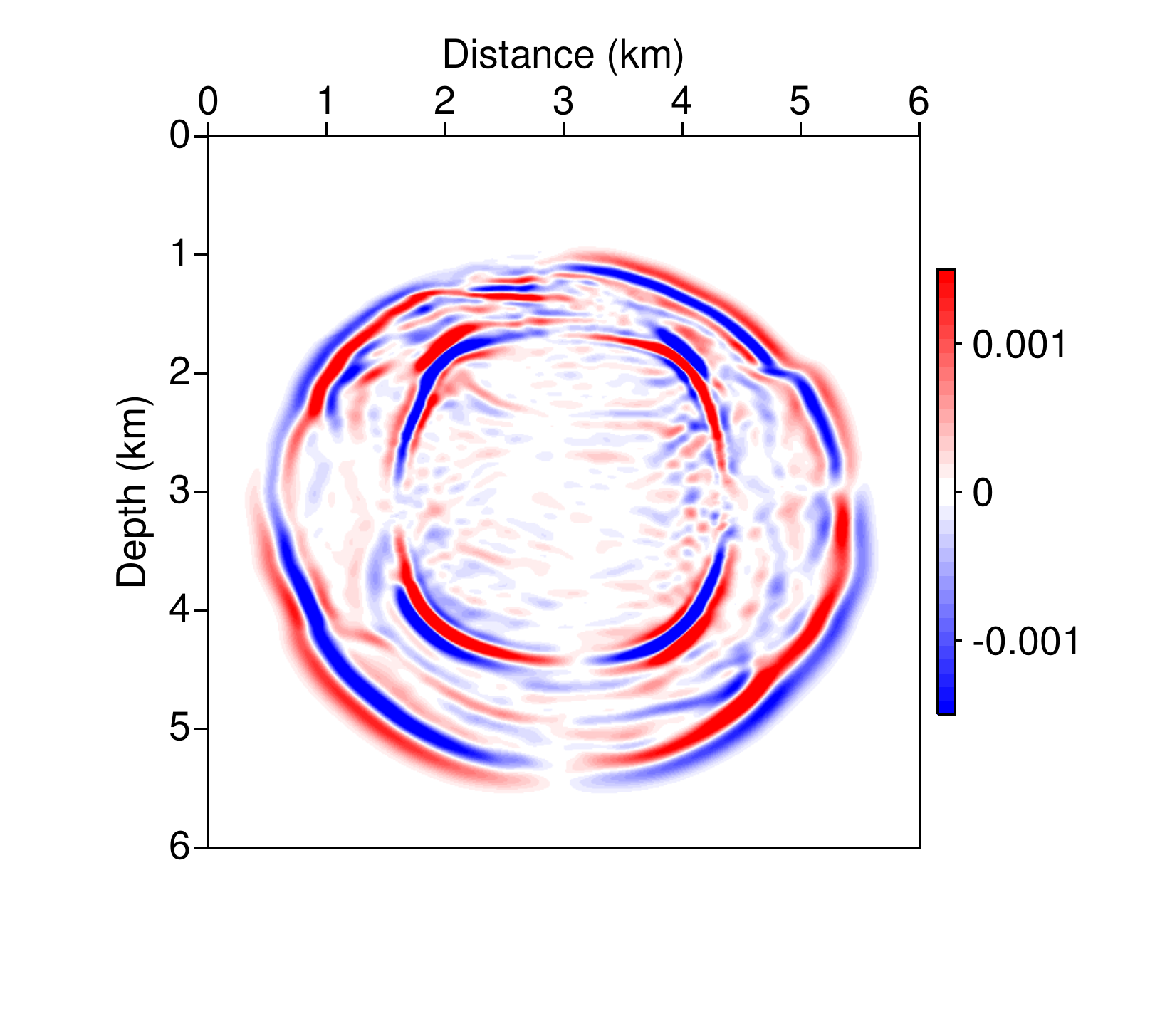}
}
\subfigure[]{
\label{fig:random_curved_u1_fine}
\includegraphics[trim=45 55 50 0,width=0.4\textwidth]{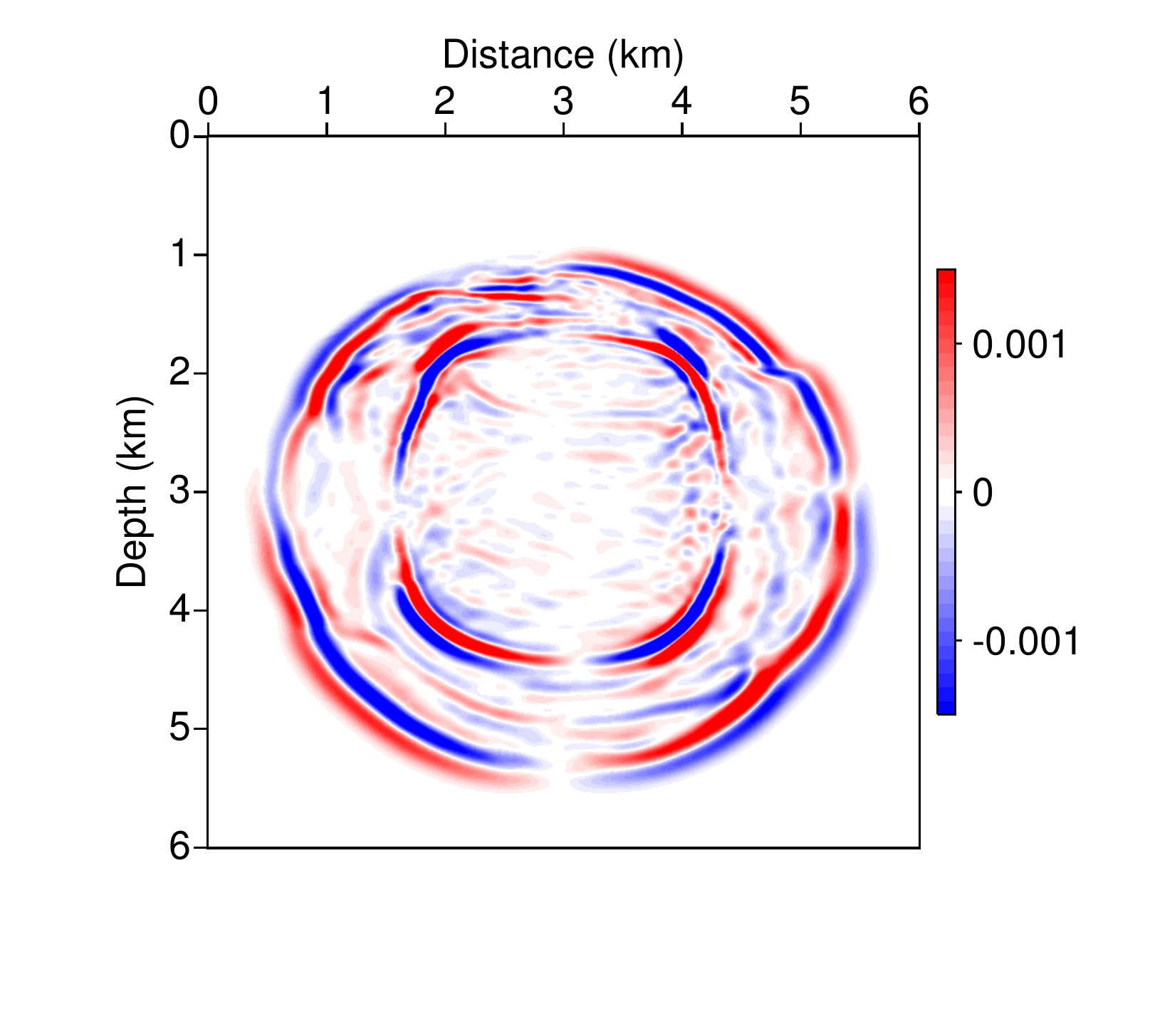}
}
\subfigure[]{
\label{fig:random_curved_u1_i10b10}
\includegraphics[trim=45 55 50 0,width=0.4\textwidth]{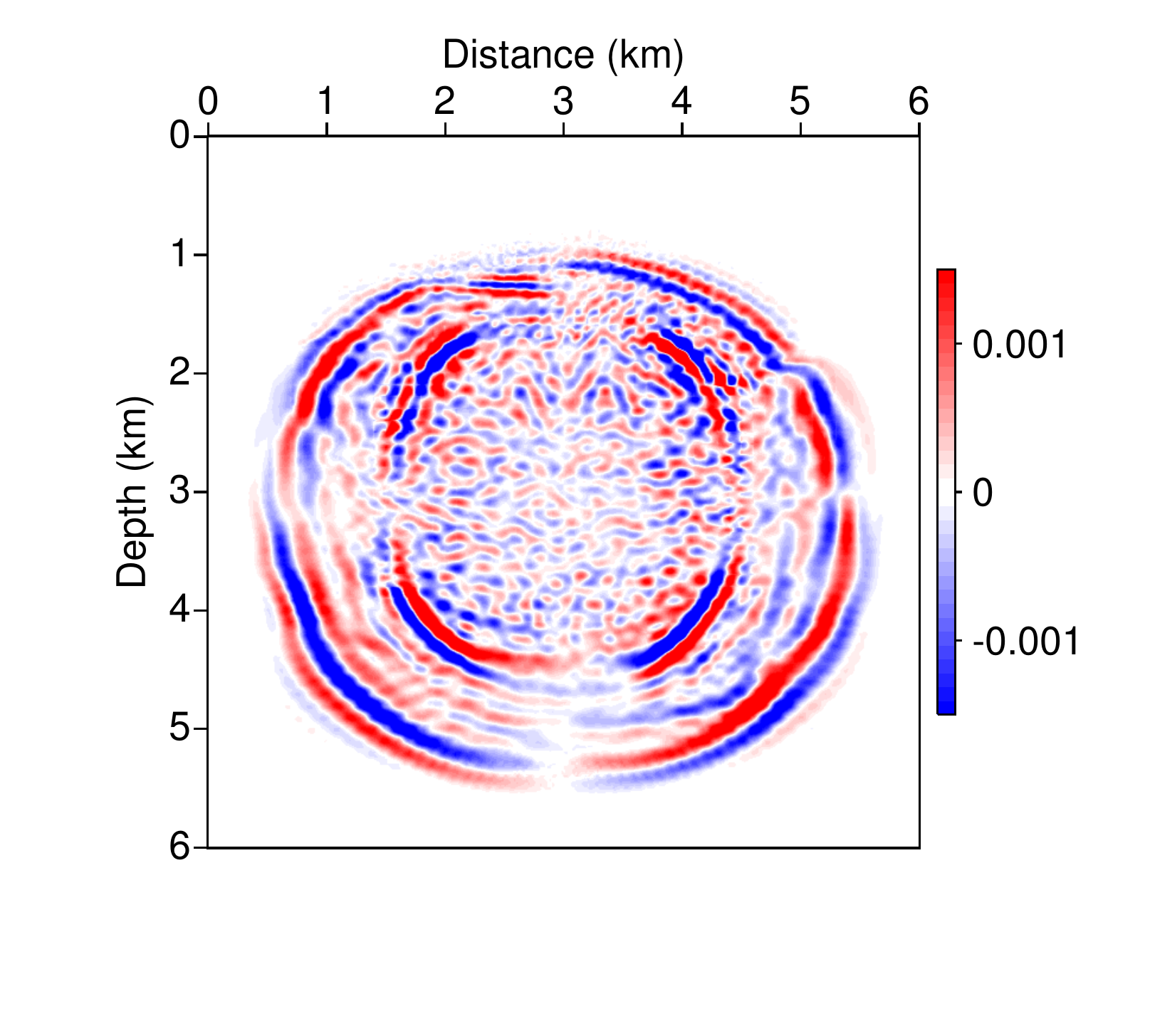}
}
\subfigure[]{
\label{fig:random_curved_u1_i20b10}
\includegraphics[trim=45 55 50 0,width=0.4\textwidth]{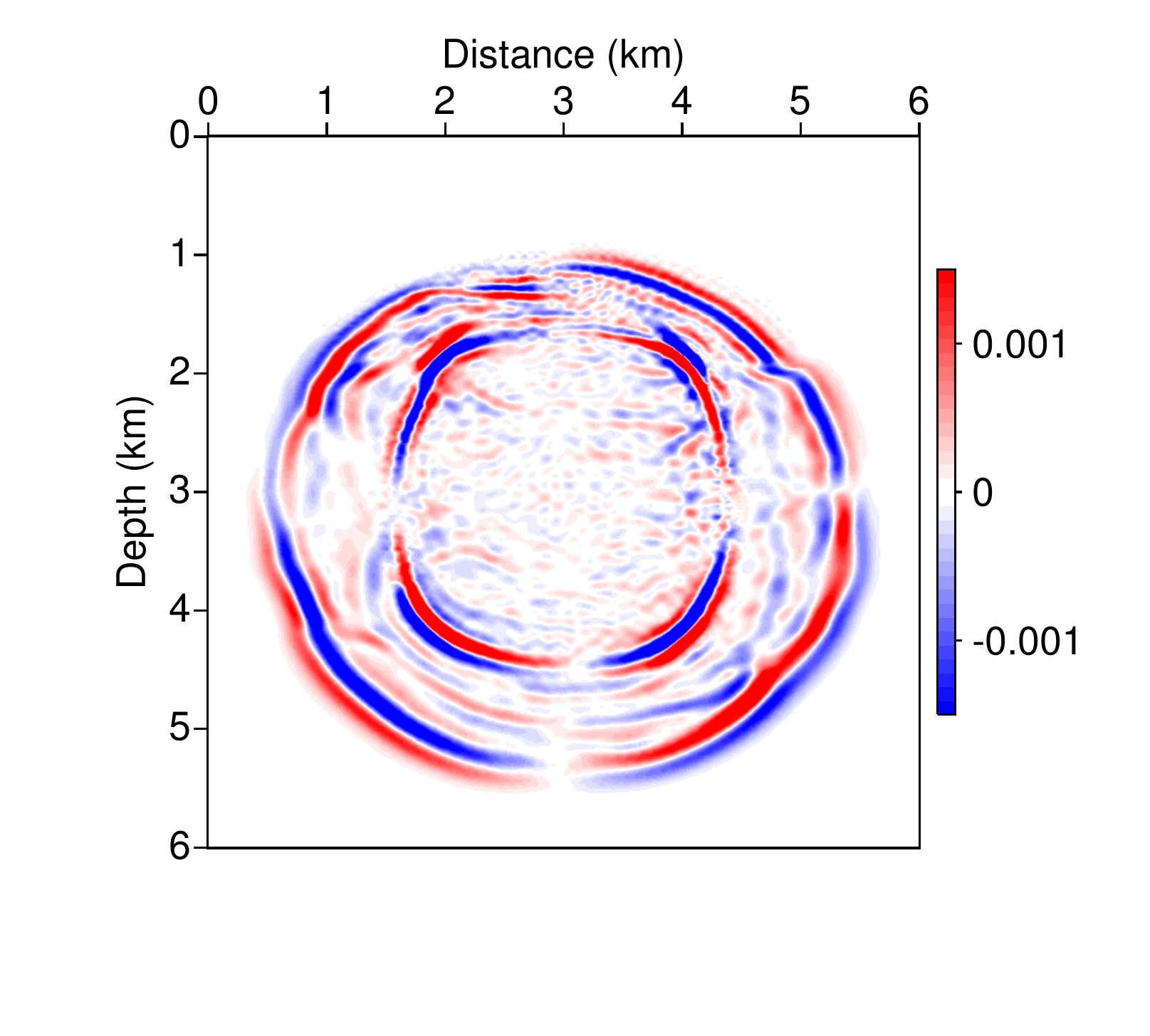}
}
\subfigure[]{
\label{fig:random_curved_u1_i30b20}
\includegraphics[trim=45 55 50 0,width=0.4\textwidth]{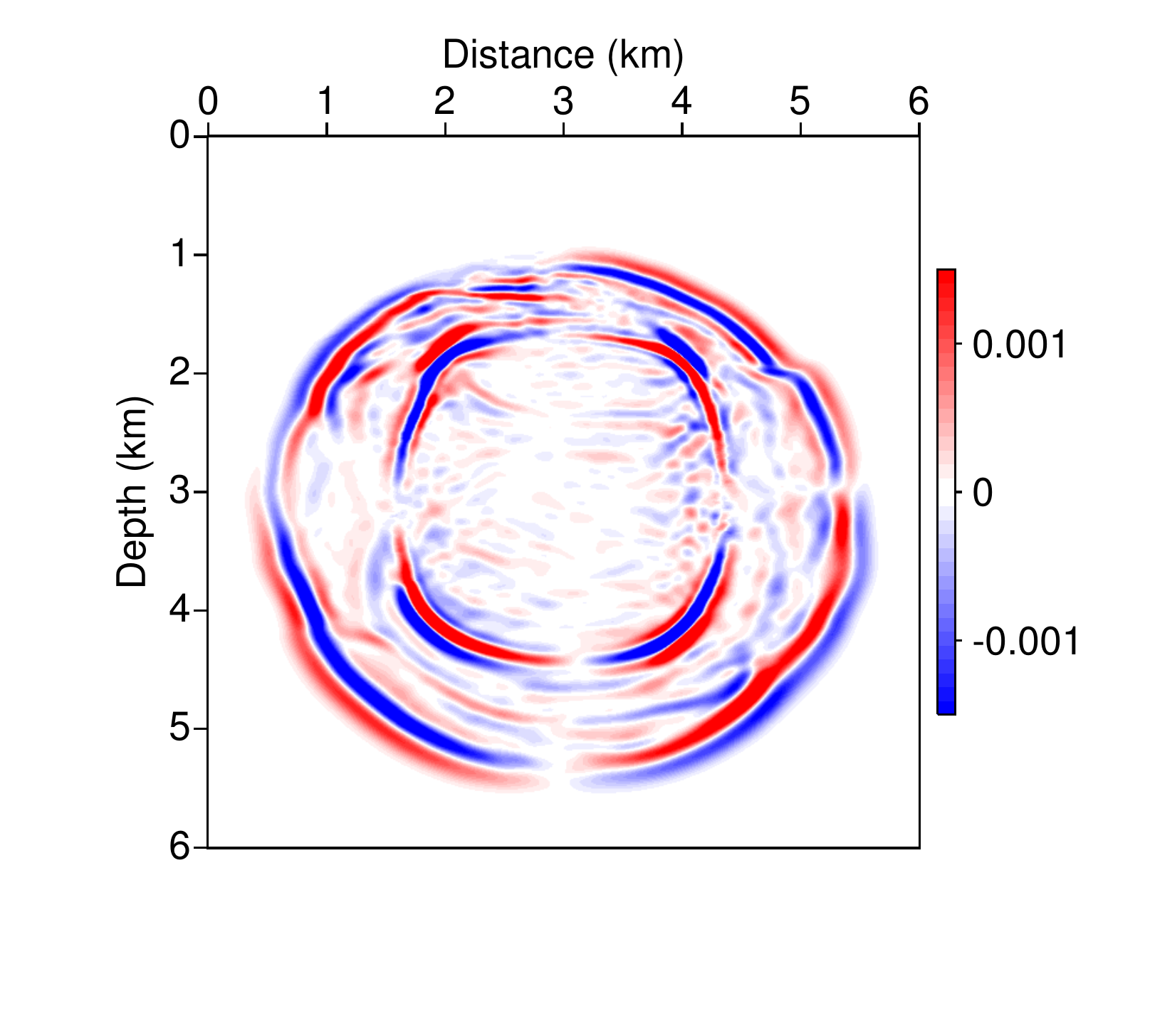}
}
\subfigure[]{
\label{fig:random_curved_u1_i30b30}
\includegraphics[trim=45 55 50 0,width=0.4\textwidth]{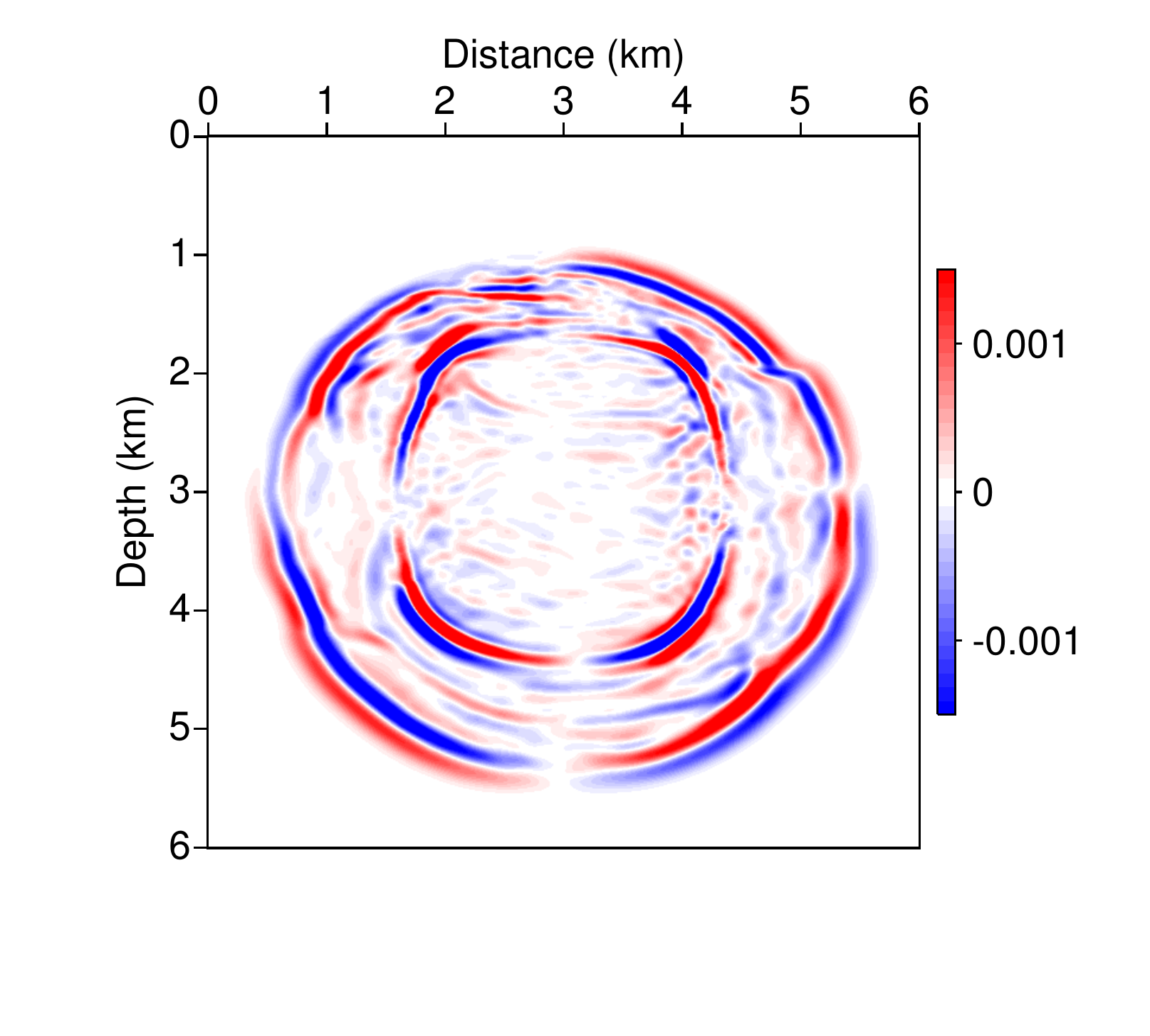}
}
\caption{$u_1$ wavefield snapshots of (a) reference solution on $600\times 600$-element mesh with 0.05~ms time step size (b) DG-FEM solution on $600\times 600$-element mesh and 0.5~ms time step size and (c)-(f) DG-GMsFEM multiscale solutions on $60\times 60$ coarse mesh with 0.5~ms time step size and with $(m_{\text{boundary}},m_{\text{interior}})=(10,10)$, $(10,20)$, $(20,30)$, $(30,30)$, respectively.}
\end{figure}

\revision{To \RLG{obtain} a reference solution, we solve the elastic wave equation with DG-FEM on the finely discretized $600\times 600$-element model, with a much smaller time sampling interval 0.05~ms (total time steps turns to be 16500). This small time step size can make the dispersion parameter $r=v\Delta t/\Delta h$ ($v$ is the phase velocity and $\Delta h$ is the grid size) be 1/10 of the dispersion parameter using $\Delta t=0.5$~ms, and therefore the solution can be adequately accurate to serve as a reference solution, which is shown in Figure \ref{fig:random_curved_u1_reference}.} 
\revision{We then calculate the DG-FEM solution on the $600\times 600$-element mesh with $\Delta t=0.5$~ms, and the solution is shown in Figure  \ref{fig:random_curved_u1_fine}. The result in Table \ref{tab:random_curved_dg} tells us that the DG-FEM solution has about $2.74\times 10^{-3}$ relative error with respect to the reference solution, although the DOF is same.}

For multiscale modeling, we discretize the model with $60\times 60$ coarse elements, and therefore the coarse element is 100~m in each direction, containing $10\times 10$ fine elements. \revision{Again, we apply \RLG{the} oversampling technique for  DG-GMsFEM, with 5 element oversampling}. We adopt a penalty parameter \oldrevision{$\gamma=1.0$ for DG-FEM} and $\gamma=5.0$ for \oldrevision{all the DG-GMsFEM} \revision{simulations}. Figures \ref{fig:random_curved_u1_i10b10}--\ref{fig:random_curved_u1_i30b30} are solutions from \revision{our} DG-GMsFEM. The number of boundary and interior basis functions range\RLG{s} from 10 to 30, respectively, indicated by $(m_{\text{boundary}},m_{\text{interior}})$. The wavefield contains complicated direct and reflected waves from curved reflectors, as well as waves scattered from the random heterogeneities. \oldrevision{Visual inspection shows that the wavefield snapshots with $(10,10)$ and $(10,20)$ basis functions contain obvious numerical dispersion. \RLG{Specifically}, there \RLG{is noise} ahead of the qP- and qSV-wavefronts, i.e., faster and slower wavefronts, at these two basis function combinations, which are non-causal artifacts due to the insufficient information provided by the small number of multiscale basis functions}. \oldrevision{\RLG{This noise disappears} in} wavefield snapshots with $(20,30)$ and $(30,30)$ basis functions, \oldrevision{which} are almost the same as that in Figure \ref{fig:random_curved_u1_reference}, i.e., the \revision {reference} solution. 

We only show four different combinations of boundary and interior basis in the wavefield snapshots. However, to further quantify the relation between the number of basis functions with the relative error as well as other quantities, we summarize more results in Table \ref{tab:random_curved_dg}. In our tests, the case with fewest basis functions, i.e., $(10,10)$, also has maximum relative error, about $8.45\times 10^{-1}$. With more and more basis functions, this error reduces to $5.02\times 10^{-3}$ when using 30 boundary basis functions and 40 interior basis functions, which is almost the same as the error of DG-FEM on fine mesh with 0.5~ms time step. At the same time, the degrees of freedom increases from $7.20\times 10^4$ to $2.52\times 10^5$, which is still much less than \revision{that of DG-FEM system} with degrees of freedom $2.88\times 10^6$. The CPU time for the time stepping part ($T_{\text{modeling}}$) of GMsFEM is shorter than that of the DG-FEM. \revision{Even in the case of $(30,40)$ basis functions, $T_{\text{modeling}}$ is still about 70\% of that with DG-FEM on the fine mesh}. The CPU time of calculating more basis functions and preprocessing the global matrices is naturally longer. However, we have to remark that this calculation is one-time \oldrevision{and can be parallelized}, as is the case in the first model. 
\revision{These results show that our DG-GMsFEM is able to achieve \RLG{the} same level of accuracy as DG-FEM, yet both the computational time and the DOF are notably less than those of DG-FEM.}

\begin{table}
\centering
\begin{tabular}{|c|c|c|c|c|c|c|}
\hline 
Type & $m_{\text{II-interior}}$ & $m_{\text{II-boundary}}$ & DOF & $e(\mathbf{u})$ & $T_{\text{basis}}$ (s) & $T_{\text{modeling}}$ (s)\tabularnewline
\hline 
\hline 
DG-FEM reference & - & - & 2.88E6 & - & - & 5698.71\tabularnewline
\hline 
DG-FEM & - & - & 2.88E6 & 2.7409E-3 & - & 527.90\tabularnewline
\hline 
DG-GMsFEM & 10 & 10 & 7.20E4 & 8.4515E-1 &  2010.75  & 38.59\tabularnewline
\hline 
DG-GMsFEM & 20 & 10 & 1.08E5 & 2.2344E-1 &  2105.33 & 75.93\tabularnewline
\hline 
DG-GMsFEM &10 & 20 & 1.08E5 & 1.4509E-1 &  2141.74 & 77.30\tabularnewline
\hline 
DG-GMsFEM &20 & 20 & 1.44E5 & 3.6960E-2 &  2349.60 & 130.80\tabularnewline
\hline 
DG-GMsFEM &30 & 20 & 1.80E5 & 1.1472E-2 &  2546.61 & 208.75\tabularnewline
\hline 
DG-GMsFEM &20 & 30 & 1.80E5 & 1.2237E-2 &  2444.47 & 207.57\tabularnewline
\hline 
DG-GMsFEM &30 & 30 & 2.16E5 & 6.2483E-3 &  2696.59 & 282.74\tabularnewline
\hline 
DG-GMsFEM &40 & 20 & 2.16E5 & 6.3878E-3 &  2974.15 & 283.96\tabularnewline
\hline 
DG-GMsFEM & 20 & 40 & 2.16E5 & 1.0703E-2 &  2475.37 & 284.66\tabularnewline
\hline 
DG-GMsFEM & 40 & 30 & 2.52E5 & 5.0195E-3 &  2761.66 & 373.94\tabularnewline
\hline
\end{tabular}
\caption{The relation between number of basis functions and the relative error,
as well as the DOF and calculation time. $m_{\text{II-boundary}}$
and $m_{\text{II-interior}}$ are the numbers of boundary and interior
basis functions of DG-GMsFEM, respectively, $e(\mathbf{u})$ is the $L^2$-norm error, as defined in the text, $T_{\text{basis}}$ is the CPU time of calculating the multiscale
basis functions, and $T_{\text{modeling}}$
is the CPU of calculating the wavefield, i.e., all the time steps. The first line represents the reference solution on $600\times 600$-element mesh calculated with 0.05~ms time step, the second line is DG-FEM solution on $600\times 600$-element mesh with time step size 0.5~ms, and all  other lines represent DG-GMsFEM with $60\times 60$-element mesh with time step 0.5~ms. }
\label{tab:random_curved_dg}
\end{table}

\subsection{Heterogeneous model: adaptive assignment of number of basis functions}
\begin{figure}
\centering
\subfigure[]{
\label{fig:Marmousi2_vp_part}
\includegraphics[trim=50 50 20 0,width=0.41\textwidth]{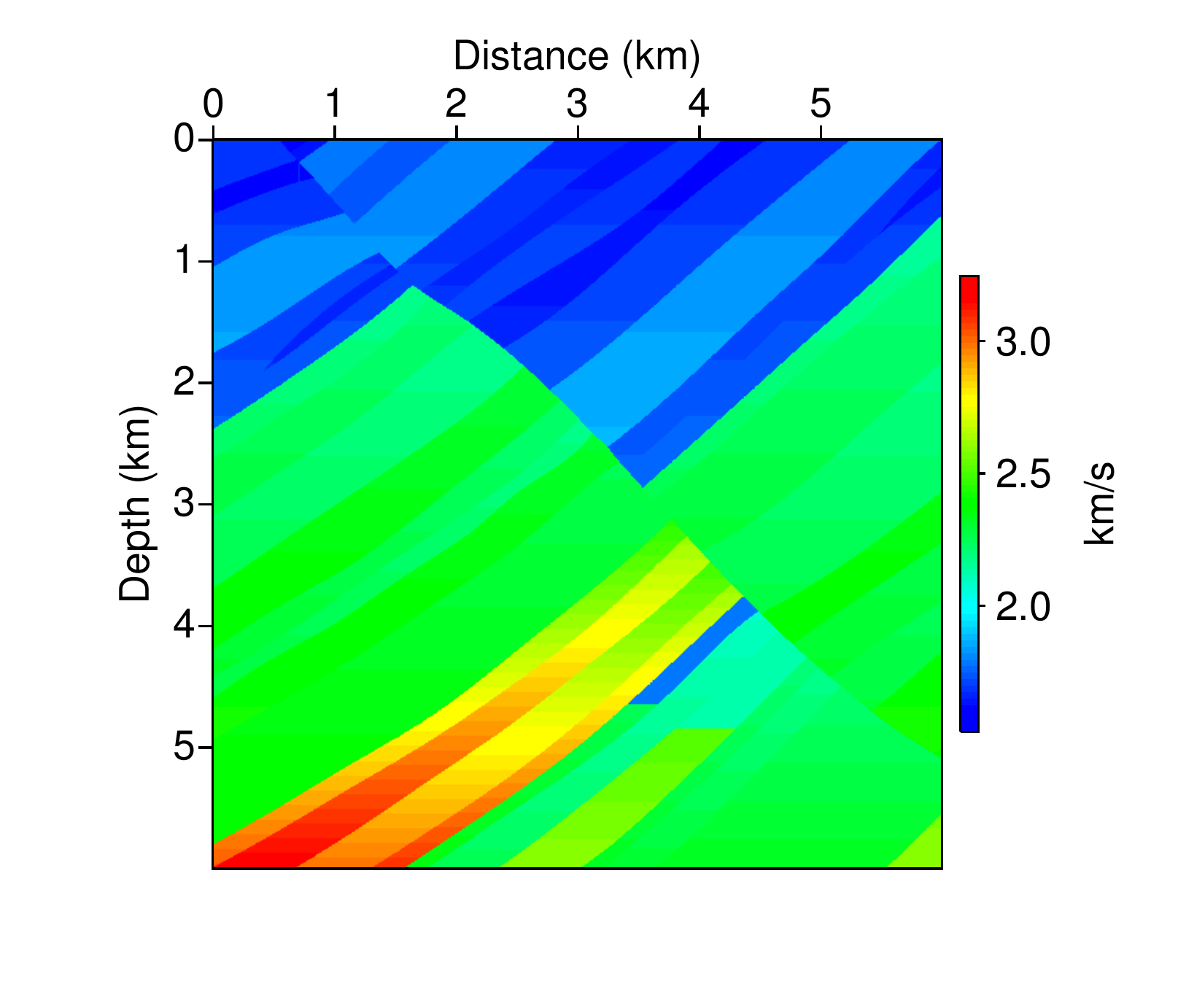}
}
\subfigure[]{
\label{fig:Marmousi2_vs_part}
\includegraphics[trim=50 50 20 0,width=0.41\textwidth]{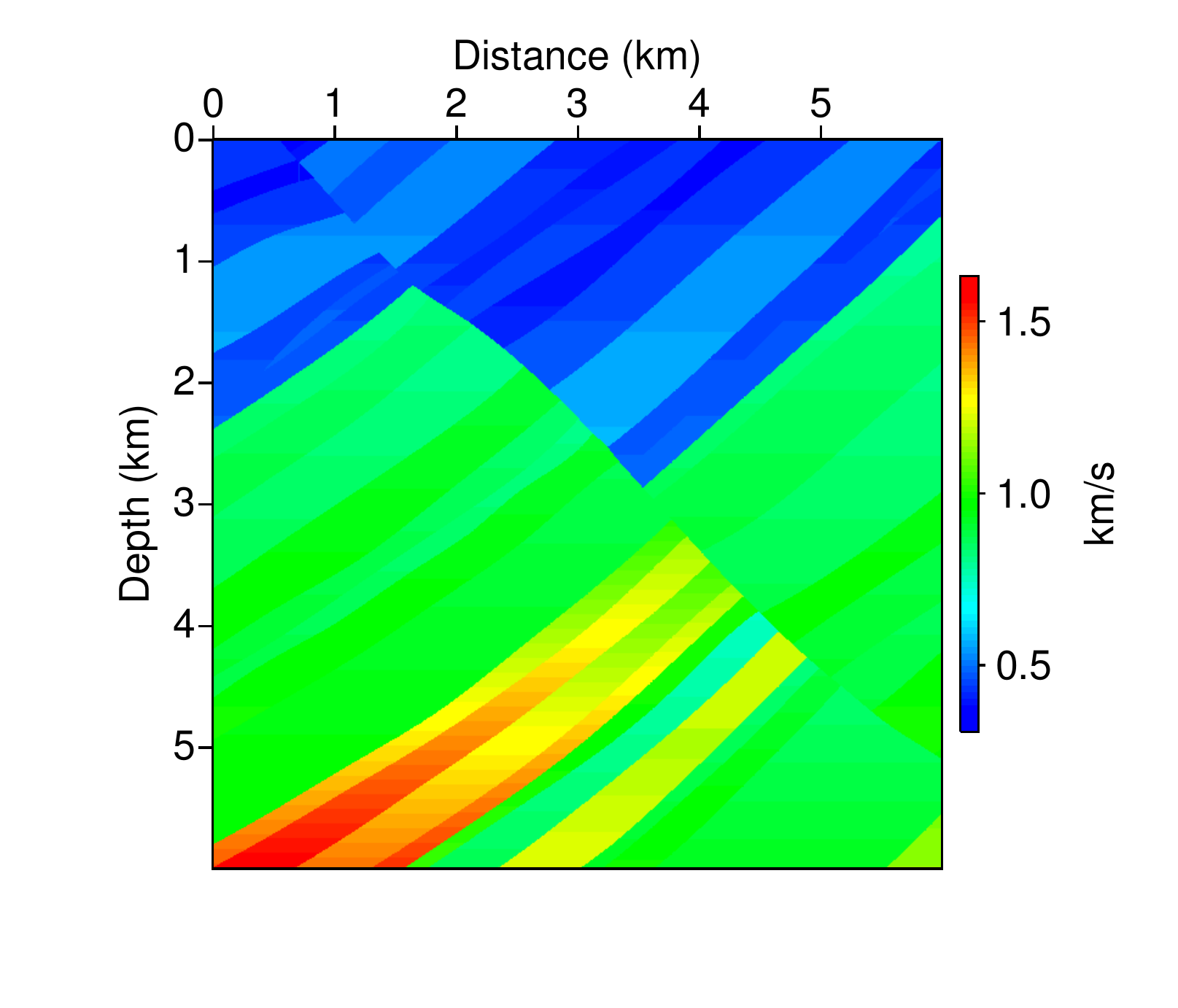}
}
\subfigure[]{
\label{fig:Marmousi2_rho_part}
\includegraphics[trim=50 50 20 0,width=0.41\textwidth]{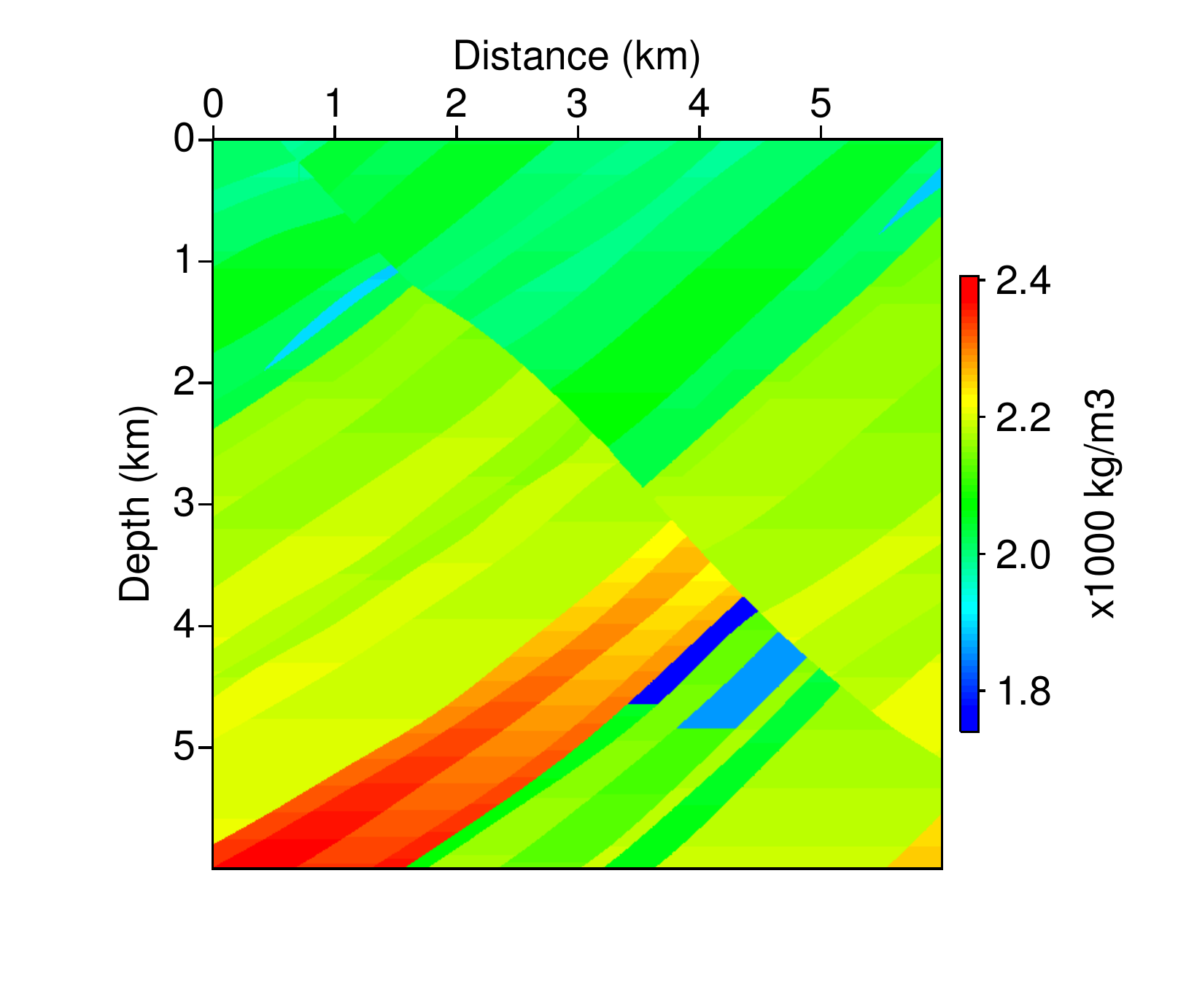}
}
\caption{The (a) P- and (b) S-wave velocity model, and (c) density model cropped from Marmousi 2 elastic model. }
\end{figure}

We use a subset of the Marmousi~2 elastic model \cite{Martin-etal_2006} to design a test model
to illustrate the process of choosing the number
of basis functions in coarse elements. Specifically, we used the model parameter grids to define the spatial distribution
of properties, but arbitrarily changed the original spatial sample interval from 1.25~m to 10~m to produce a model
with a larger size scale.
Figures \ref{fig:Marmousi2_vp_part}, \ref{fig:Marmousi2_vs_part} and \ref{fig:Marmousi2_rho_part} show the 
P- and S-wave velocity, and density of the chosen part of the model. 
The \RLG{number of elements} in each direction is 600, and we intend to solve the elastic wave equation in this model with a coarse mesh composed of $30\times 30$ coarse elements, which means that each coarse element contains $20\times 20$ fine elements, and the coarse element size is 200~m in each direction. One important motivation for using this model is that the velocity in the upper part of the model is clearly slower than that in the lower part, and therefore we want to test the speed up in computation time by using fewer bases in the lower portion where wavelength is longer. We also set damping boundary conditions at all four boundaries. The source is a Ricker wavelet with central frequency 5~Hz, placed at $(3,2)$~km, \revision{and we set $\eta=5$}. We have used a penalty parameter $\gamma=5.0$ in DG-FEM simulation and $\gamma=100.0$ in DG-GMsFEM simulations. \oldrevision{Also, we use a time step of 0.5~ms, and implement totally 3000 time steps, i.e., 1.5~s.}

\begin{figure}
\centering
\subfigure[]{
\label{fig:Marmousi2_basis_map_interior}
\includegraphics[trim=50 50 20 0,width=0.41\textwidth]{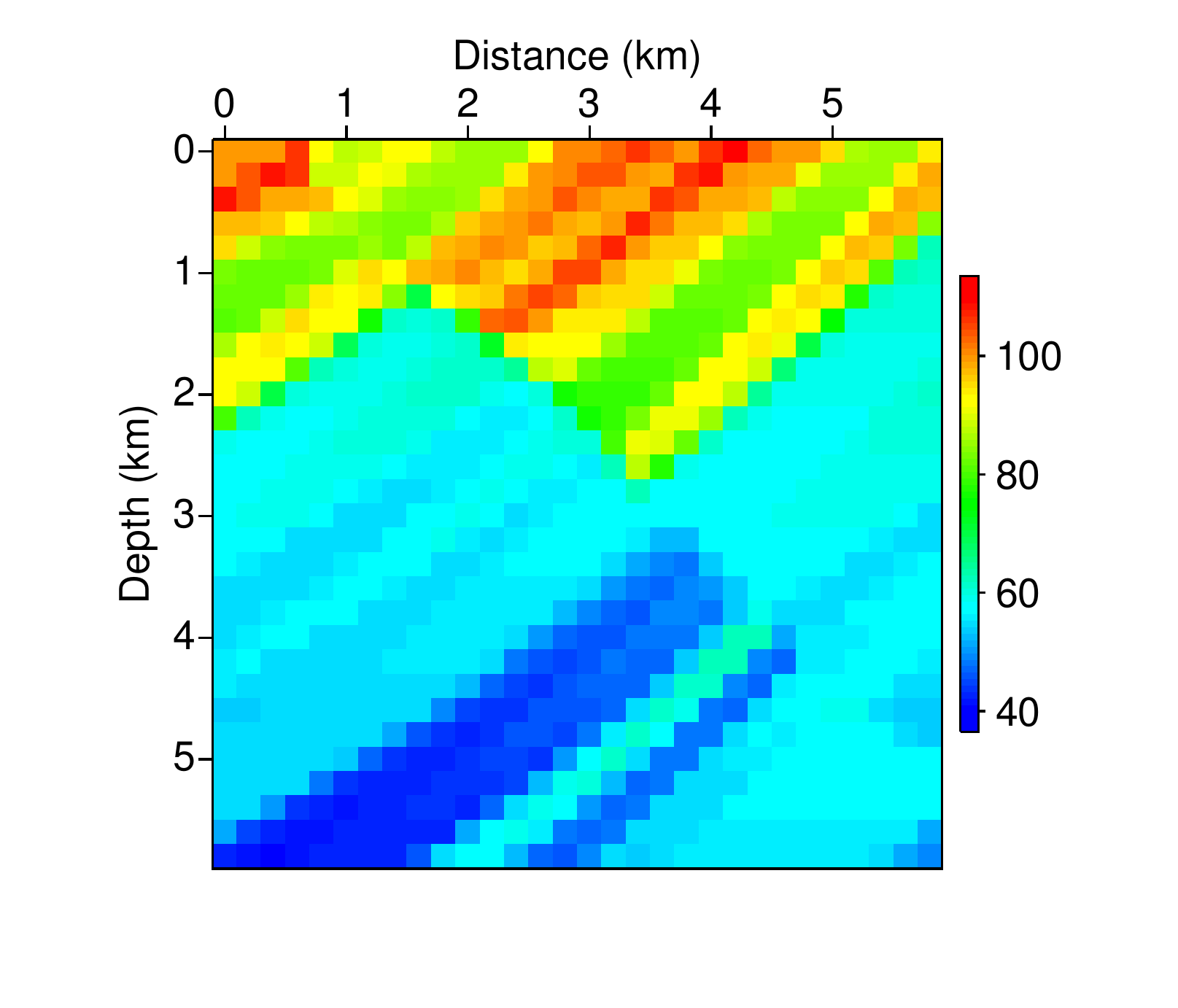}
}
\subfigure[]{
\label{fig:Marmousi2_basis_map_boundary}
\includegraphics[trim=50 50 20 0,width=0.41\textwidth]{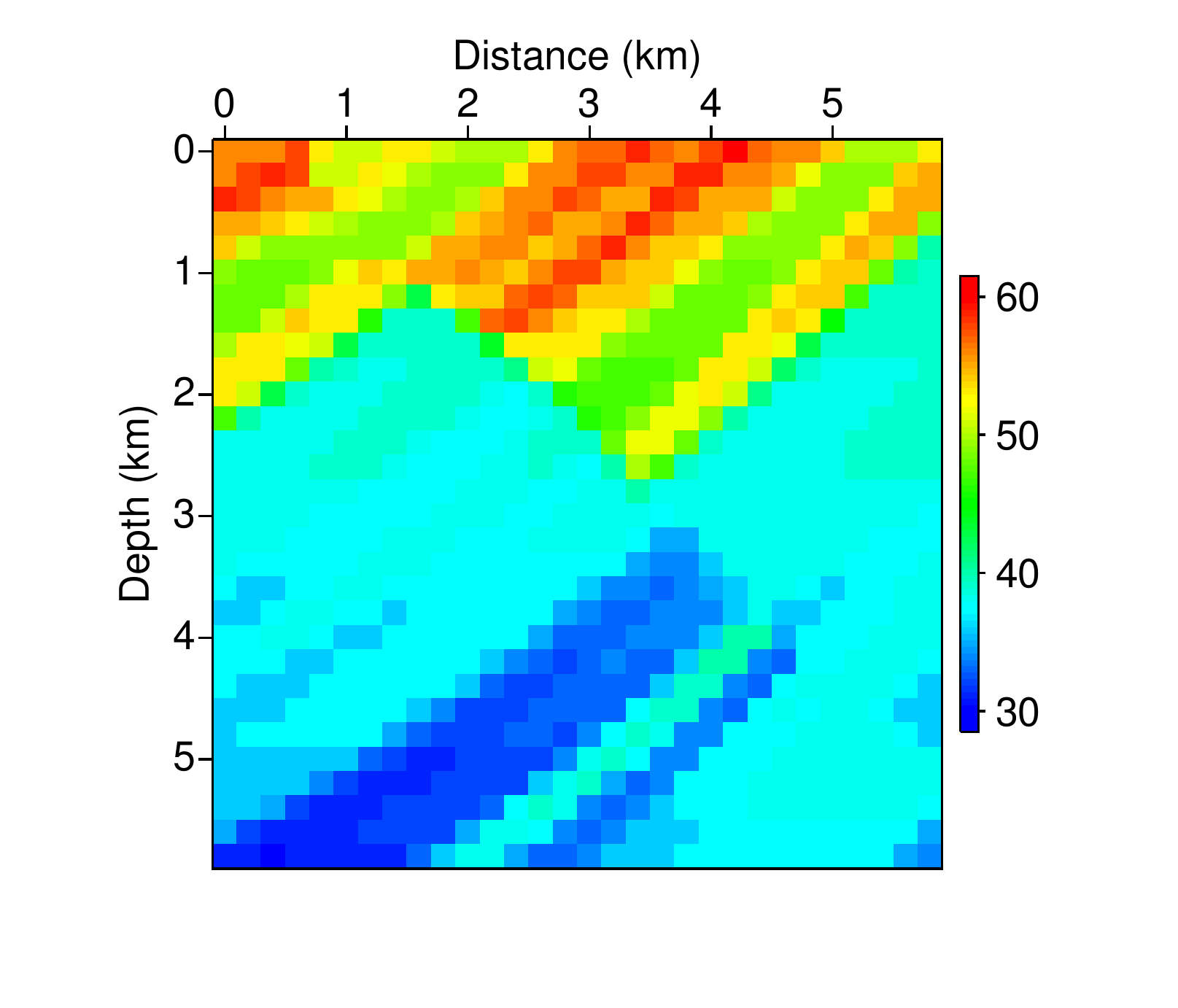}
}
\caption{Number of (a) interior and (b) boundary basis functions calculated based on the S-wave velocity.}
\end{figure}

We first calculate the number of multiscale basis functions based on the method we introduced in the Implementation part. The number of interior and boundary basis functions are shown in Figures \ref{fig:Marmousi2_basis_map_interior} and \ref{fig:Marmousi2_basis_map_boundary}. We can see that this map is consistent with our expectation that the near surface part, where the velocity is slower, needs more basis functions, and the very lower part of the model requires a much smaller number of basis functions. 

\begin{figure}
\centering
\subfigure[]{
\label{fig:Marmousi_u1_dg}
\includegraphics[trim=50 50 20 0,width=0.42\textwidth]{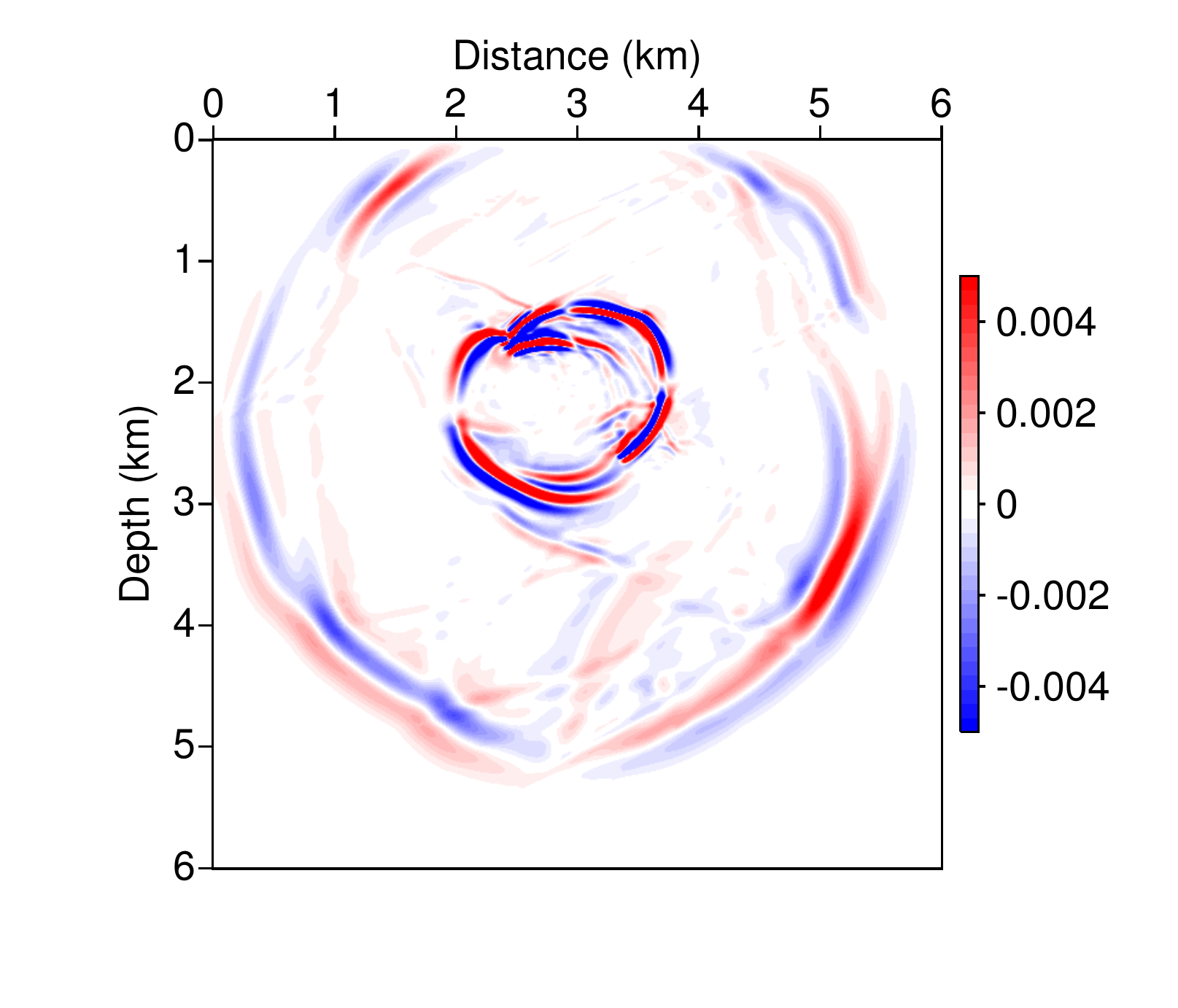}
}
\subfigure[]{
\label{fig:Marmousi_u1_70}
\includegraphics[trim=50 50 20 0,width=0.42\textwidth]{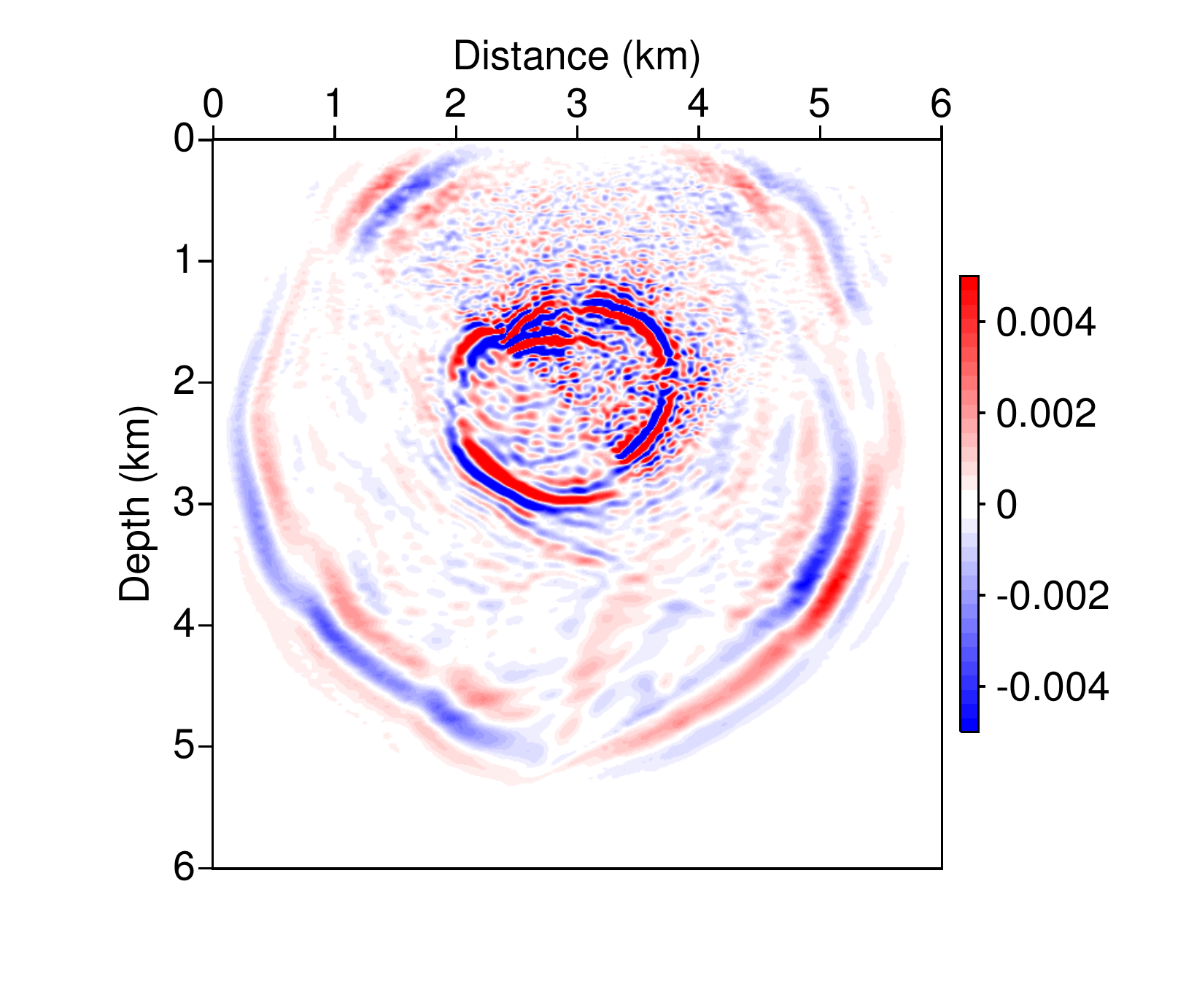}
}
\subfigure[]{
\label{fig:Marmousi_u1_170}
\includegraphics[trim=50 50 20 0,width=0.42\textwidth]{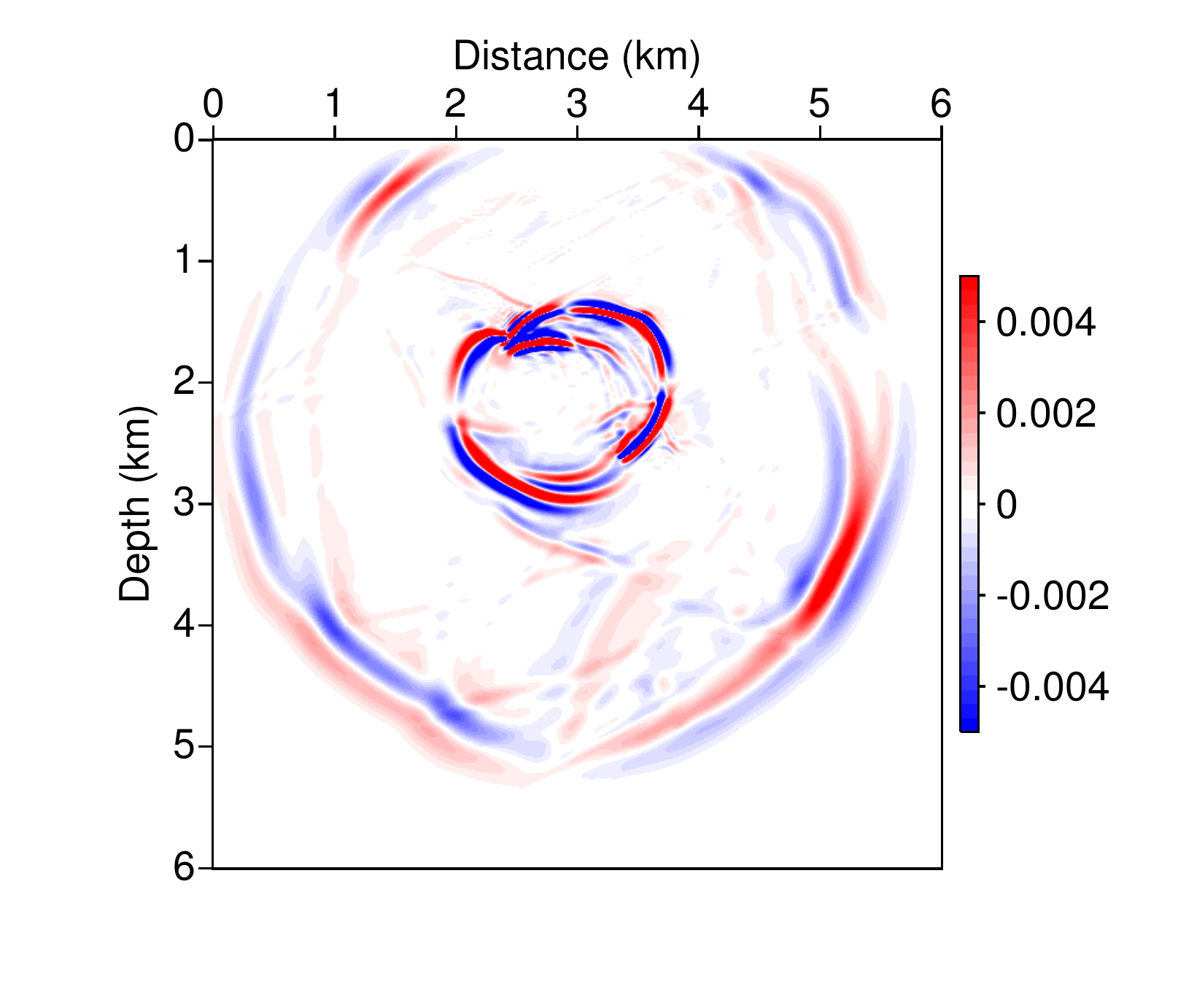}
}
\subfigure[]{
\label{fig:Marmousi_u1_adp}
\includegraphics[trim=50 50 20 0,width=0.42\textwidth]{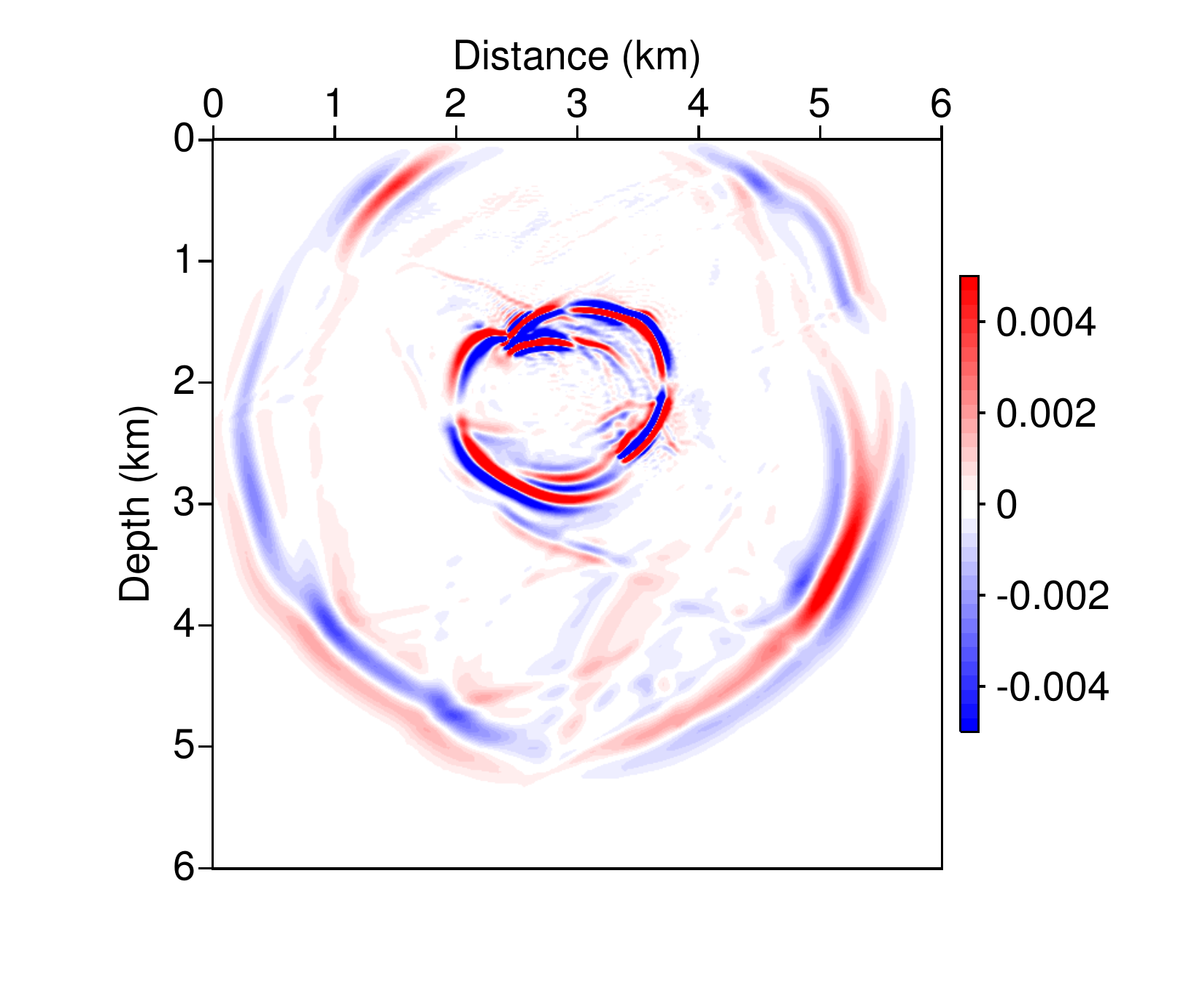}
}
\caption{$u_1$ wavefield snapshots at 1.5~s from (a) conventional DG-FEM, (b) DG-GMsFEM with a total of 70 type II basis functions and (c) DG-GMsFEM with a total of 170 type II basis functions, and (d) with adaptive assignment of a total of 70 to 170 type II basis functions. }
\end{figure}

We now compare the wavefield solutions. As in previous examples, we set the \oldrevision{DG-FEM} solution as the reference solution, and the $u_1$ wavefield snapshot at 1.5~s is shown in Figures \ref{fig:Marmousi_u1_dg}. The computation time of the DG-FEM solution is about 915~s. This computation time is longer than that in the second example, since \revision{the number of time steps is larger than that in the second example, and} we use thicker damping layers to absorb the outgoing waves at the boundaries. Meanwhile, Figures \ref{fig:Marmousi_u1_70} shows the $u_1$ wavefield snapshot solved from DG-GMsFEM with \revision{40 type II interior basis functions and 30 type II boundary} basis functions \revision{in each coarse element}. There \oldrevision{is} obvious dispersion of the S-wave due to the lack of adequate basis functions in these coarse blocks. The computation time is about 206~s, and the $L^2$-norm error of this multiscale solution with respect to DG-FEM solution is $6.74\times 10^{-1}$ (see Table \ref{tab:marmousi_dg}). We further adopt \revision{110 type II interior basis functions and 60 type II boundary} basis functions \revision{in each coarse element to get the GMsFEM solution} in Figure \ref{fig:Marmousi_u1_170}, which takes about 956~s to finish all the time steps in the simulation, with about $3.28\times 10^{-2}$ $L^2$-norm error. This solution is more accurate than the multiscale solution with a total of 70 type II basis functions, due to the suppression of S-wave dispersion with more basis functions, \revision{but the computational time is also more than 4 times longer}. We now implement the DG-GMsFEM with different numbers of basis functions in each coarse element according to the result shown in Figures \ref{fig:Marmousi2_basis_map_interior} and \ref{fig:Marmousi2_basis_map_boundary}, which takes about 424~s, with $L^2$-norm error of about $6.97\times 10^{-2}$. The error is larger than that using a total of 170 basis functions in all coarse elements, but still on the same level, and \revision{the simulation} uses \revision{less than} half of the computation time \revision{of GMsFEM with a total of 170 basis functions for each coarse element}. \revision{Similar \RLG{to} the computational time comparison, the DOF of the adaptive GMsFEM system is in the middle of the two GMsFEM systems, as shown in Table \ref{tab:marmousi_dg}.} We then know that by assigning different numbers of basis functions for each coarse element according to the magnitude of average time difference in the coarse block can help to reduce the computation time \revision{as well as the DOF}. 

\begin{table}
\centering
\begin{tabular}{|c|c|c|c|c|c|c|}
\hline 
Type & $m_{\text{II-interior}}$ & $m_{\text{II-boundary}}$ & DOF & $e(\mathbf{u})$ & $T_{\text{basis}}$ (s) & $T_{\text{modeling}}$ (s)\tabularnewline
\hline 
\hline 
DG-FEM reference & - & - & 2.88E6 & - & - & 914.69\tabularnewline
\hline 
DG-GMsFEM & 40 & 30 & 6.30E4 & 6.7385E-1 &  1168.73 & 206.21\tabularnewline
\hline 
DG-GMsFEM & 110 & 60 & 1.53E5 & 3.2768E-2 & 1925.42 & 956.21\tabularnewline
\hline 
Adaptive DG-GMsFEM & $40\sim 110$ & $30\sim 60$ & 9.72E4 & 6.9713E-2 &  1469.74 & 423.56\tabularnewline
\hline 
\end{tabular}
\caption{A comparison of DOF, computational time and accuracy between DG-GMsFEM and
adaptive DG-GMsFEM. }
\label{tab:marmousi_dg}
\end{table}

\section{Discussions}
In the proceeding sections, we have introduced the GMsFEM for elastic wave propagation in heterogeneous, anisotropic media in both CG and DG formulations and have explained how to construct the multiscale basis functions. The CG and DG formulations 
both have strengths and weaknesses.
The CG formulation does not require tuning of penalty parameters. 
The DG formulation, on the 
other hand,
requires a suitable choice of penalty parameters. Furthermore, \oldrevision{CG-GMsFEM requires the mesh discretization to be conformal, while} DG-GMsFEM naturally allows for non-conformal meshes, and is therefore more flexible in handling complex medium property variations in practice.
\oldrevision{The numerical results we have presented in the proceeding section show that both CG-GMsFEM and DG-GMsFEM can reduce the DOF of system as well as the computation time, and can produce approximate solutions for elastic wave equation in heterogeneous, anisotropic media with decent accuracy. However, the more multiscale basis functions used, the longer the computation time will become, since the increase of the number of basis functions will make the relevant matrices, such as the stiffness, mass and damping matrices, less sparse, which will introduce heavier computational burden for the time stepping, especially for CG-GMsFEM. }
\oldrevision{Therefore} in the future, we may investigate some possible improvements for CG-GMsFEM to reduce the computational burden in mass matrix \oldrevision{inversion}, so that the total time consumption of CG-GMsFEM could be reduced. \oldrevision{We also aim to develop parallel implementation of the DG-GMsFEM.}

\section{Conclusions}
We have developed a generalized multiscale finite-element method for elastic wave propagation in heterogeneous, anisotropic media, both in continuous Galerkin and discontinuous Galerkin formulations. This method is a significant extension of the similar methodology for acoustic wave equation. We explore two ways to compute the multiscale basis functions, one from linear elasticity eigenvalue problem, the other from two separate local spectral problems that are related to the boundaries and interior of coarse blocks. These multiscale basis functions can effectively capture the finer scale information of the model, and allow us to use much fewer degrees of freedom than the corresponding system of the modeling problem using conventional finite-element methods, to implement the seismic wave simulation. We designed \oldrevision{three} examples to verify the effectiveness of our method, and \oldrevision{found} that the accuracy of the multiscale solution is closely related to the number of bases used in modeling. The level of accuracy can be controlled by varying this number, which can be important in applications where a more approximate result is acceptable. 

\section{Acknowledgements}
The project is supported by Saudi Aramco. 
\revision{The research of E.T.C. is supported by Hong Kong RGC General Research Fund (Project: 400411)
and CUHK Direct Grant for Research 2014-15.} \RLG{R.L.G. and K.G. were partially supported by the U.S. Department of Energy under Grant No. DE-FG03-00ER15034.} \revision{We greatly appreciate the anonymous reviewers for their suggestions and comments that have greatly improved the quality of the manuscript.}

\bibliographystyle{abbrv}
\section*{References}
\bibliography{refs}

\section*{Appendix A: The solution to local problem \ref{eq:typeii_1} }
We illustrate in this appendix that the solutions for local problem \ref{eq:typeii_1} can be achieved in a \revision{single step}. To construct the boundary bases described in section 2.2.2., we need first solve $dp$ local problems in equation \ref{eq:typeii_1} with Dirichlet boundary conditions $\mathbf{u}=\boldsymbol{\delta}_j$, where $d$ is the number of dimensions, and $p$ is the number of boundary nodes. Without loss of generality, we take $d=2$. For the $j$-th boundary node, the corresponding local problem can be expressed in the matrix form as:
\begin{equation} \label{eq:original_local_problem}
\mathbf{A} \mathbf{x} = \mathbf{b},
\end{equation}
where 
\begin{align}
\mathbf{A}&=\int_K \boldsymbol{\sigma}(\boldsymbol{\gamma}):\boldsymbol{\varepsilon}(\boldsymbol{\eta}) d\mathbf{x}, \label{eq:coef_A_matrix}\\
\mathbf{b} &=(0,0,\cdots,1_{(j)},0_{(j)},\cdots,0,0)^{\mathrm{T}} \quad \text{or} \quad (0,0,\cdots,0_{(j)},1_{(j)},\cdots,0,0)^{\mathrm{T}},
\end{align}
where $1_{(j)}$ and $0_{(j)}$ represent the $\delta$ function for the $j$-th boundary node. As we have described, for each boundary node, we have two solutions, i.e., $\mathbf{x}_{j,1}$ and ${\mathbf{x}_{j,2}}$, and therefore we have to solve $2p$ local problems with equation \ref{eq:original_local_problem}. However, this series of local problems can be solved in a \revision{single step}. We replace the original expression \ref{eq:original_local_problem} with the following form:
\begin{equation}
\mathbf{A}\mathbf{X} = \mathbf{B}, \label{eq:new_local_problem}
\end{equation}
where $\mathbf{A}$ still follows the form in equation \ref{eq:coef_A_matrix}, but
\begin{equation}
\mathbf{B}=\left(
\begin{array}{cc|cc|c|cc|c|cc}
1_{(1)} & 0_{(1)} & 0 & 0 & \cdots & 0 & 0 & \cdots & 0 & 0\\
0_{(1)} & 1_{(1)} & 0 & 0 & \cdots & 0 & 0 & \cdots & 0 & 0\\
0 & 0 & 1_{(2)} & 0_{(2)} & \cdots & 0 & 0 & \cdots & 0 & 0\\
0 & 0 & 0_{(2)} & 1_{(2)} & \cdots & 0 & 0 & \cdots & 0 & 0\\
\vdots & \vdots & \vdots & \vdots & \cdots & \vdots & \vdots & \cdots & \vdots & \vdots\\
\vdots & \vdots & \vdots & \vdots & \cdots & 1_{(j)} & 0_{(j)} & \cdots & \vdots & \vdots\\
\vdots & \vdots & \vdots & \vdots & \cdots & 0_{(j)} & 1_{(j)} & \cdots & \vdots & \vdots\\
\vdots & \vdots & \vdots & \vdots & \cdots & \vdots & \vdots & \cdots & \vdots & \vdots\\
0 & 0 & 0 & 0 & \cdots & 0 & 0 & \cdots & 1_{(p)} & 0_{(p)}\\
0 & 0 & 0 & 0 & \cdots & 0 & 0 & \cdots & 0_{(p)} & 1_{(p)}
\end{array}
\right),
\end{equation}
and therefore every two columns of $\mathbf{B}$ represents the Dirichlet boundary condition for one boundary node. By solving the above equation with the built-in left division function in MATLAB, or any other mature linear algebra packages in C/C++ or FORTRAN, we get $\mathbf{X}$, and it is obvious that every two columns of matrix $\mathbf{X}$ correspond with the local problem solution of one boundary node. In this way, we get all the local problem solutions by solving only one equation instead of solving $2p$ local problems. For 3D problem, we can construct similar local problems and solve them in the same way. 

\end{document}